\documentclass[10pt]{amsart}

\usepackage{graphicx}

\usepackage{amscd}
\usepackage{amsthm}
\usepackage{amsmath}
\usepackage{xypic}
\newtheorem{thm}{Theorem}[section]
\newtheorem{df}{Definition}[section]
\newtheorem{lem}{Lemma}

\begin{document}

\title[A Generalization of Connes-Kreimer Hopf Algebra]
{A Generalization of Connes-Kreimer Hopf Algebra}

\author[J.Byun]{Jungyoon Byun}

\begin{abstract}
``Bonsai'' Hopf algebras, introduced here, are generalizations of
Connes-Kreimer Hopf algebras, which are motivated by Feynman
diagrams and renormalization. We show that we can find operad
structure on the set of bonsais. We introduce a new differential on
these bonsai Hopf algebras, which is inspired by the tree
differential. The cohomologies of these are computed here, and the
relationship of this differential with the appending operation $*$
of Connes-Kreimer Hopf algebras is investigated.

\end{abstract}

\maketitle

\section{Motivation}

In \cite{Kr}, Kreimer discovered a Hopf algebra structure on Feynman
diagrams and the forest formula of perturbative quantum field
theory. In \cite{CK}, Connes and Kreimer suggested the
representation of Feynman graphs using rooted tree diagrams and
represented the Hopf algebra structure with the notion of `cuts' of
tree diagrams. That expression is as following: let us consider a
Feynman diagram in $\phi ^3$ theory as in Figure \ref{Fig:1};

\begin{figure}[h]

\begin{picture}(50,65)

\put(0,0){\line(0,1){10}}
\put(0,10){\line(1,2){25}}
\put(0,10){\line(-1,2){25}}
\put(-25,60){\line(1,0){50}}
\put(-25,60){\line(-2,1){10}}
\put(25,60){\line(2,1){10}}

\end{picture}

\caption{}           \label{Fig:1}

\end{figure}

This is a 1-loop graph.
Now let us look at another loop having subloops in Figure \ref{Fig:2};

\begin{figure}[h]

\begin{picture}(100,130)

\put(0,0){\line(0,1){10}}
\put(0,10){\line(1,2){50}}
\put(0,10){\line(-1,2){50}}
\put(-50,110){\line(1,0){100}}
\put(-50,110){\line(-2,1){10}}
\put(50,110){\line(2,1){10}}
\put(-15,40){\line(1,0){30}}
\put(-40,90){\line(1,2){10}}
\put(40,90){\line(-1,2){10}}
\put(-5,20){\line(1,0){10}}

\put(0,90){1}
\put(-40,95){2}
\put(40,95){2}
\put(0,30){2}
\put(0,10){3}
\end{picture}

\caption{}        \label{Fig:2}

\end{figure}

In Kreimer's expression of a Feynman diagram using decorated rooted
trees (\cite{CK}), if the loop of Figure \ref{Fig:1} is labeled 1
(In Kreimer's context, this label indicates a specific shape of
loop. So, later in this paper, if every loop in Feynman diagram has
the same shape, we do not need this label.), the loop of Figure
\ref{Fig:2} is expressed as in Figure \ref{Fig:3}. In Figure
\ref{Fig:2}, the loops labeled 2 are immediate subloops of loop 1,
and the loop 3 is an immediate subloop of the lower loop 2 and not
of loop 1.

\begin{figure}[h]

\begin{picture}(60,60)(0,-50)

\put(0,0){\circle*{2}}
\put(0,0){1}
\put(0,0){\line(0,-1){20}}
\put(0,-20){\circle*{2}}
\put(0,-26){1}
\put(0,0){\line(1,-2){10}}
\put(10,-20){\circle*{2}}
\put(10,-26){1}
\put(0,0){\line(-1,-2){10}}
\put(-10,-20){\circle*{2}}
\put(-10,-26){1}
\put(-10,-20){\line(0,-1){20}}
\put(-10,-40){\circle*{2}}
\put(-10,-46){1}
\end{picture}

\caption{}             \label{Fig:3}

\end{figure}

In Connes and Kreimer's context, we call a connected rooted tree,
which corresponds to a connected Feynman graph, a {\em tree}
and we call a diagram of trees
having more than one connected component a {\em forest}.

The Connes-Kreimer Hopf Algebra ${\mathcal H}_K$
is a Hopf algebra with forests of rooted trees as basis elements
(See section \ref{sec:bonsai} for details).

In Figure \ref{Fig:2}, the author observed that
the biggest loop cannot include
more than 3  immediate subloops of the shape of Figure \ref{Fig:1}.
Hence, in the tree diagram, the vertex labeled 1 cannot have
more than 3 subsidiary vertices labeled 1,
and so the rooted tree of Figure \ref{Fig:3} cannot have a ramification number
(or arity, branch number) greater than 3 at the root.

{\em So, in the $\phi^3$ theory in which the only allowed loop is that of
Figure \ref{Fig:1},
the corresponding tree diagrams are forbidden
to have ramification number greater than 3.}
The theory of such ramification number bounded trees is our main interest
in this paper.
We will call them {\em bonsais}.

For a more precise description of Feynman diagrams,
 let us consider the positions of subloops in a loop.
For the loops having subloops like Figure \ref{Fig:2}, in the
context of \cite{CK2}, sometimes we need to indicate which subloop
is shrinking and what position  is available for a subloop. For
that, we label each corner of the loop in Figure \ref{Fig:11} and
change that loop into a tree as shown in Figure \ref{Fig:11}, by
expressing a subloop as a subsidiary vertex in the tree diagram and
attaching the labels representing the subloop positions to the
edges.

\begin{figure}[h]

\begin{picture}(120,65)

\put(0,0){\line(0,1){10}}
\put(0,10){\line(1,2){25}}
\put(0,10){\line(-1,2){25}}
\put(-25,60){\line(1,0){50}}
\put(-25,60){\line(-2,1){10}}
\put(-15,50){1}
\put(25,60){\line(2,1){10}}
\put(-5,20){\line(1,0){10}}
\put(-20,50){\line(1,2){5}}
\put(15,50){2}
\put(0,20){3}

\put(50,30){$\rightarrow$}

\put(90,40){\circle*{2}}
\put(90,40){1}
\put(90,40){\line(1,-2){10}}
\put(100,20){\circle*{2}}
\put(100,14){1}
\put(95,30){3}
\put(90,40){\line(-1,-2){10}}
\put(80,20){\circle*{2}}
\put(80,14){1}
\put(85,30){1}

\end{picture}
\caption{}               \label{Fig:11}
\end{figure}

The tree diagram of Figure \ref{Fig:11}
asssigns the numbers of the occupied corners
in the big loop to edges of the tree.
Note that there is no edge numbered 2.
This means there is no subloop on the corner 2.
We easily see that, in this expression, the left tree of Figure \ref{Fig:12}
is allowed but the right tree of Figure \ref{Fig:12} is not.

\begin{figure}[h]

\begin{picture}(120,60)(0,-40)

\put(0,0){\circle*{2}}
\put(0,0){1}
\put(0,0){\line(1,-1){20}}
\put(20,-20){\circle*{2}}
\put(20,-26){2}
\put(15,-15){3}
\put(0,0){\line(-1,-1){20}}
\put(-20,-20){\circle*{2}}
\put(-20,-26){1}
\put(-15,-15){1}
\put(0,0){\line(0,-1){20}}
\put(0,-20){\circle*{2}}
\put(0,-26){1}
\put(0,-15){2}
\put(-20,-40){(allowed)}

\put(100,0){\circle*{2}}
\put(100,0){1}
\put(100,0){\line(1,-1){20}}
\put(120,-20){\circle*{2}}
\put(120,-26){2}
\put(115,-15){1}
\put(100,0){\line(-1,-1){20}}
\put(80,-20){\circle*{2}}
\put(80,-26){1}
\put(85,-15){1}
\put(100,0){\line(0,-1){20}}
\put(100,-20){\circle*{2}}
\put(100,-26){1}
\put(100,-15){2}
\put(80,-40){(forbidden)}

\end{picture}

\caption{}              \label{Fig:12}

\end{figure}

\section{Main results}

\begin{df}
We define a new Hopf algebra which has the same operations as in the
Connes-Kreimer Hopf Algebra, and whose basis elements are forests of
{\em trees having ramification numbers at each vertex smaller than
or equal to $m$ and under each vertex $v$, each subsidiary edge of
$v$ has labels from 1,2,...,$m$ without duplication}. We call this
Hopf algebra the {\em $m$-bonsai Hopf algebra} ${\mathcal H}_{b,m}$.
In ${\mathcal H}_{b,m}$, each tree is called a {\em $m$-bonsai}.
\end{df}

As in \cite{CK}, we can show that

\begin{thm}
 ${\mathcal H}_{b,m}$ is a Hopf algebra.
\end{thm}

As in \cite{CK}, when we define an {\em appending operation }
\begin{eqnarray}
T * T'=\Sigma (\mbox{a bonsai obtained by connecting the root of  $T$}\notag\\
         \mbox{ to a vertex $v$ of $T'$ with one edge, where}\notag\\
         \mbox{ the added edge has every possible label}), \notag
\end{eqnarray}
(An example of the $*$ operation is in Figure \ref{Fig:4'})

\begin{thm}
The operation $*$ is pre-Lie, and we have
${\mathcal H}_{b,m}={\mathcal U}({\mathcal L})^{\vee}$,
where $V^{\vee}$ is the dual of $V$.
\end{thm}

In the $m$-bonsai Hopf algebra, the set of $m$-bonsais has a
structure of an operad, thus there is a natural analog of the tree
differential (as in \cite{MSS}). We call it the {\em
vertex-appending differential} $\partial$ (Definition
\ref{Def:vadiff}).

Then, mainly using the K{\" u}nneth theorem,
we can calculate the cohomology groups of $\partial$ as:

\begin{thm}
In $m$-bonsai,
\begin{eqnarray}
H^i({\mathcal H}_{b,m},\partial)= \left\{ \begin{array}{ll}
k^{\frac{(mn)!}{((m-1)n+1)!n!}} & \mbox{if } i=(2m-1)n+1, n \geq 0 \\
0                               & \mbox{otherwise.}
\end{array} \right. \notag
\end{eqnarray}
\end{thm}

Here, $\frac{(mn)!}{((m-1)n+1)!n!}$ is the number of rooted trees
consisting of $n$ of $m$-corollas, which is called the `$m$-Catalan
number'. When $m=2$, this number is just the Catalan number.
Representatives of $H^i$ are $\Sigma($a bonsai obtained by appending
edges to all tips of a rooted tree every vertex of which except tips
has ramification number $m$, one edge to each tip, with every
possible label$)$.

When we define $T_1 *_1 T_2$,
which is the deviation from $\partial$ being a derivation of $*$ as
\begin{eqnarray}
T_1 *_1 T_2 =
(\partial T_1) * T_2 + T_1 * (\partial T_2) - \partial(T_1 * T_2) , \notag
\end{eqnarray}
we have

\begin{thm}
With coefficients mod 2,

\begin{eqnarray}
T_1 *_1 T_2
&=&\Sigma(\mbox{a bonsai obtained by connecting a tip $v$ of $T_2$}\notag\\
& & \mbox{and the root of $T_1$ with one-edge, and attaching }\notag\\
& & \mbox{an edge to $v$, added edge have every admissible label})\notag\\
&+&\Sigma(\mbox{a bonsai obtained by connecting a non-tip of $T_2$}\notag\\
& & \mbox{and the root of  $T_1$ with two-edge ladder, having} \notag\\
& & \mbox{every possible label}) \notag
\end{eqnarray}

and

\begin{eqnarray}
\partial(T_1 *_1 T_2) = (\partial T_1) *_1 T_2 + T_1 *_1 (\partial T_2).
\notag
\end{eqnarray}

\end{thm}
We have an example in Figure \ref{Fig:7'}.

\begin{figure}[h]

\begin{picture}(200,120)(60,-90)

\put(0,0){\circle*{2}}
\put(0,0){\line(-1,-3){5}}
\put(0,0){\line(0,-1){15}}
\put(0,0){\line(1,-3){5}}
\put(-5,-15){\circle*{2}}
\put(0,-15){\circle*{2}}
\put(5,-15){\circle*{2}}
\put(5,-10){$*_1$}
\put(15,0){\circle*{2}}
\put(15,0){\line(0,-1){15}}
\put(15,-15){\circle*{2}}
\put(15,-10){1}

\put(25,-10){=}

\put(35,15){\circle*{2}}
\put(35,15){\line(0,-1){15}}
\put(35,5){1}
\put(35,0){\circle*{2}}
\put(35,0){\line(0,-1){15}}
\put(35,-10){1}
\put(35,0){\line(1,-1){15}}
\put(50,-15){\circle*{2}}
\put(45,-10){2}
\put(35,-15){\circle*{2}}
\put(35,-15){\line(-1,-3){5}}
\put(35,-15){\line(0,-1){15}}
\put(35,-15){\line(1,-3){5}}
\put(30,-30){\circle*{2}}
\put(35,-30){\circle*{2}}
\put(40,-30){\circle*{2}}

\put(55,-10){+}

\put(65,15){\circle*{2}}
\put(65,15){\line(0,-1){15}}
\put(65,5){1}
\put(65,0){\circle*{2}}
\put(65,0){\line(0,-1){15}}
\put(65,-10){1}
\put(65,0){\line(1,-1){15}}
\put(80,-15){\circle*{2}}
\put(75,-10){3}
\put(65,-15){\circle*{2}}
\put(65,-15){\line(-1,-3){5}}
\put(65,-15){\line(0,-1){15}}
\put(65,-15){\line(1,-3){5}}
\put(60,-30){\circle*{2}}
\put(65,-30){\circle*{2}}
\put(70,-30){\circle*{2}}

\put(85,-10){+}

\put(95,15){\circle*{2}}
\put(95,15){\line(0,-1){15}}
\put(95,5){1}
\put(95,0){\circle*{2}}
\put(95,0){\line(0,-1){15}}
\put(95,-10){2}
\put(95,0){\line(1,-1){15}}
\put(110,-15){\circle*{2}}
\put(105,-10){1}
\put(95,-15){\circle*{2}}
\put(95,-15){\line(-1,-3){5}}
\put(95,-15){\line(0,-1){15}}
\put(95,-15){\line(1,-3){5}}
\put(90,-30){\circle*{2}}
\put(95,-30){\circle*{2}}
\put(100,-30){\circle*{2}}

\put(115,-10){+}

\put(125,15){\circle*{2}}
\put(125,15){\line(0,-1){15}}
\put(125,5){1}
\put(125,0){\circle*{2}}
\put(125,0){\line(0,-1){15}}
\put(125,-10){2}
\put(125,0){\line(1,-1){15}}
\put(140,-15){\circle*{2}}
\put(135,-10){3}
\put(125,-15){\circle*{2}}
\put(125,-15){\line(-1,-3){5}}
\put(125,-15){\line(0,-1){15}}
\put(125,-15){\line(1,-3){5}}
\put(120,-30){\circle*{2}}
\put(125,-30){\circle*{2}}
\put(130,-30){\circle*{2}}

\put(145,-10){+}

\put(155,15){\circle*{2}}
\put(155,15){\line(0,-1){15}}
\put(155,5){1}
\put(155,0){\circle*{2}}
\put(155,0){\line(0,-1){15}}
\put(155,-10){3}
\put(155,0){\line(1,-1){15}}
\put(170,-15){\circle*{2}}
\put(165,-10){1}
\put(155,-15){\circle*{2}}
\put(155,-15){\line(-1,-3){5}}
\put(155,-15){\line(0,-1){15}}
\put(155,-15){\line(1,-3){5}}
\put(150,-30){\circle*{2}}
\put(155,-30){\circle*{2}}
\put(160,-30){\circle*{2}}

\put(175,-10){+}

\put(185,15){\circle*{2}}
\put(185,15){\line(0,-1){15}}
\put(185,5){1}
\put(185,0){\circle*{2}}
\put(185,0){\line(0,-1){15}}
\put(185,-10){3}
\put(185,0){\line(1,-1){15}}
\put(200,-15){\circle*{2}}
\put(195,-10){2}
\put(185,-15){\circle*{2}}
\put(185,-15){\line(-1,-3){5}}
\put(185,-15){\line(0,-1){15}}
\put(185,-15){\line(1,-3){5}}
\put(180,-30){\circle*{2}}
\put(185,-30){\circle*{2}}
\put(190,-30){\circle*{2}}

\put(25,-70){+}

\put(50,-45){\circle*{2}}
\put(50,-45){\line(-1,-2){10}}
\put(40,-65){\circle*{2}}
\put(45,-55){1}
\put(50,-45){\line(1,-2){10}}
\put(60,-65){\circle*{2}}
\put(55,-55){2}
\put(60,-65){\line(0,-1){20}}
\put(60,-85){\circle*{2}}
\put(62,-81){1}
\put(60,-85){\line(-1,-3){5}}
\put(60,-85){\line(0,-1){15}}
\put(60,-85){\line(1,-3){5}}
\put(55,-100){\circle*{2}}
\put(60,-100){\circle*{2}}
\put(65,-100){\circle*{2}}

\put(65,-70){+}

\put(90,-45){\circle*{2}}
\put(90,-45){\line(-1,-2){10}}
\put(80,-65){\circle*{2}}
\put(85,-55){1}
\put(90,-45){\line(1,-2){10}}
\put(100,-65){\circle*{2}}
\put(95,-55){2}
\put(100,-65){\line(0,-1){20}}
\put(100,-85){\circle*{2}}
\put(102,-81){2}
\put(100,-85){\line(-1,-3){5}}
\put(100,-85){\line(0,-1){15}}
\put(100,-85){\line(1,-3){5}}
\put(95,-100){\circle*{2}}
\put(100,-100){\circle*{2}}
\put(105,-100){\circle*{2}}

\put(105,-70){+}

\put(130,-45){\circle*{2}}
\put(130,-45){\line(-1,-2){10}}
\put(120,-65){\circle*{2}}
\put(125,-55){1}
\put(130,-45){\line(1,-2){10}}
\put(140,-65){\circle*{2}}
\put(135,-55){2}
\put(140,-65){\line(0,-1){20}}
\put(140,-85){\circle*{2}}
\put(142,-81){3}
\put(140,-85){\line(-1,-3){5}}
\put(140,-85){\line(0,-1){15}}
\put(140,-85){\line(1,-3){5}}
\put(135,-100){\circle*{2}}
\put(140,-100){\circle*{2}}
\put(145,-100){\circle*{2}}

\put(145,-70){+}

\put(170,-45){\circle*{2}}
\put(170,-45){\line(-1,-2){10}}
\put(160,-65){\circle*{2}}
\put(165,-55){1}
\put(170,-45){\line(1,-2){10}}
\put(180,-65){\circle*{2}}
\put(175,-55){3}
\put(180,-65){\line(0,-1){20}}
\put(180,-85){\circle*{2}}
\put(182,-81){1}
\put(180,-85){\line(-1,-3){5}}
\put(180,-85){\line(0,-1){15}}
\put(180,-85){\line(1,-3){5}}
\put(175,-100){\circle*{2}}
\put(180,-100){\circle*{2}}
\put(185,-100){\circle*{2}}

\put(185,-70){+}

\put(210,-45){\circle*{2}}
\put(210,-45){\line(-1,-2){10}}
\put(200,-65){\circle*{2}}
\put(205,-55){1}
\put(210,-45){\line(1,-2){10}}
\put(220,-65){\circle*{2}}
\put(215,-55){3}
\put(220,-65){\line(0,-1){20}}
\put(220,-85){\circle*{2}}
\put(222,-81){2}
\put(220,-85){\line(-1,-3){5}}
\put(220,-85){\line(0,-1){15}}
\put(220,-85){\line(1,-3){5}}
\put(215,-100){\circle*{2}}
\put(220,-100){\circle*{2}}
\put(225,-100){\circle*{2}}

\put(225,-70){+}

\put(250,-45){\circle*{2}}
\put(250,-45){\line(-1,-2){10}}
\put(240,-65){\circle*{2}}
\put(245,-55){1}
\put(250,-45){\line(1,-2){10}}
\put(260,-65){\circle*{2}}
\put(255,-55){3}
\put(260,-65){\line(0,-1){20}}
\put(260,-85){\circle*{2}}
\put(262,-81){3}
\put(260,-85){\line(-1,-3){5}}
\put(260,-85){\line(0,-1){15}}
\put(260,-85){\line(1,-3){5}}
\put(255,-100){\circle*{2}}
\put(260,-100){\circle*{2}}
\put(265,-100){\circle*{2}}

\end{picture}

\caption{}                 \label{Fig:7'}

\end{figure}

Now, let us consider another Hopf algebra,
having the same operations
but the trees having ramification numbers
at each vertex smaller than or equal to $m$
but no edge labels,
and call it {\em clear-edged $m$-bonsai Hopf algebra} ${\mathcal H}_{c,m}$.
(In other words, a clear-edged $m$-bonsai
is an $m$-bonsai without edge labels.)

Clear-edged $m$-bonsai Hopf algebras still represent Feynman graphs,
actually more physically relevant, and also appear in the tree
diagrams of ``open-closed homotopy algebra(OCHA)''(\cite{KS}).

Then we can define the {\em vertex-appending differential} similarly to the
case of $m$-bonsai.
For example, in planar clear-edged 3-bonsai,
we can get an example like Figure \ref{Fig:8'}.

\begin{figure}[h]

\begin{picture}(200,140)(0,-130)

\put(10,-10){\circle*{4}}

\put(25,-10){$\rightarrow \qquad 0$}

\put(10,-40){\circle*{4}}
\put(10,-40){\line(0,-1){20}}

\put(10,-60){\circle*{4}}

\put(25,-50){$\rightarrow$}

\put(60,-40){\circle*{4}}
\put(60,-40){\line(-1,-2){10}}
\put(50,-60){\circle{4}}

\put(60,-40){\line(1,-2){10}}
\put(70,-60){\circle*{4}}

\put(80,-50){$-$}

\put(100,-40){\circle*{4}}
\put(100,-40){\line(-1,-2){10}}
\put(90,-60){\circle*{4}}

\put(100,-40){\line(1,-2){10}}
\put(110,-60){\circle{4}}

\put(115,-50){$= \quad 0$}

\put(10,-80){\circle*{4}}
\put(10,-80){\line(-1,-2){10}}
\put(0,-100){\circle*{4}}

\put(10,-80){\line(1,-2){10}}
\put(20,-100){\circle*{4}}

\put(25,-90){$\rightarrow$}

\put(60,-80){\circle*{4}}
\put(60,-80){\line(-1,-2){10}}
\put(50,-100){\circle{4}}
\put(60,-80){\line(0,-1){20}}
\put(60,-100){\circle*{4}}
\put(60,-80){\line(1,-2){10}}
\put(70,-100){\circle*{4}}

\put(80,-90){$-$}

\put(100,-80){\circle*{4}}
\put(100,-80){\line(-1,-2){10}}
\put(90,-100){\circle*{4}}
\put(100,-80){\line(0,-1){20}}
\put(100,-100){\circle{4}}
\put(100,-80){\line(1,-2){10}}
\put(110,-100){\circle*{4}}

\put(120,-90){$+$}

\put(140,-80){\circle*{4}}
\put(140,-80){\line(-1,-2){10}}
\put(130,-100){\circle*{4}}
\put(140,-80){\line(0,-1){20}}
\put(140,-100){\circle*{4}}
\put(140,-80){\line(1,-2){10}}
\put(150,-100){\circle{4}}

\put(160,-90){$=$}

\put(180,-80){\circle*{4}}
\put(180,-80){\line(-1,-2){10}}
\put(170,-100){\circle*{4}}
\put(180,-80){\line(0,-1){20}}
\put(180,-100){\circle*{4}}
\put(180,-80){\line(1,-2){10}}
\put(190,-100){\circle*{4}}

\put(10,-120){\circle*{4}}
\put(10,-120){\line(-1,-2){10}}
\put(0,-140){\circle*{4}}

\put(10,-120){\line(0,-1){20}}
\put(10,-140){\circle*{4}}

\put(10,-120){\line(1,-2){10}}
\put(20,-140){\circle*{4}}

\put(25,-130){$\rightarrow \qquad 0$}

\end{picture}

\caption{}                 \label{Fig:8'}

\end{figure}

The cohomology groups of the vertex-appending differential in
clear-edged bonsai are not as easy to calculate as in $m$-bonsai and
we have just partial results as follows:

We first define a specific form of bonsai $S$ called ``seedling''
(Definition \ref{Def:clearseedling}),
and then we define the complexes $({\mathbf C}^{S,*},\partial)$
Then we can show that the cohomology of the whole bonsai complex
$H^i = \bigoplus H^i(S)$, where the sum is over all seedlings.

By the definition of seedling,
 when $S_1$, $S_2$,..., $S_n$ are seedlings, the new bonsai $S$
obtained by appending the roots of each $S_i$'s to a single new root
is a seedling again. There is an example in Figure \ref{Fig:12'}

\begin{figure}[h]

\begin{picture}(100,50)(60,-50)

\put(170,-10){$S=$}

\put(220,0){\line(0,-1){20}}
\put(220,-20){\circle*{3}}
\put(220,0){\circle*{3}}
\put(220,0){\line(-1,-1){20}}
\put(200,-20){\circle*{3}}
\put(220,0){\line(1,-1){20}}
\put(240,-20){\circle*{3}}

\put(200,-20){\line(0,-1){30}} \put(200,-35){\circle*{3}}
\put(200,-50){\circle*{3}}

\put(220,-20){\line(0,-1){15}}
\put(220,-35){\circle*{3}}

\put(240,-20){\line(0,-1){15}}
\put(240,-35){\circle*{3}}

\put(0,-10){$S_1=$}
\put(30,0){\circle*{3}}
\put(30,0){\line(0,-1){30}}
\put(30,-15){\circle*{3}}
\put(30,-30){\circle*{3}}
\put(55,-10){$S_2=$}
\put(85,0){\circle*{3}}
\put(85,0){\line(0,-1){20}}
\put(85,-20){\circle*{3}}
\put(105,-10){$S_3=$}
\put(135,0){\circle*{3}}
\put(135,0){\line(0,-1){20}}
\put(135,-20){\circle*{3}}

\end{picture}

\caption{}                 \label{Fig:12'}

\end{figure}

On the way to find the relationship of $H(S)$ and
$H(S_1)$,...,$H(S_n)$, we have a new definition of a bonsai called
{\em grafting seedling}$gs(n;T_1,...,T_{n+1};S_1,...,S_n)$
(Definition \ref{def:gs}), a complex $\{{\mathbf
K}^i(gs(n;T_1,...,T_{n+1};S_1,...,S_n))\}$ (Definition \ref{def:K})
and

\begin{thm}
When $H^i(gs(n;T_1,...,T_{n+1};S_1,...,S_n))$ is the $i$-th cohomology
group of the complex ${\mathbf K}^i(gs(n;T_1,...,T_{n+1};S_1,...,S_n))$,
the $i$-th cohomology group $H^i$ of clear-edged $m$-bonsai is
$H^i = \underset{\mbox{$S$ is a grafting seedling}}{\bigoplus} H^i(S)$,
and
\begin{eqnarray}
H^{i}(gs(n;T_1,...,T_{n+1};S_1,...,S_n))
=\bigoplus_{j_1+...+j_n=i-m} [H^{i_1}(S_1)\otimes ... \otimes
H^{i_n}(S_n)]^{\oplus N} \notag
\end{eqnarray}
where $N$ is combinatorially all determined and
\begin{eqnarray}
P=degT_1+...+degT_n. \notag
\end{eqnarray}
\end{thm}

Finally, as in the case of $m$-bonsai, we have again

\begin{thm}
For any clear-edged $m$-bonsai $T_1$ and $T_2$, $\partial(T_1 *_1
T_2) = (\partial T_1) *_1 T_2 + T_1 *_1 \partial T_2$ with
coefficients mod 2.
\end{thm}

\section{Bonsai Hopf Algebra} \label{sec:bonsai}

As seen in the last section, loops in Feynman diagrams
of a specific theory have a maximum number
of immediate subloops.
In the example of the last section, the maximum number is 3 and each edge of
the tree diagram corresponding to a Feynman diagram has label 1, 2 or 3.

From this motivation, we define

\begin{df}
A {\em simple cut} of rooted tree is a cut of edges such that at any
vertex of T, the path between it and the root has at most one cut,
$P_c(T)$ is the part of $T$ cut off by $c$ and $R_c(T)$ is the part
of $T$ remaining after cut $c$.
\end{df}

\begin{df}
Let ${\mathcal H}_{b,m}$ be the vector space having
as its basis the forests consisting of trees
whose vertices have ramification numbers $\leq m$
and whose edges are labeled by numbers in $1,2,...,m$.

We equip this ${\mathcal H}_{b,m}$ with operations, as in \cite{CK},
\begin{align}
\mbox{(multiplication)   } &m(T_1T_2...T_m,S_1S_2...S_n)=T_1...T_mS_1...S_n
                         \label{E:mult}\\
                       &\mbox{  ($T_i$, $S_j$ are trees, $m$ is commutative)}
                         \notag \\
\mbox{(diagonal)   }       &\Delta(T)=T \otimes 1
                         + \sum_c P_c(T) \otimes R_c(T)
                         \mbox{    ($T$ is a tree)}  \label{E:diag}   \\
                       &\Delta(T_1...T_n) = \Delta(T_1)...\Delta(T_n) \notag\\
\mbox{(antipode)   }   &S(T)=-\sum_c S(P_c(T))R_c(T)
                         \mbox{    ($T$ is a tree)}  \label{E:anti}  \\
                       &S(1)=1, S(v)= -v,
                         \mbox{    (where $v$ is the one-vertex bonsai)}
                         \notag\\
                       &S(T_1...T_n)=S(T_n)...S(T_1)       \notag
\end{align}
where $c$ runs over simple cuts of $T$ including $c=\emptyset$, and
a counit function
\begin{eqnarray}
\quad \epsilon: {\mathcal H}_{b,m} \rightarrow
                {\mathcal H}_{b,m}
\mbox{  such that } \epsilon(1)=1 \mbox{ and } \epsilon(f)=0
\mbox{ if a forest $f \ne 1$.} \label{E:epsi}
\end{eqnarray}
We call the rooted tree $T$ an {\em $m$-bonsai} and
${\mathcal H}_{b,m}$ the {\em $m$-bonsai Hopf algebra}.
\end{df}
It will be proved in the next section that this vector space
${\mathcal H}_{b,m}$ is actually a Hopf algebra.

\begin{df}
Sometimes we will ignore the positions of subloops in Feynman graphs
and use trees without labels on edges.
Then the trees in the forests corresponding to the Feynman graphs
have no label on their edges.
In this case, we denote the vector space having a basis consisting
of forests of planar trees as
${\mathcal H}_{c,m}$, where $m$
is the maximum of ramification number of each vertex in the trees of the
forests in ${\mathcal H}_{c,m}$.
We equip ${\mathcal H}_{c,m}$ with operations
\eqref{E:mult}-\eqref{E:epsi} in Definition 3.2.
Then we call that Hopf algebra
{\em planar clear-edged $m$-bonsai Hopf algebra}.
\end{df}

\section{Basic Results Related to Hopf Algebras }

In order to develop a basic theorem related to Lie algebras,
let us adapt \cite{CK} to our bonsai language and get some basic results.

In order to prove that our Hopf algebras are actually Hopf algebras
and derive some algebraic results, let us give another expression
of bonsai Hopf algebras and their elements.

First we give

\begin{df}
For a bonsai $T$, $deg(T)$ is the number of vertices of $T$.
\end{df}

For each $p$, we let $\Sigma_p$ be the set of bonsai $T$ such that
$deg(T) \leq p$ with the restriction of ramification numbers by $m$,
and let ${\mathcal H}_p$ be the polynomial commutative algebra
generated by the symbols,

\begin{eqnarray}
\delta_T, \qquad T \in \Sigma_p.              \label{E:1}
\end{eqnarray}

We define a coproduct on ${\mathcal H}_p$ by,

\begin{eqnarray}
\Delta\delta_T = \delta_T \otimes 1
+ \sum_{c}(\prod_{P_c(T)} \delta_{T_i}) \otimes \delta _{R_c (T)}, \label{E:2}
\end{eqnarray}
where the last sum is over all simple cuts including $c=\emptyset$,
while the product $\prod_{P_c(T)}$ is over the cut branches.
Sometimes $\prod_{P_c(T)}\delta_{T_i}$ is written $\delta_{P_c(T)}$.
The antipodal map $S$ is given as

\begin{eqnarray}
S(1)=1
\end{eqnarray}
\begin{eqnarray}
S(\delta_T)= -\delta_T - \sum_{\mbox{simple cuts $c \ne \emptyset$ of $T$}}
              S(\delta_{P_c(T)}) \delta _{R_c(T)}.
\end{eqnarray}

We let ${\mathcal H}_{b,m}=\bigcup {\mathcal H}_p$
and extend the maps on ${\mathcal H}_p$ to ${\mathcal H}_{b,m}$.\\
Coassociativity of $\Delta$ and $m((S \otimes id)\Delta)=\epsilon$
can be shown just by introducing the notion of double cuts of $T$.
But in order to emphasize the algebraic aspect of the new definition,
let us give another proof of the following theorem.

\begin{thm}
$\Delta$ is coassociative.
\end{thm}

\begin{proof}
It is enough to check
\begin{eqnarray}
(id \otimes \Delta) \Delta \delta_T = (\Delta \otimes id) \Delta
\delta_T
 \qquad \forall T \in \Sigma_p.                    \label{E:5}
\end{eqnarray}
where $T$ is a tree in ${\mathcal H}_{b,m}$. Define $L_T:{\mathcal
H}_{b,m} \to {\mathcal H}_{b,m}$ as follows; Let $T_1'$,...,$T_n'$
be the subsidiary branches of the root of $T$ in $T$. Let $T_{n_i}$
be a subtree of $T_{n_i}'$ whose root is the root of $T_{n_i}'$.
Define $T'$ to be the tree obtained by appending $T_{n_i}$ 's to a
new root $*$ and the edge connecting $*$ and the root of $T_{n_i}$
is labeled the same as the edge connecting the root of $T$ and the
root of $T_{n_i}'$. Then
$L_T(\delta_{T_{n_1}}...\delta_{T_{n_p}})=\delta_{T'}$. If some
$T_j$ is not a subsidiary branch of the root in $T$,
$L_T(\delta_{T_1}...\delta_{T_n})$ is 0. (An example of this
notation is in Figure \ref{Fig:20'}.)

\begin{figure}[h]

\begin{picture}(100,230)(60,-230)

\put(10,-50){$T=$}

\put(70,-30){\line(0,-1){20}}
\put(60,-45){1}
\put(70,-50){\circle*{3}}
\put(70,-30){\circle*{3}}
\put(70,-30){\line(-1,-1){20}}
\put(70,-45){2}
\put(50,-50){\circle*{3}}
\put(70,-30){\line(1,-1){20}}
\put(80,-45){3}
\put(90,-50){\circle*{3}}

\put(50,-50){\line(-1,-2){10}}
\put(45,-60){1}
\put(40,-70){\circle*{3}}
\put(50,-50){\line(0,-1){20}}
\put(50,-60){2}
\put(50,-70){\circle*{3}}
\put(50,-50){\line(1,-2){10}}
\put(55,-60){3}
\put(60,-70){\circle*{3}}

\put(70,-50){\line(0,-1){20}}
\put(70,-60){1}
\put(70,-70){\circle*{3}}
\put(70,-50){\line(1,-2){10}}
\put(75,-60){2}
\put(80,-70){\circle*{3}}

\put(90,-50){\line(0,-1){20}}
\put(90,-60){2}
\put(90,-70){\circle*{3}}
\put(90,-50){\line(1,-2){10}}
\put(95,-60){3}
\put(100,-70){\circle*{3}}

\put(10,-100){$T'_1=$}
\put(45,-90){\circle*{3}}
\put(45,-90){\line(-1,-2){10}}
\put(40,-100){1}
\put(35,-110){\circle*{3}}
\put(45,-90){\line(0,-1){20}}
\put(45,-100){2}
\put(45,-110){\circle*{3}}
\put(45,-90){\line(1,-2){10}}
\put(50,-100){3}
\put(55,-110){\circle*{3}}

\put(65,-100){$T'_2=$}
\put(100,-90){\circle*{3}}
\put(100,-90){\line(-1,-2){10}}
\put(95,-100){1}
\put(90,-110){\circle*{3}}
\put(100,-90){\line(1,-2){10}}
\put(105,-100){2}
\put(110,-110){\circle*{3}}

\put(125,-100){$T'_3=$}
\put(160,-90){\circle*{3}}
\put(160,-90){\line(-1,-2){10}}
\put(155,-100){2}
\put(150,-110){\circle*{3}}
\put(160,-90){\line(1,-2){10}}
\put(165,-100){3}
\put(170,-110){\circle*{3}}

\put(10,-130){$T_1=$}
\put(45,-120){\circle*{3}}
\put(45,-120){\line(-1,-2){10}}
\put(40,-130){1}
\put(35,-140){\circle*{3}}
\put(45,-120){\line(0,-1){20}}
\put(45,-130){2}
\put(45,-140){\circle*{3}}

\put(65,-130){$T_2=$}
\put(100,-130){\circle*{3}}

\put(125,-130){$T_3=$}
\put(160,-120){\circle*{3}}
\put(160,-120){\line(0,-1){20}}
\put(160,-130){2}
\put(160,-140){\circle*{3}}

\put(10,-160){$L_T(\delta _{T_1} \delta _{T_2} \delta _{T_3}) = \delta _{T'}$}

\put(10,-190){$T'=$}

\put(70,-170){\line(0,-1){20}}
\put(60,-185){1}
\put(70,-190){\circle*{3}}
\put(70,-170){\circle*{3}}
\put(70,-170){\line(-1,-1){20}}
\put(70,-185){2}
\put(50,-190){\circle*{3}}
\put(70,-170){\line(1,-1){20}}
\put(80,-185){3}
\put(90,-190){\circle*{3}}

\put(50,-190){\line(-1,-2){10}}
\put(45,-200){1}
\put(40,-210){\circle*{3}}
\put(50,-190){\line(0,-1){20}}
\put(50,-200){2}
\put(50,-210){\circle*{3}}

\put(90,-190){\line(0,-1){20}}
\put(90,-200){2}
\put(90,-210){\circle*{3}}

\put(10,-230){$t_n=$}
\put(45,-220){\circle*{3}}
\put(45,-220){\line(-1,-2){10}}
\put(40,-230){1}
\put(35,-240){\circle*{3}}
\put(45,-220){\line(0,-1){20}}
\put(45,-230){2}
\put(45,-240){\circle*{3}}
\put(45,-220){\line(1,-2){10}}
\put(50,-230){3}
\put(55,-240){\circle*{3}}

\end{picture}

\caption{}                 \label{Fig:20'}

\end{figure}

First let us show that
\begin{eqnarray}
\Delta\circ L_T(a)=L_T(a) \otimes 1 + (id \otimes L_T) \circ \Delta(a)
   \label{E:6}
\end{eqnarray}
where $a=\delta_{T_1}\delta_{T_2}...\delta_{T_n}$ and $T_1,...,T_n$
are all subsidiary branches of the root of $T$ in $T$
so that $L_T(a)=\delta_T$.
From \eqref{E:2}, we get
\begin{eqnarray}
\Delta(L_T(a))-L_T(a) \otimes 1 =\sum_{c}( \prod_{P_c}
                  \delta_{T'_i}) \otimes \delta_{R_c},    \label{E:8}
\end{eqnarray}
where all simple cuts of $T$ (including $c=\emptyset$) are allowed. Moreover,
\begin{eqnarray}
\Delta(a)=\prod_{i=1}^{n} (\delta_{T_i} \otimes 1+
          \sum_{c_i} (\prod_{P_{c_i}} \delta_{T''_{i_j}}) \otimes
              \delta_{R_{c_i}}),   \label{E:9}
\end{eqnarray}
where again all simple cuts $c_i$ of $T_i$ are allowed.

Let $t_n$ be the corolla  with root $*$ and $n$ other vertices $v_i$
labeled by $i_1,...,i_n$, where $i_j$ is the label of the edge in
$T$ connecting the root of $T$ and the vertex of $T_j'$, all
directly connected to the root $*$, as in Figure \ref{Fig:21'}.

\begin{figure}[h]

\begin{picture}(160,70)(10,-60)

\put(50,-30){$t_n=$}

\put(120,-20){\circle*{3}}
\put(120,-15){$*$}
\put(120,-20){\line(-2,-1){40}}
\put(100,-30){$i_1$}
\put(80,-40){\circle*{3}}
\put(80,-50){$v_1$}
\put(120,-20){\line(-1,-1){20}}
\put(115,-30){$i_2$}
\put(100,-40){\circle*{3}}
\put(100,-50){$v_2$}
\put(120,-37){......}
\put(120,-20){\line(2,-1){40}}
\put(140,-30){$i_n$}
\put(160,-40){\circle*{3}}
\put(160,-50){$v_n$}

\end{picture}

\caption{}              \label{Fig:21'}

\end{figure}

We view $t_n$ in an obvious way as a subgraph of the tree $T$, where
$*$ is the root of $T$ and the vertex $v_i$ is the root of $T_i$,
i.e., we can get $T$ by attaching the root of $T_i$ to the vertex
$v_i$ of $t_n$. Given a simple cut $c$ of $T$ we get by restriction
to the corolla subgraph $t_n \subset T$ a cut of $t_n$. It is
characterized by the subset $I=\{i|(*,v_i) \in c\} \subset
\{1,...,m\}$. The simple cut c is uniquely determined by the
restriction $c_i$ of $c$ to each subtree $T_i'$. Thus the simple
cuts $c_i$ of $T$ are in one to one correspondence with the various
terms of the expression \eqref{E:9}, namely the $\prod_{k \in I}
(\delta_{T_k} \otimes 1) \prod_{i \in \{1,...,m\} - I} \
\prod_{P_{c_i}} \delta_{T''_{i_j}} \otimes \delta_{R_{c_i}}$. So,
applying $id \otimes L$ to \eqref{E:9} and comparing with
\eqref{E:8}, we get \eqref{E:6}.

Now let us show \eqref{E:6} by induction. We have,
\begin{eqnarray}
\Delta\delta_{\bullet} = \delta_{\bullet} \otimes 1
                       + 1 \otimes \delta_{\bullet}
\end{eqnarray}
where $\bullet$ is the one-vertex bonsai,
so that ${\mathcal H}_1$ is coassociative.
Let us assume that ${\mathcal H}_n$
is coassociative and prove it for ${\mathcal H}_{n+1}$.
It is enough to check \eqref{E:5} for the generators $\delta_T$,
with $deg(T) \leq n+1$.
We have $\delta_T=L_T(\delta_{T_1}\delta_{T_2}...\delta_{T_m})=L_T(a)$
where the degrees of all $T_j$ are $\leq n$, i.e. $a \in {\mathcal H}_n$.
Using \eqref{E:6} we can replace $\Delta\delta_T$ by
\begin{eqnarray}
L_T(a) \otimes 1 +(id \otimes L_T) \Delta(a)
\end{eqnarray}
where $\Delta$ is the coassociative coproduct in ${\mathcal H}_n$.

The first term of \eqref{E:5} is then:
\begin{eqnarray}
(id \otimes \Delta)(L_T(a) \otimes 1 +(id \otimes L_T) \Delta(a)) \\
=L_T(a)\otimes 1 \otimes 1 + \Sigma a_{(1)} \otimes \Delta \circ L_T
a_{(2)} \notag
\end{eqnarray}
where $\Delta(a)=\Sigma a_{(1)} \otimes a_{(2)}$, which by
\eqref{E:6} gives
\begin{eqnarray}
L_T(a) \otimes 1 \otimes 1 +\Sigma a_{(1)} \otimes L_T a_{(2)}
\otimes 1 +\Sigma a'_{(1)} \otimes a'_{(2)} \otimes L_T a'_{(3)}
\end{eqnarray}
where
\begin{eqnarray}
(\Delta \otimes id)\Delta(a) = (id \otimes \Delta)\Delta(a) =
\Sigma a'_{(1)} \otimes a'_{(2)} \otimes a'_{(3)}.
\end{eqnarray}
by induction hypothesis on $n$, since $a \in {\mathcal H}_n$.

 The second term of \eqref{E:5} is
$\Delta \circ L(a) \otimes 1 +\Sigma \Delta a_{(1)} \otimes L
a_{(2)}$, which by \eqref{E:6} gives,
\begin{eqnarray}
L(a) \otimes 1 \otimes 1 + \Sigma a_{(1)} \otimes L a_{(2)} \otimes 1
+ \Sigma  a'_{(1)} \otimes a'_{(2)} \otimes L a'_{(3)}.
\end{eqnarray}
Thus we conclude that $\Delta$ is coassociative.
\end{proof}

\begin{thm}
$m((S\otimes id)\Delta)=\epsilon$.
\end{thm}

\begin{proof}
We have $m((S \otimes id)\Delta)(1)=m(S \otimes id)(1 \otimes
1)=S(1)1=1=\epsilon(1)$. And when $\delta_T \ne 1$,
\begin{align}
m((S \otimes id)\Delta)(\delta_T)&=&
           m((S \otimes id)(\delta_T \otimes 1 + \sum_{\mbox{simple cuts }c}
           \delta_{P_c(T)} \otimes \delta_{R_c(T)})) \\
                                 &=&
           S(\delta_T) + m(\sum_{\mbox{simple cuts }c}
           S(\delta_{P_c(T)}) \otimes \delta_{R_c(T)}) \notag \\
                                 &=&
           S(\delta_T) +\sum_{\mbox{simple cuts }c}
           S(\delta_{P_c(T)}) \delta_{R_c(T)} \notag \\
                                 &=&0      \notag
\end{align}
where the last equality is by the definition of the antipodal map
$S$.
\end{proof}

\section{Lie Algebra ${\mathcal L}^1$}

Let ${\mathcal L}^1 \subset {\mathcal H}_{b,m}^{\vee}$ be the linear
space having basis $\{Z_T| T \in {\mathcal H}_{b,m}\mbox{ is a
tree}\}$, where $\delta _T$ is defined as
\begin{eqnarray}
\langle Z_T, \delta_T \rangle =1
\end{eqnarray}
and
\begin{eqnarray}
\langle Z_T,P(\delta_{T_i}) \rangle =(\partial /\partial \delta_T
P)(0)
\end{eqnarray}
for each rooted tree $T$.

We introduce an operation on ${\mathcal L}^1$ by
\begin{eqnarray}
Z_{T_1}*Z_{T_2}=\sum_{T} n(T_1,T_2;T)Z_T,
\end{eqnarray}
where the integer $n(T_1,T_2;T)$ is determined
as the number of simple cuts $c$ with cardinality $|c|=1$ of bonsai $T$
such that the cut branch is $T_1$ while the remaining trunk is $T_2$.

With a notational abuse such as $T=Z_T$, we have an example of $*$
in Figure \ref{Fig:4'}.

\begin{figure}[h]

\begin{picture}(200,110)(60,-100)

\put(0,-5){\circle*{2}}

\put(5,-10){*}

\put(20,0){\circle*{2}}

\put(20,0){\line(1,-2){10}}
\put(30,-20){\circle*{2}}

\put(25,-10){3}
\put(20,0){\line(-1,-2){10}}
\put(10,-20){\circle*{2}}

\put(15,-10){2}

\put(35,-10){=}

\put(100,0){\circle*{2}}

\put(100,0){\line(1,-2){10}}
\put(110,-20){\circle*{2}}

\put(105,-10){3}
\put(100,0){\line(-1,-2){10}}
\put(90,-20){\circle*{2}}

\put(95,-10){2}
\put(100,0){\line(-2,-1){40}}
\put(60,-20){\circle*{2}}
\put(80,-10){1}

\put(35,-50){+}

\put(60,-40){\circle*{2}}

\put(60,-40){\line(1,-2){10}}
\put(70,-60){\circle*{2}}

\put(65,-50){3}
\put(60,-40){\line(-1,-2){10}}
\put(50,-60){\circle*{2}}

\put(55,-50){2}
\put(50,-60){\line(0,-1){20}}
\put(52,-76){1}
\put(50,-80){\circle*{2}}

\put(75,-50){+}

\put(100,-40){\circle*{2}}

\put(100,-40){\line(1,-2){10}}
\put(110,-60){\circle*{2}}

\put(105,-50){3}
\put(100,-40){\line(-1,-2){10}}
\put(90,-60){\circle*{2}}

\put(95,-50){2}
\put(90,-60){\line(0,-1){20}}
\put(92,-76){2}
\put(90,-80){\circle*{2}}

\put(115,-50){+}

\put(140,-40){\circle*{2}}

\put(140,-40){\line(1,-2){10}}
\put(150,-60){\circle*{2}}

\put(145,-50){3}
\put(140,-40){\line(-1,-2){10}}
\put(130,-60){\circle*{2}}

\put(135,-50){2}
\put(130,-60){\line(0,-1){20}}
\put(130,-80){\circle*{2}}
\put(132,-76){3}

\put(155,-50){+}

\put(180,-40){\circle*{2}}

\put(180,-40){\line(1,-2){10}}
\put(190,-60){\circle*{2}}

\put(185,-50){3}
\put(180,-40){\line(-1,-2){10}}
\put(170,-60){\circle*{2}}

\put(175,-50){2}
\put(190,-60){\line(0,-1){20}}
\put(190,-80){\circle*{2}}
\put(192,-76){1}

\put(195,-50){+}

\put(220,-40){\circle*{2}}

\put(220,-40){\line(1,-2){10}}
\put(230,-60){\circle*{2}}

\put(225,-50){3}
\put(220,-40){\line(-1,-2){10}}
\put(210,-60){\circle*{2}}

\put(215,-50){2}
\put(230,-60){\line(0,-1){20}}
\put(230,-80){\circle*{2}}
\put(232,-76){2}

\put(235,-50){+}

\put(260,-40){\circle*{2}}

\put(260,-40){\line(1,-2){10}}
\put(270,-60){\circle*{2}}

\put(265,-50){3}
\put(260,-40){\line(-1,-2){10}}
\put(250,-60){\circle*{2}}

\put(255,-50){2}
\put(270,-60){\line(0,-1){20}}
\put(270,-80){\circle*{2}}
\put(272,-76){3}

\end{picture}

\caption{}                 \label{Fig:4'}

\end{figure}

In this section, we will show that
${\mathcal L}^1$ is a Lie algebra and
the Hopf algebra ${\mathcal H}_{b,m}$
is the dual of the enveloping algebra of ${\mathcal L}^1$.

\begin{thm}
$deg(T)$ defines a grading of the Lie algebra ${\mathcal L}^1$.
\end{thm}
\begin{proof}
If we write  $Z_{T_1} * Z_{T_2} = \Sigma \ Z_T$ ,then the number of
vertices in $T$ is the sum of numbers of vertices in $T_1$ and
$T_2$.
\end{proof}

\begin{df}
We define the bracket
$[Z_{T_1},Z_{T_2}]=Z_{T_1}*Z_{T_2}-Z_{T_2}*Z_{T_1}$.
\end{df}

\begin{thm}
a) The bracket of the previous definition makes ${\mathcal L}^1$ a
Lie algebra.

b) The Hopf algebra ${\mathcal H}_{b,m}$
is the dual of the enveloping algebra of the Lie algebra ${\mathcal L}^1$.
\end{thm}

First we define the associator
\begin{eqnarray}
A(T_1,T_2,T_3):=Z_{T_1}*(Z_{T_2}*Z_{T_3})-(Z_{T_1}*Z_{T_2})*Z_{T_3}.
               \label{E:17}
\end{eqnarray}
and see
\begin{lem}
$A(T_1,T_2,T_3)=\Sigma n(T_1,T_2,T_3;T)Z_T$, where the integer
$n(T_1, T_2, T_3;T)$ is the number of simple cuts $c$ of $T$ such
that the number of elements $|c|$ of $c$ is  2 and the two branches
cut out from $T_3$ by $c$ are $T_1$, $T_2$ while the remaining trunk
$R_c(T)=T_3$.
\end{lem}
\begin{proof}

When we evaluate \eqref{E:17} against $Z_T$ we get the coefficient,
\begin{eqnarray}
\sum_{T'} n(T_1,T';T)n(T_2,T_3;T')
- \sum_{T''} n(T_1,T_2;T'')n(T'',T_3;T).          \label{E:18}
\end{eqnarray}
The first term corresponds to pairs of cuts, $c$, $c'$ of $T$ with
$|c|=|c'|=1$ and where $c'$ is a cut of $R_c(T)$. These pairs of
cuts fall into two classes either $c \cup c'$ is an admissible cut
or it is not. The second sum corresponds to pairs of cuts $c_1$,
$c'_1$ of T such that $|c_1|=|c'_1|=1$, $R_{c_1}(T)=T_3$ and $c'_1$
is a cut of $P_{c_1}(T)$. In such a case $c_1 \cup c'_1$ is never an
admissible cut so the difference \eqref{E:18} amounts to subtracting
from the first sum the pairs $c$, $c'$ such that $c \cup c'$ is not
an admissible cut. This gives,
\begin{eqnarray}
A(T_1,T_2,T_3)=\sum_{T} n(T_1,T_2,T_3;T)Z_T
\end{eqnarray}
where $n(T_1,T_2,T_3;T)$ is the number of admissible cuts $c$
of $T$ of cardinality 2
such that the two cut branches are $T_1$ and $T_2$ and $T_3$ is
the remaining trunk.
\end{proof}
Now for the theorem, we have
\begin{proof}
a) By the lemma, it is clear that
\begin{eqnarray}
A(T_1,T_2,T_3)=A(T_2,T_1,T_3).
\end{eqnarray}
Now compute $[[Z_{T_1},Z_{T_2}],Z_{T_3}]+[[Z_{T_2},Z_{T_3}],Z_{T_1}]
  +[[Z_{T_3},Z_{T_1}],Z_{T_2}]$. We can write it as a sum of 12 terms,
\begin{align}
  &(T_1*T_2)*T_3 -(T_2*T_1)*T_3 -T_3*(T_1*T_2) +T_3*(T_2*T_1)        \\
+ &(T_2*T_3)*T_1 -(T_3*T_2)*T_1 -T_1*(T_2*T_3) +T_1*(T_3*T_2) \notag \\
+ &(T_3*T_1)*T_2 -(T_1*T_3)*T_2 -T_2*(T_3*T_1) +T_2*(T_1*T_3) \notag \\
  &                                                           \notag \\
= &-A(T_1,T_2,T_3) +A(T_2,T_1,T_3) -A(T_3,T_1,T_2) +A(T_3,T_2,T_1) \notag \\
  &-A(T_2,T_3,T_1) +A(T_1,T_3,T_2) =0                               \notag
\end{align}

b) Let $P$ and $Q$ be polynomials of $\delta_T$'s. By the definition
of $Z_T$, $Z_T$ vanishes when paired with any monomial
$\delta^{n_1}_{T_1}...\delta^{n_k}_{T_k}$ except when this monomial
is $\delta_T$. Since $P \rightarrow P(0)$ is the counit $\epsilon$
of ${\mathcal H}_{b,m}$ and since $Z_T$ satisfies
\begin{align}
\langle Z_T, PQ \rangle &= (\partial /\partial \delta_T PQ)(0)  \label{E:24}\\
                        &= (\partial /\partial \delta_T P)(0)Q(0)
                         + P(0)(\partial /\partial \delta_T Q)(0)  \notag \\
                        &= \langle Z_T, P \rangle \epsilon (Q)
                         + \epsilon (P) \langle Z_T, Q \rangle ,   \notag
\end{align}
it follows that the coproduct of $Z_T$ is,
\begin{eqnarray}
\Delta Z_T = Z_T \otimes 1 +1 \otimes Z_T.        \label{E:25}
\end{eqnarray}
The product of two elements of ${\mathcal H}_{b,m}^{\vee}$
is defined by
\begin{eqnarray}
\langle Z_1 Z_2,P \rangle =\langle Z_1 \otimes Z_2 , \Delta P \rangle.
\end{eqnarray}
Since the commutator of two derivations is still a derivation,
the subspace of
${\mathcal H}_{b,m}^{\vee}$ satisfying
\eqref{E:25} is stable under bracket.
What remains is to show that
\begin{eqnarray}
Z_{T_1}Z_{T_2}-Z_{T_2}Z_{T_1}=[Z_{T_1},Z_{T_2}],  \label{E:27}
\end{eqnarray}
where $[Z_{T_1},Z_{T_2}]=Z_{T_1}*Z_{T_2}-Z_{T_2}*Z_{T_1}$ by definition. \\
Let ${\mathcal H}_0=\mbox{Ker } \epsilon$ be the augmentation ideal of
${\mathcal H}_{b,m}$.
By definition of $\Delta$,
\begin{eqnarray}
\Delta \delta_T = \delta_T \otimes 1 + 1 \otimes \delta_T + R_T
\end{eqnarray}
where $R_T \in {\mathcal H}_0 \otimes {\mathcal H}_0$. In fact, we
have
\begin{eqnarray}
R_{T}= \sum_{c} \delta_{T'_c} \otimes \delta_{T_c}
\end{eqnarray}
modulo $({\mathcal H}_{0})^{2} \otimes {\mathcal H}_0$, where $c$
varies among single cuts of the bonsai tree $T$, where $T_c$ is the
trunk of $T$ that contains the root, and $T'_c$ is the tree which
remains. When we compute
\begin{eqnarray}
\langle Z_{T_1} Z_{T_2} , \delta_T \rangle =
      \langle Z_{T_1} \otimes Z_{T_2}, \Delta \delta_T \rangle ,
\end{eqnarray}
the only part which contributes comes from $R_T$ and it counts the
number of ways of obtaining $T$ from $T_1$ and $T_2$, which gives
\eqref{E:27}.

If a map $f$ satisfies
\begin{align}
\langle f, PQ \rangle &= \langle f, P \rangle \epsilon (Q)
                         + \epsilon (P) \langle f, Q \rangle ,  \label{E:coprod}
\end{align}
$f$ is determined by $f(\delta_T) = \langle f, \delta_T \rangle$'s
and each of them is a scalar. Since $f(\delta_T)=\Sigma_{T'}
f(\delta_{T'}) Z_{T'}(\delta_T)$, $f$ has the form $\Sigma
f(\delta_T) Z_T$. Hence $\{Z_T\}$ is a basis of the subspace of
${\mathcal H}_{b,m}^{\vee}$ consisting of the vectors $f$ satisfying
\eqref{E:coprod}.

Since every $Z_T$ satisfies \eqref{E:coprod} by \eqref{E:24} and $f
\in {\mathcal H}_{b,m}^{\vee}$  satisfies \eqref{E:coprod} if and
only if  $f$ satisfies
\begin{eqnarray}
\Delta f = f \otimes 1 +1 \otimes f,        \label{E:25}
\end{eqnarray}
we have ${\mathcal L}^1=Prim({\mathcal H}_{b,m}^{\vee})$ and they
are isomorphic as Lie algebras. Since ${\mathcal H}_{b,m}^{\vee}$ is
connected and cocommutative, by the Milnor-Moore theorem, ${\mathcal
H}_{b,m}^{\vee} ={\mathcal U}(Prim({\mathcal H}_{b,m}^{\vee}))
={\mathcal U}({\mathcal L}^1)$ and so ${\mathcal H}_{b,m}={\mathcal
U}({\mathcal L}^1)^{\vee}$.
\end{proof}

\section{Operad of $m$-Bonsai }

Now let us consider operad theory with respect to the
$m$-bonsai Hopf algebra structure.
As seen in the last section, for trees $T, T' \in {\mathcal H}_{b,m}$
we can define $T*T'$ and this is a (left) pre-Lie operation.
The map $T \mapsto Z_T$ is a pre-Lie isomorphism
from the space spanned by trees to ${\mathcal L}^1$.
In ${\mathcal H}_{b,m}$,
we denote this ${\mathcal L}^1$ as ${\mathcal L}_{b,m}$.
We will sometimes allow a notational abuse such as $T=Z_T$ from now on.

Let us start from a rudimentary idea.
Every bonsai in ${\mathcal H}_{b,m}$ has a unique form
in which for each vertex,
its subsidiary edges are arranged
so that lower edge-label is on the left of higher edge-label
as in Figure \ref{Fig:30}.

\begin{figure}[h]

\begin{picture}(120,100)(0,-70)

\put(0,0){\circle*{2}}

\put(0,0){\line(-1,-1){20}}
\put(-20,-20){\circle*{2}}
\put(-15,-15){3}
\put(0,0){\line(0,-1){20}}
\put(0,-20){\circle*{2}}
\put(0,-15){2}
\put(0,0){\line(1,-1){20}}
\put(20,-20){\circle*{2}}
\put(15,-15){1}

\put(20,-20){\circle*{2}}
\put(20,-20){\line(-1,-2){10}}
\put(10,-40){\circle*{2}}
\put(12,-35){3}
\put(20,-20){\line(1,-2){10}}
\put(30,-40){\circle*{2}}
\put(28,-35){1}

\put(30,-40){\circle*{2}}
\put(30,-40){\line(-1,-2){10}}
\put(20,-60){\circle*{2}}
\put(22,-55){2}
\put(30,-40){\line(1,-2){10}}
\put(40,-60){\circle*{2}}
\put(38,-55){1}

\put(80,-25){$\rightarrow$}

\put(0,0){\circle*{2}}

\put(150,0){\line(-1,-1){20}}
\put(130,-20){\circle*{2}}
\put(135,-15){1}
\put(150,0){\line(0,-1){20}}
\put(150,-20){\circle*{2}}
\put(150,-15){2}
\put(150,0){\line(1,-1){20}}
\put(170,-20){\circle*{2}}
\put(165,-15){3}

\put(130,-20){\circle*{2}}
\put(130,-20){\line(-1,-2){10}}
\put(120,-40){\circle*{2}}
\put(122,-35){1}
\put(130,-20){\line(1,-2){10}}
\put(140,-40){\circle*{2}}
\put(138,-35){3}

\put(120,-40){\circle*{2}}
\put(120,-40){\line(-1,-2){10}}
\put(110,-60){\circle*{2}}
\put(112,-55){1}
\put(120,-40){\line(1,-2){10}}
\put(130,-60){\circle*{2}}
\put(128,-55){2}

\end{picture}

\caption{}              \label{Fig:30}

\end{figure}

We can number the possible positions in the bonsai of Figure
\ref{Fig:30} to append other bonsais as in Figure \ref{Fig:31}, the
example in ${\mathcal L}_{b,3}$ (the orders of possible appending
positions are underlined). When the labeling of Figure \ref{Fig:31}
is changed into that of Figure \ref{Fig:32}, then the numbering of
possible appending positions is also changed.

\begin{figure}[h]

\begin{picture}(160,160)(-70,-150)

\put(0,0){\circle*{4}}

\put(0,0){\line(-1,-1){40}}
\put(-40,-40){\circle*{4}}
\put(-30,-30){1}
\put(0,0){\line(0,-1){40}}
\put(0,-40){\circle*{4}}
\put(0,-30){2}
\put(0,0){\line(1,-1){40}}
\put(40,-40){\circle*{4}}
\put(30,-30){3}

\put(-40,-40){\circle*{4}}
\put(-40,-40){\line(-1,-1){40}}
\put(-80,-80){\circle*{4}}
\put(-70,-70){1}
\put(-40,-40){\line(1,-1){40}}
\put(0,-80){\circle*{4}}
\put(-10,-70){3}

\put(-80,-80){\circle*{4}}
\put(-80,-80){\line(-1,-1){40}}
\put(-120,-120){\circle*{4}}
\put(-110,-110){1}
\put(-80,-80){\line(0,-1){40}}
\put(-80,-120){\circle*{4}}
\put(-80,-110){2}

\put(-120,-120){\line(-1,-1){10}}
\put(-130,-140){$\underline{0}$}
\put(-120,-120){\line(0,-1){10}}
\put(-120,-140){$\underline{1}$}
\put(-120,-120){\line(1,-1){10}}
\put(-110,-140){$\underline{2}$}

\put(-80,-120){\line(-1,-1){10}}
\put(-90,-140){$\underline{3}$}
\put(-80,-120){\line(0,-1){10}}
\put(-80,-140){$\underline{4}$}
\put(-80,-120){\line(1,-1){10}}
\put(-70,-140){$\underline{5}$}

\put(-80,-80){\line(1,-1){10}}
\put(-70,-100){$\underline{6}$}

\put(-40,-40){\line(0,-1){10}}
\put(-40,-60){$\underline{7}$}

\put(0,-80){\line(-1,-1){10}}
\put(-10,-100){$\underline{8}$}
\put(0,-80){\line(0,-1){10}}
\put(0,-100){$\underline{9}$}
\put(0,-80){\line(1,-1){10}}
\put(10,-100){$\underline{10}$}

\put(0,-40){\line(-1,-1){10}}
\put(-15,-60){$\underline{11}$}
\put(0,-40){\line(0,-1){10}}
\put(-3,-60){$\underline{12}$}
\put(0,-40){\line(1,-1){10}}
\put(10,-60){$\underline{13}$}

\put(40,-40){\line(-1,-1){10}}
\put(25,-60){$\underline{14}$}
\put(40,-40){\line(0,-1){10}}
\put(38,-60){$\underline{15}$}
\put(40,-40){\line(1,-1){10}}
\put(50,-60){$\underline{16}$}

\end{picture}

\caption{}              \label{Fig:31}

\end{figure}

\begin{figure}[h]

\begin{picture}(160,160)(-70,-150)

\put(0,0){\circle*{4}}

\put(0,0){\line(-1,-1){40}}
\put(-40,-40){\circle*{4}}
\put(-30,-30){1}
\put(0,0){\line(0,-1){40}}
\put(0,-40){\circle*{4}}
\put(0,-30){2}
\put(0,0){\line(1,-1){40}}
\put(40,-40){\circle*{4}}
\put(30,-30){3}

\put(-40,-40){\circle*{4}}
\put(-40,-40){\line(0,-1){40}}
\put(-40,-80){\circle*{4}}
\put(-40,-70){2}
\put(-40,-40){\line(1,-1){40}}
\put(0,-80){\circle*{4}}
\put(-10,-70){3}

\put(-40,-80){\circle*{4}}
\put(-40,-80){\line(-1,-1){40}}
\put(-80,-120){\circle*{4}}
\put(-70,-110){1}
\put(-40,-80){\line(0,-1){40}}
\put(-40,-120){\circle*{4}}
\put(-40,-110){2}

\put(-80,-120){\line(-1,-1){10}}
\put(-90,-140){$\underline{1}$}
\put(-80,-120){\line(0,-1){10}}
\put(-80,-140){$\underline{2}$}
\put(-80,-120){\line(1,-1){10}}
\put(-70,-140){$\underline{3}$}

\put(-40,-120){\line(-1,-1){10}}
\put(-50,-140){$\underline{4}$}
\put(-40,-120){\line(0,-1){10}}
\put(-40,-140){$\underline{5}$}
\put(-40,-120){\line(1,-1){10}}
\put(-30,-140){$\underline{6}$}

\put(-40,-80){\line(1,-1){10}}
\put(-30,-100){$\underline{7}$}

\put(-40,-40){\line(-1,-1){10}}
\put(-50,-60){$\underline{0}$}

\put(0,-80){\line(-1,-1){10}}
\put(-10,-100){$\underline{8}$}
\put(0,-80){\line(0,-1){10}}
\put(0,-100){$\underline{9}$}
\put(0,-80){\line(1,-1){10}}
\put(10,-100){$\underline{10}$}

\put(0,-40){\line(-1,-1){10}}
\put(-15,-60){$\underline{11}$}
\put(0,-40){\line(0,-1){10}}
\put(-3,-60){$\underline{12}$}
\put(0,-40){\line(1,-1){10}}
\put(10,-60){$\underline{13}$}

\put(40,-40){\line(-1,-1){10}}
\put(25,-60){$\underline{14}$}
\put(40,-40){\line(0,-1){10}}
\put(38,-60){$\underline{15}$}
\put(40,-40){\line(1,-1){10}}
\put(50,-60){$\underline{16}$}

\end{picture}

\caption{}              \label{Fig:32}

\end{figure}

Then, by taking the standard form of bonsai and ordering the
possible positions of appending, we can get the transform of a
bonsai into the broomstick diagram used in \cite{MSS} like Figure
\ref{Fig:33}, again in ${\mathcal L}_{b,3}$.

\begin{figure}[h]

\begin{picture}(160,170)(0,-160)

\put(-90,-30){$T_2=$}

\put(0,0){\circle*{4}}

\put(0,0){\line(-1,-1){40}}
\put(-40,-40){\circle*{4}}
\put(-30,-30){1}
\put(0,0){\line(0,-1){40}}
\put(0,-40){\circle*{4}}
\put(0,-30){2}
\put(0,0){\line(1,-1){40}}
\put(40,-40){\circle*{4}}
\put(30,-30){3}

\put(-40,-40){\line(-1,-1){10}}
\put(-55,-60){$\underline{0}$}
\put(-40,-40){\line(0,-1){10}}
\put(-43,-60){$\underline{1}$}
\put(-40,-40){\line(1,-1){10}}
\put(-30,-60){$\underline{2}$}

\put(0,-40){\line(-1,-1){10}}
\put(-15,-60){$\underline{3}$}
\put(0,-40){\line(0,-1){10}}
\put(-3,-60){$\underline{4}$}
\put(0,-40){\line(1,-1){10}}
\put(10,-60){$\underline{5}$}

\put(40,-40){\line(-1,-1){10}}
\put(25,-60){$\underline{6}$}
\put(40,-40){\line(0,-1){10}}
\put(38,-60){$\underline{7}$}
\put(40,-40){\line(1,-1){10}}
\put(50,-60){$\underline{8}$}

\put(60,-30){$\rightarrow$}

\put(120,0){\line(0,-1){20}}

\put(120,-20){\line(2,-1){40}}
\put(120,-20){\line(-2,-1){40}}
\put(80,-40){\line(1,0){80}}

\put(84,-40){\line(0,-1){10}}
\put(93,-40){\line(0,-1){10}}
\put(102,-40){\line(0,-1){10}}
\put(111,-40){\line(0,-1){10}}
\put(120,-40){\line(0,-1){10}}
\put(129,-40){\line(0,-1){10}}
\put(138,-40){\line(0,-1){10}}
\put(147,-40){\line(0,-1){10}}
\put(156,-40){\line(0,-1){10}}

\put(82,-60){0}
\put(91,-60){1}
\put(100,-60){2}
\put(109,-60){3}
\put(118,-60){4}
\put(127,-60){5}
\put(136,-60){6}
\put(145,-60){7}
\put(154,-60){8}

\put(-90,-100){$T_1=$}

\put(0,-70){\circle*{4}}

\put(0,-70){\line(-1,-1){40}}
\put(-40,-110){\circle*{4}}
\put(-30,-100){1}

\put(0,-70){\line(1,-1){40}}
\put(40,-110){\circle*{4}}
\put(30,-100){3}

\put(-40,-110){\line(-1,-1){10}}
\put(-55,-130){$\underline{0}$}
\put(-40,-110){\line(0,-1){10}}
\put(-43,-130){$\underline{1}$}
\put(-40,-110){\line(1,-1){10}}
\put(-30,-130){$\underline{2}$}

\put(0,-70){\line(0,-1){10}}
\put(0,-90){$\underline{3}$}

\put(40,-110){\line(-1,-1){10}}
\put(25,-130){$\underline{4}$}
\put(40,-110){\line(0,-1){10}}
\put(38,-130){$\underline{5}$}
\put(40,-110){\line(1,-1){10}}
\put(50,-130){$\underline{6}$}

\put(60,-100){$\rightarrow$}

\put(120,-70){\line(0,-1){20}}

\put(120,-90){\line(2,-1){40}}
\put(120,-90){\line(-2,-1){40}}
\put(80,-110){\line(1,0){80}}

\put(84,-110){\line(0,-1){10}}
\put(96,-110){\line(0,-1){10}}
\put(108,-110){\line(0,-1){10}}
\put(120,-110){\line(0,-1){10}}
\put(132,-110){\line(0,-1){10}}
\put(144,-110){\line(0,-1){10}}
\put(156,-110){\line(0,-1){10}}

\put(82,-130){0}
\put(94,-130){1}
\put(106,-130){2}
\put(118,-130){3}
\put(130,-130){4}
\put(142,-130){5}
\put(154,-130){6}

\end{picture}

\caption{}              \label{Fig:33}

\end{figure}

So, we can define $T_1 \circ_i T_2$ as appending $T_1$ to $T_2$ at
the $i$-th appending position of $T_2$, and for $T_2$ and $T_1$ in
Figure \ref{Fig:33}, $T_1 \circ_4 T_2$ is given as in Figure
\ref{Fig:34}.

\begin{figure}[h]

\begin{picture}(160,170)(0,-160)

\put(-105,-70){$T_1 \circ_4 T_2=$}

\put(0,0){\circle*{4}}

\put(0,0){\line(-1,-1){40}}
\put(-40,-40){\circle*{4}}
\put(-30,-30){1}
\put(0,0){\line(0,-1){40}}
\put(0,-40){\circle*{4}}
\put(0,-30){2}
\put(0,0){\line(1,-1){40}}
\put(40,-40){\circle*{4}}
\put(30,-30){3}

\put(-40,-40){\line(-1,-1){10}}
\put(-55,-60){$\underline{0}$}
\put(-40,-40){\line(0,-1){10}}
\put(-43,-60){$\underline{1}$}
\put(-40,-40){\line(1,-1){10}}
\put(-30,-60){$\underline{2}$}

\put(0,-40){\line(-1,-1){10}}
\put(-15,-60){$\underline{3}$}
\put(0,-40){\line(0,-1){30}}
\put(0,-60){2}

\put(0,-40){\line(1,-1){10}}
\put(10,-60){$\underline{11}$}

\put(40,-40){\line(-1,-1){10}}
\put(25,-60){$\underline{12}$}
\put(40,-40){\line(0,-1){10}}
\put(38,-60){$\underline{13}$}
\put(40,-40){\line(1,-1){10}}
\put(50,-60){$\underline{14}$}

\put(67,-70){$\rightarrow$}

\put(120,0){\line(0,-1){20}}

\put(120,-20){\line(2,-1){40}}
\put(120,-20){\line(-2,-1){40}}
\put(80,-40){\line(1,0){80}}

\put(84,-40){\line(0,-1){10}}
\put(93,-40){\line(0,-1){10}}
\put(102,-40){\line(0,-1){10}}
\put(111,-40){\line(0,-1){10}}
\put(120,-40){\line(0,-1){30}}
\put(129,-40){\line(0,-1){10}}
\put(138,-40){\line(0,-1){10}}
\put(147,-40){\line(0,-1){10}}
\put(156,-40){\line(0,-1){10}}

\put(82,-60){0}
\put(91,-60){1}
\put(100,-60){2}
\put(109,-60){3}

\put(120,-60){11}
\put(131,-60){12}
\put(143,-60){13}
\put(154,-60){14}

\put(0,-70){\circle*{4}}

\put(0,-70){\line(-1,-1){40}}
\put(-40,-110){\circle*{4}}
\put(-30,-100){1}

\put(0,-70){\line(1,-1){40}}
\put(40,-110){\circle*{4}}
\put(30,-100){3}

\put(-40,-110){\line(-1,-1){10}}
\put(-55,-130){$\underline{4}$}
\put(-40,-110){\line(0,-1){10}}
\put(-43,-130){$\underline{5}$}
\put(-40,-110){\line(1,-1){10}}
\put(-30,-130){$\underline{6}$}

\put(0,-70){\line(0,-1){10}}
\put(0,-90){$\underline{7}$}

\put(40,-110){\line(-1,-1){10}}
\put(25,-130){$\underline{8}$}
\put(40,-110){\line(0,-1){10}}
\put(38,-130){$\underline{9}$}
\put(40,-110){\line(1,-1){10}}
\put(50,-130){$\underline{10}$}

\put(120,-70){\line(0,-1){20}}

\put(120,-90){\line(2,-1){40}}
\put(120,-90){\line(-2,-1){40}}
\put(80,-110){\line(1,0){80}}

\put(84,-110){\line(0,-1){10}}
\put(96,-110){\line(0,-1){10}}
\put(108,-110){\line(0,-1){10}}
\put(120,-110){\line(0,-1){10}}
\put(132,-110){\line(0,-1){10}}
\put(144,-110){\line(0,-1){10}}
\put(156,-110){\line(0,-1){10}}

\put(82,-130){4}
\put(94,-130){5}
\put(106,-130){6}
\put(118,-130){7}
\put(130,-130){8}
\put(142,-130){9}
\put(154,-130){10}

\put(167,-70){$=$}

\put(220,-40){\line(0,-1){20}}

\put(220,-60){\line(2,-1){40}}
\put(220,-60){\line(-2,-1){40}}
\put(180,-80){\line(1,0){80}}

\put(184,-80){\line(0,-1){10}}
\put(193,-80){\line(0,-1){10}}
\put(202,-80){\line(0,-1){10}}
\put(211,-80){\line(0,-1){10}}

\put(229,-80){\line(0,-1){10}}
\put(238,-80){\line(0,-1){10}}
\put(247,-80){\line(0,-1){10}}
\put(256,-80){\line(0,-1){10}}

\put(182,-100){0}
\put(191,-100){1}
\put(200,-100){2}
\put(210,-100){......}
\put(231,-100){12}
\put(243,-100){13}
\put(254,-100){14}

\end{picture}

\caption{}              \label{Fig:34}

\end{figure}

Then obviously, this $\circ_i$ satisfies the definition of
{\em pre-Lie system} of \cite{G}
(It is called {\em nonsymmetric pseudo-operad} in \cite{MSS},
but it has a difference in the convention of grading).
When we use the pseudo-operad later, we will give an extra definition,
which we give here;

\begin{df}
When $\{V_i\}$ is a graded module over a field $k$ and
$\circ_i=\circ_i(m,n):V_m \otimes V_n \to V_{m+n}$
is an operation satisfying; when $f^m$, $g^n$ and $h^p$ are in $V_m$,
$V_n$ and $V_p$ respectively,
\begin{eqnarray}
h^p \circ_j (g^n \circ_i f^m)=
\left\{ \begin{array}{ll}
g^n \circ_{i+p} (h^p \circ_j f^m) & \mbox{ if }0 \leq j \leq i-1 \\
(h^p \circ_{j-i} g^n) \circ_i f^m & \mbox{ if }i \leq j \leq n+i
\end{array} \right.
\end{eqnarray}
then $\{\{V_i\}, \circ_i\}$ is called a {\em (left) pre-Lie system}.
\end{df}
(In \cite{G} the right pre-Lie system is defined, but we define and
use the left pre-lie system. This is mainly intended for the theory
related to Hopf algebra we will argue later.) By the broomstick
diagrams shown in Figure \ref{Fig:33}-\ref{Fig:34}, we have

\begin{df}
When $W_{m,n}$ is the vector subspace of ${\mathcal L}_{b,m}$
generated by the trees having the number $n$ of possible appending
positions, and $T_1 \circ_i T_2$ is appending $T_1$ to $T_2$ at the
appending position $i$ of $T_2$, then $\{\{W_{m,n}\}, \circ_i\}$ is
a left pre-Lie system. It is called {\em $m$-bonsai pre-Lie system}.
For trees $T$ which are basis elements of $W_{m,n}$, $n$ is called
the appending degree of $T$, and denoted $deg_{ap}(T)$.
\end{df}
(Graphically, a basis element of $W_{m,n}$ has the broomstick
representation like Figure \ref{Fig:35}.)

\begin{figure}[h]

\begin{picture}(160,70)(40,-60)

\put(120,0){\line(0,-1){20}}

\put(120,-20){\line(2,-1){40}}
\put(120,-20){\line(-2,-1){40}}
\put(80,-40){\line(1,0){80}}

\put(84,-40){\line(0,-1){10}}
\put(93,-40){\line(0,-1){10}}
\put(102,-40){\line(0,-1){10}}
\put(111,-40){\line(0,-1){10}}
\put(156,-40){\line(0,-1){10}}

\put(82,-60){0}
\put(91,-60){1}
\put(100,-60){2}
\put(109,-60){3}
\put(118,-60){......}
\put(154,-60){$n$}

\end{picture}

\caption{}              \label{Fig:35}

\end{figure}

\section{Branch-fixed Differential}

In the next several sections, following the oracle of \cite{MSS}, we
will define some complexes related to bonsais. To get the analogy of
the {\em cobar complex} and the {\em tree differential} in Section
3.1 of \cite{MSS}, first let us give an order of edges of a bonsai
as in Figure \ref{Fig:38}, i.e., starting from the root, sweeping
around the bonsai counterclockwise and numbering the edges.
\begin{figure}[h]

\begin{picture}(160,160)(-70,-150)

\put(0,0){\circle*{4}}

\put(0,0){\line(-1,-1){40}}
\put(-40,-40){\circle*{4}}
\put(-30,-30){$\underline{1}, 1$}
\put(0,0){\line(0,-1){40}}
\put(0,-40){\circle*{4}}

\put(0,-30){$\underline{6}, 2$}
\put(0,0){\line(1,-1){40}}
\put(40,-40){\circle*{4}}
\put(30,-30){$\underline{7}, 3$}

\put(-40,-40){\circle*{4}}
\put(-40,-40){\line(-1,-1){40}}
\put(-80,-80){\circle*{4}}
\put(-70,-70){$\underline{2}, 1$}
\put(-40,-40){\line(1,-1){40}}
\put(0,-80){\circle*{4}}
\put(-10,-70){$\underline{5}, 3$}

\put(-80,-80){\circle*{4}}
\put(-80,-80){\line(-1,-1){40}}
\put(-120,-120){\circle*{4}}
\put(-110,-110){$\underline{3}, 1$}
\put(-80,-80){\line(0,-1){40}}
\put(-80,-120){\circle*{4}}
\put(-80,-110){$\underline{4}, 2$}

\put(-120,-120){\line(-1,-1){10}}

\put(-120,-120){\line(0,-1){10}}

\put(-120,-120){\line(1,-1){10}}

\put(-80,-120){\line(-1,-1){10}}

\put(-80,-120){\line(0,-1){10}}

\put(-80,-120){\line(1,-1){10}}

\put(-80,-80){\line(1,-1){10}}

\put(-40,-40){\line(0,-1){10}}

\put(0,-80){\line(-1,-1){10}}

\put(0,-80){\line(0,-1){10}}

\put(0,-80){\line(1,-1){10}}

\put(0,-40){\line(-1,-1){10}}

\put(0,-40){\line(0,-1){10}}

\put(0,-40){\line(1,-1){10}}

\put(40,-40){\line(-1,-1){10}}

\put(40,-40){\line(0,-1){10}}

\put(40,-40){\line(1,-1){10}}

\put(-50,-110){(underlined numbers are orders of edges and}
\put(-50,-120){non-underlined numbers are edge-labels of 3-bonsai)}

\end{picture}

\caption{}              \label{Fig:38}

\end{figure}
We call this order the {\em traversing order}.
In the traversing order, $e_{k,l_k}$
is a vector representing the $k$-th edge of a tree $T$,
such that $1 \leq l_k \leq m$ is the edge-label of the $k$- th edge.

Second, let us define a vector space ${\mathbf C}^n$ having basis $T
\otimes e_{1, l_1} \wedge ... \wedge e_{k,l_k}$, where $T$ is a
$m$-bonsai (not forest) having $n$ edges and the pairs ${k,l_k}$ run
over the labels of edges of $T$. (If $T$ is a vertex, i.e. a
connected bonsai without any edge, then $e_{1,l_1} \wedge ... \wedge
e_{k,l_k}$ is the constant unit 1.) For later use, we denote this
$e_{1,l_1} \wedge ... \wedge e_{k,l_k}$ as $det(T)$ and call it the
{\em determinant term} of $T$, and call $T \otimes det(T)$ a {\em
determinanted bonsai}. So the basis element corresponding to the
bonsai $T$ of Figure \ref{Fig:38} is $T \otimes e_{1,1} \wedge
e_{2,1} \wedge e_{3,1} \wedge e_{4,2}
    \wedge e_{5,3} \wedge e_{6,2} \wedge e_{7,3} = T \otimes det(T)$.
Let us denote ${\mathbf C} = \oplus {\mathbf C}^m$. Once we define
$d$, we call this complex $({\mathbf C},d)$ the {\em bonsai cobar
complex} after the {\em cobar complex} of \cite{MSS}.

Third, let us define a map $d^i:{\mathbf C}^i \to {\mathbf C}^{i+1}$
and show that $d^{i+1} \circ d^i=0$ as follows;

\begin{df} \label{def:branch-fixed}
Let $T$ be an $m$-bonsai. Let $T'$ be a bonsai such that we can
obtain $T$ by contracting an edge $e'$ from $T'$ and the following
conditions are satisfied;

i) $T'$ does not have more branching vertices
(i.e., vertices which have the ramification numbers $>$1) than $T$,

ii) $e'$ is not attatched to a branching vertex of $T$ so that $e'$
becomes a subsidiary edge of that branching vertex.

We call this $T'$ a {\em branch-fixed extension} of $T$.
\end{df}

For example, for the 3-bonsai $T$ in Figure \ref{Fig:39},
$T_1$, $T_3$, $T_5$ and $T_6$ are all branch-fixed extension of $T$,
but $T_2$(violating i)) and $T_4$(violating ii)) are not.

\begin{figure}[h]

\begin{picture}(200,200)(0,-200)

\put(-35,-20){$T=$}

\put(0,0){\circle*{2}}
\put(0,0){\line(-1,-2){10}}
\put(-10,-20){\circle*{2}}
\put(-10,-20){\line(0,-1){20}}
\put(-8,-30){1}
\put(-10,-40){\circle*{2}}

\put(-8,-15){1}
\put(0,0){\line(1,-2){10}}
\put(10,-20){\circle*{2}}
\put(8,-15){2}

\put(10,-20){\circle*{2}}
\put(10,-20){\line(-1,-2){10}}
\put(0,-40){\circle*{2}}
\put(2,-35){1}
\put(10,-20){\line(1,-2){10}}
\put(20,-40){\circle*{2}}
\put(18,-35){3}

\put(-35,-90){$T_1=$}

\put(0,-70){\line(0,1){20}}
\put(0,-50){\circle*{2}}
\put(-5,-60){$e$}
\put(2,-60){1}
\put(0,-70){\circle*{2}}
\put(0,-70){\line(-1,-2){10}}
\put(-10,-90){\circle*{2}}
\put(-10,-90){\line(0,-1){20}}
\put(-8,-100){1}
\put(-10,-110){\circle*{2}}
\put(-8,-85){1}
\put(0,-70){\line(1,-2){10}}
\put(10,-90){\circle*{2}}
\put(8,-85){2}

\put(10,-90){\circle*{2}}
\put(10,-90){\line(-1,-2){10}}
\put(0,-110){\circle*{2}}
\put(2,-105){1}
\put(10,-90){\line(1,-2){10}}
\put(20,-110){\circle*{2}}
\put(18,-105){3}

\put(35,-90){$T_2=$}

\put(70,-70){\circle*{2}}
\put(70,-70){\line(-1,-2){10}}
\put(60,-90){\circle*{2}}
\put(62,-85){1}
\put(70,-70){\line(1,-2){10}}
\put(80,-90){\circle*{2}}
\put(78,-85){2}

\put(60,-90){\line(-1,-2){10}}
\put(50,-110){1}
\put(50,-110){\circle*{2}}

\put(60,-90){\line(0,-1){20}}
\put(55,-105){$e$}
\put(62,-105){2}
\put(60,-110){\circle*{2}}

\put(80,-90){\circle*{2}}
\put(80,-90){\line(-1,-2){10}}
\put(70,-110){\circle*{2}}
\put(72,-105){1}
\put(80,-90){\line(1,-2){10}}
\put(90,-110){\circle*{2}}
\put(88,-105){3}

\put(105,-90){$T_3=$}

\put(140,-70){\circle*{2}}
\put(140,-70){\line(-1,-2){10}}
\put(130,-90){\circle*{2}}
\put(132,-85){1}

\put(130,-90){\line(0,-1){20}}
\put(132,-100){1}
\put(130,-110){\circle*{2}}

\put(140,-70){\line(1,-2){10}}
\put(150,-90){\circle*{2}}
\put(148,-85){2}

\put(150,-90){\line(0,-1){20}}
\put(145,-100){$e$}
\put(152,-100){1}

\put(150,-110){\circle*{2}}
\put(150,-110){\line(-1,-2){10}}
\put(140,-130){\circle*{2}}
\put(142,-125){1}
\put(150,-110){\line(1,-2){10}}
\put(160,-130){\circle*{2}}
\put(158,-125){3}

\put(175,-90){$T_4=$}

\put(210,-70){\circle*{2}}
\put(210,-70){\line(-1,-2){10}}
\put(200,-90){\circle*{2}}
\put(202,-85){1}
\put(200,-90){\line(0,-1){20}}
\put(202,-100){1}
\put(200,-110){\circle*{2}}

\put(210,-70){\line(1,-2){10}}
\put(220,-90){\circle*{2}}
\put(218,-85){2}
\put(211,-85){$e$}

\put(220,-90){\line(0,-1){20}}
\put(222,-100){2}

\put(220,-110){\circle*{2}}
\put(220,-110){\line(-1,-2){10}}
\put(210,-130){\circle*{2}}
\put(212,-125){1}
\put(220,-110){\line(1,-2){10}}
\put(230,-130){\circle*{2}}
\put(228,-125){3}

\put(-35,-170){$T_5=$}

\put(0,-150){\line(0,1){20}}
\put(-5,-140){$e$}
\put(0,-130){\circle*{2}}
\put(2,-140){2}
\put(0,-150){\circle*{2}}
\put(0,-150){\line(-1,-2){10}}
\put(-10,-170){\circle*{2}}
\put(-8,-165){1}
\put(-10,-170){\line(0,-1){20}}
\put(-8,-180){1}
\put(-10,-190){\circle*{2}}

\put(0,-150){\line(1,-2){10}}
\put(10,-170){\circle*{2}}
\put(8,-165){2}

\put(10,-170){\circle*{2}}
\put(10,-170){\line(-1,-2){10}}
\put(0,-190){\circle*{2}}
\put(2,-185){1}
\put(10,-170){\line(1,-2){10}}
\put(20,-190){\circle*{2}}
\put(18,-185){3}

\put(35,-170){$T_6=$}

\put(70,-150){\circle*{2}}
\put(70,-150){\line(-1,-2){10}}
\put(60,-170){\circle*{2}}
\put(62,-165){1}
\put(60,-170){\line(0,-1){20}}
\put(62,-180){1}
\put(60,-190){\circle*{2}}

\put(70,-150){\line(1,-2){10}}
\put(80,-170){\circle*{2}}
\put(78,-165){3}

\put(75,-180){$e$}
\put(80,-170){\line(0,-1){20}}
\put(82,-180){2}

\put(80,-190){\circle*{2}}
\put(80,-190){\line(-1,-2){10}}
\put(70,-210){\circle*{2}}
\put(72,-205){1}
\put(80,-190){\line(1,-2){10}}
\put(90,-210){\circle*{2}}
\put(88,-205){3}

\end{picture}

\caption{}              \label{Fig:39}

\end{figure}

Then we define $d^i:C^i \to C^{i+1}$ as following; when $T \in {\mathbf C}^i$,

\begin{eqnarray}
d^i(T) = \sum T' \otimes e'_{j,l_j} \wedge e'_{1,l_1} \wedge ... \wedge
                                         \widehat{e'_{j,l_j}}
                                  \wedge ... \wedge e'_{i+1,l_{i+1}}
\end{eqnarray}
where the sum runs over $T'$, which is a branch-fixed extension of $T$
having an edge $e$ added to $T$ and that $e$ is denoted $e'_{j,l_j}$
in the edge-ordering of $T'$.

\begin{thm} \label{Thm:lediff}
$d^{i+1} \circ d^i = 0$.
\end{thm}

\begin{proof}
Suppose $T''$ is a branch-fixed extension of a bonsai $T'$
with added edge $e''$,
which is a branch-fixed extension of $T$ with added edge $e'$.
Then, when $e''$ is $e''_{j,l_j}$ and $e'$ is $e''_{k,l_k}$
in the edge-ordering of $T''$ and
$d^{i+1} \circ d^i (T)$ is wrote
$\sum S   \otimes f_{1,p_1}   \wedge ...\wedge f_{i+2,p_{i+2}}$
where $S$ runs over the bonsais obtained by attaching two edges
as given in i) and ii) in Definition \ref{def:branch-fixed}
and $f_{i,p_i}$'s are the edges of $S$,
$T'' \otimes e''_{1,l_1} \wedge ...\wedge e''_{i+2,l_{i+2}}$
can be obtained only in two ways;

i) adding $e'$ first to $T$: then the component of
$T'' \otimes e''_{1,l_1} \wedge ... \wedge e''_{i+2,l_{1+2}}$ is
$T'' \otimes e''_{j,l_j} \wedge e''_{k,l_k} \wedge e''_{1,l_1}
\wedge ... \wedge \widehat{e''_{j,l_j}} \wedge ... \wedge
\widehat{e''_{k,l_k}} \wedge ... \wedge e''_{i+2,l_{i+2}}$.

ii) adding $e''$ first to $T$: then the component of
$T'' \otimes e''_{1,l_1} \wedge ... \wedge e''_{i+2,l_{1+2}}$ is
$T'' \otimes e''_{k,l_k} \wedge e''_{j,l_j} \wedge e''_{1,l_1}
\wedge ... \wedge \widehat{e''_{j,l_j}} \wedge ... \wedge
\widehat{e''_{k,l_k}} \wedge ... \wedge e''_{i+2,l_{i+2}}$.

Since the orders of $e''_{k,l_k}$ and $e''_{j,l_j}$
are different in the wedge products,
the sum of two terms in i) and ii) is 0,
and this is true for all components of $d^{i+1} \circ d^i (T)$.
Hence $d^{i+1} \circ d^i = 0$.
\end{proof}

We call this boundary map $d^i$ the {\em branch-fixed differential}.
A simple example is given in Figure \ref{Fig:40}.
We will study the cohomology of this $\{d^i\}$,
but before that, following \cite{MSS},
let us see an important property of this bonsai complex in the next section.

\begin{figure}[h]

\begin{picture}(200,90)(0,-90)

\put(0,-10){$d^0($}
\put(20,-5){\circle*{2}}
\put(25,-10){$\otimes 1)=e'$}
\put(65,0){\circle*{2}}
\put(65,0){\line(0,-1){15}}
\put(65,-15){\circle*{2}}
\put(67,-10){$1 \otimes e_{1,1}$}

\put(100,-10){$\quad +\quad e'$}
\put(140,0){\circle*{2}}
\put(140,0){\line(0,-1){15}}
\put(140,-15){\circle*{2}}
\put(142,-10){$2 \otimes e_{1,2}$}

\put(0,-45){$d^1($}
\put(20,-45){$e'$}
\put(30,-35){\circle*{2}}
\put(30,-35){\line(0,-1){15}}
\put(30,-50){\circle*{2}}
\put(32,-45){$1 \otimes e_{1,1})=$}
\put(95,-42){\circle*{2}}
\put(95,-42){\line(0,1){15}}
\put(95,-42){\line(0,-1){15}}
\put(95,-27){\circle*{2}}
\put(95,-57){\circle*{2}}
\put(85,-39){$e''$}
\put(85,-52){$e'$}
\put(97,-39){1}
\put(97,-52){1}
\put(105,-45){$\otimes e_{1,1} \wedge e_{2,1}$}

\put(160,-45){+}

\put(185,-42){\circle*{2}}
\put(185,-42){\line(0,1){15}}
\put(185,-42){\line(0,-1){15}}
\put(185,-27){\circle*{2}}
\put(185,-57){\circle*{2}}
\put(175,-39){$e'$}
\put(175,-52){$e''$}
\put(187,-39){1}
\put(187,-52){1}
\put(195,-45){$\otimes e_{2,1} \wedge e_{1,1}$}

\put(72,-85){$+$}
\put(95,-82){\circle*{2}}
\put(95,-82){\line(0,1){15}}
\put(95,-82){\line(0,-1){15}}
\put(95,-67){\circle*{2}}
\put(95,-97){\circle*{2}}
\put(85,-79){$e''$}
\put(85,-92){$e'$}
\put(97,-79){2}
\put(97,-92){1}
\put(105,-85){$\otimes e_{1,2} \wedge e_{2,1}$}

\put(160,-85){$+$}

\put(185,-82){\circle*{2}}
\put(185,-82){\line(0,1){15}}
\put(185,-82){\line(0,-1){15}}
\put(185,-67){\circle*{2}}
\put(185,-97){\circle*{2}}
\put(175,-79){$e'$}
\put(175,-92){$e''$}
\put(187,-79){1}
\put(187,-92){2}
\put(195,-85){$\otimes e_{2,2} \wedge e_{1,1}$}

\end{picture}

\caption{}              \label{Fig:40}

\end{figure}

\section{Cohomology of Branch-fixed Differential}

In this section,
we study the cohomology theory of the cochain complex $\{{\mathcal C}^i,d^i\}$,
where ${\mathcal C}^i$ is the bonsai cobar complex
and $d^i$ is the branch-fixed differential.
We will define a new kind of bonsai called {\em seedling} and a new complex
$\{{\mathbf C}^{S,j},d^{i+j}\}_{j \geq 0}$ called {\em thread}
and show that
the cohomology groups of $\{{\mathcal C}^i,d^i\}$ are the direct sum
of cohomology groups of threads $\{{\mathbf C}^{S,j},d^{i+j}\}_{j \geq 0}$.

First, let us give some definitions:

\begin{df}
A bonsai every vertex of which has the ramification number 0 or 1 is called
a {\em ladder}. In other words, a ladder is a bonsai which has no branching
vertex.
\end{df}

\begin{df}
If a bonsai $T$ has an edge, a vertex $v$
which is an end of only one edge and is not the root, is called a {\em tip}.
If a bonsai $T$ is a one-vertex bonsai, the root $v$ is a tip.
\end{df}

By the definition of $d^i$,
all terms in $d^i(T)$ are of the form $\pm T' \otimes e \wedge det(T)$,
where $T'$ runs over bonsais obtained by adding a new edge $e$ to $T$
so that i) and ii) of Definition \ref{def:branch-fixed} hold.
So $T'$ has the form of extending a subladder of $T$
which does not contain the subsidiary edges of branching vertices,
as in the example of Figure \ref{Fig:42},
boxed subladders of which are denoted $L_1,L_2,...,L_5$.

\begin{figure}[h]

\begin{picture}(200,80)(-90,-80)

\put(6,4){\line(1,0){8}}
\put(6,4){\line(0,-1){8}}
\put(10,0){\circle*{2}}
\put(14,4){\line(0,-1){8}}
\put(6,-4){\line(1,0){8}}
\put(14,-4){$L_1$}

\put(10,0){\line(-1,-2){10}}
\put(5,-10){1}

\put(-4,-16){\line(1,0){8}}
\put(-4,-16){\line(0,-1){28}}
\put(0,-20){\circle*{2}}
\put(0,-20){\line(0,-1){20}}
\put(0,-40){\circle*{2}}
\put(4,-16){\line(0,-1){28}}
\put(-4,-44){\line(1,0){8}}
\put(-35,-30){$T = L_2$}

\put(10,0){\line(1,-2){10}}
\put(15,-10){2}

\put(16,-16){\line(1,0){8}}
\put(16,-16){\line(0,-1){28}}
\put(20,-20){\circle*{2}}
\put(20,-20){\line(0,-1){20}}
\put(20,-40){\circle*{2}}
\put(24,-16){\line(0,-1){28}}
\put(16,-44){\line(1,0){8}}
\put(24,-30){$L_3$}

\put(20,-40){\line(-1,-2){10}}
\put(15,-50){1}

\put(6,-56){\line(1,0){8}}
\put(6,-56){\line(0,-1){8}}
\put(10,-60){\circle*{2}}
\put(14,-56){\line(0,-1){8}}
\put(6,-64){\line(1,0){8}}
\put(6,-74){$L_4$}

\put(20,-40){\line(1,-2){10}}
\put(25,-50){2}

\put(26,-56){\line(1,0){8}}
\put(26,-56){\line(0,-1){8}}
\put(30,-60){\circle*{2}}
\put(34,-56){\line(0,-1){8}}
\put(26,-64){\line(1,0){8}}
\put(26,-74){$L_5$}

\put(-70,-86){$L_1$ only grows upward by ii) of
             Definition \ref{def:branch-fixed}.}

\end{picture}

\caption{}              \label{Fig:42}

\end{figure}

So, the action of $d^i$ on $T \otimes det(T)$ is by extending a ladder of $T$,
getting a new edge $e$ and changing $det(T)$ into $e \wedge det(T)$.
Acting by $d^i$'s on $T \otimes det(T)$,
the possible bonsais appearing in the $d^i(T)$ are obtained
by extending a subladder of $T$ as in the example of Figure \ref{Fig:42}.

keeping this intuitive fact in mind, we have some definitions;
\begin{df}
A {\em seedling} is an $m$-bonsai all of whose vertices other than
tips are branching vertices. For example, in 2-bonsai, $S_1$ of
Figure \ref{Fig:43} is a seedling, but $S_2$ is not, because the
root vertex is not a branching vertex. In other words, a seedling is
a bonsai which cannot be obtained from another bonsai by adding an
edge so that i) and ii) of Definition \ref{def:branch-fixed} are
satisfied.

\begin{figure}[h]

\begin{picture}(200,70)(-100,-50)

\put(-70,-20){$S_1=$}
\put(-35,0){\circle*{2}}
\put(-35,0){\line(-1,-2){10}}
\put(-40,-10){1}
\put(-45,-20){\circle*{2}}
\put(-35,0){\line(1,-2){10}}
\put(-30,-10){2}
\put(-25,-20){\circle*{2}}

\put(-25,-20){\circle*{2}}
\put(-25,-20){\line(-1,-2){10}}
\put(-30,-30){1}
\put(-35,-40){\circle*{2}}
\put(-25,-20){\line(1,-2){10}}
\put(-20,-30){2}
\put(-15,-40){\circle*{2}}

\put(-45,-50){seedling}

\put(0,-20){$S_2=$}
\put(35,0){\line(0,1){20}}
\put(35,10){1}
\put(35,20){\circle*{2}}
\put(35,0){\circle*{2}}
\put(35,0){\line(-1,-2){10}}
\put(30,-10){1}
\put(25,-20){\circle*{2}}
\put(35,0){\line(1,-2){10}}
\put(40,-10){2}
\put(45,-20){\circle*{2}}

\put(45,-20){\circle*{2}}
\put(45,-20){\line(-1,-2){10}}
\put(40,-30){1}
\put(35,-40){\circle*{2}}
\put(45,-20){\line(1,-2){10}}
\put(50,-30){2}
\put(55,-40){\circle*{2}}

\put(25,-50){not seedling}

\end{picture}

\caption{}              \label{Fig:43}

\end{figure}

\end{df}

\begin{df}
Let ${\mathbf C}^{S,0}$ be the submodule of ${\mathcal C}^i$, where
$S$ is a seedling and $i$ is the number of edges of $S$, generated
by $S \otimes det(S)$. Let ${\mathbf C}^{S,j}$ ($j \geq 0$) be the
submodule of ${\mathcal C}^{i+j}$ generated by $T \otimes det(T)$,
where $T$ is an $m$-bonsai obtained by adding $j$ edges to $S$ so
that i) and ii) of Definition \ref{def:branch-fixed} are satisfied.
\end{df}

Then, every ${\mathcal C}^i$ is the direct sum of some ${\mathbf C}^{S,j}$'s
and $d({\mathbf C}^{S,j}) \subset {\mathbf C}^{S,j+1}$.
For example, in 2-bonsai, when $S_0$, $S_1$, $S_2$ and $S_3$
are as given in Figure \ref{Fig:44},
we have
\begin{eqnarray}
{\mathcal C}^0 & = & {\mathbf C}^{S_0,0},                                               \\
{\mathcal C}^1 & = & {\mathbf C}^{S_0,1},                                     \notag    \\
{\mathcal C}^2 & = & {\mathbf C}^{S_0,2} \oplus {\mathbf C}^{S_1,0},          \notag    \\
{\mathcal C}^3 & = & {\mathbf C}^{S_0,3} \oplus {\mathbf C}^{S_1,1},          \notag    \\
{\mathcal C}^4 & = & {\mathbf C}^{S_0,4} \oplus {\mathbf C}^{S_1,2} \oplus
                    {\mathbf C}^{S_0,3} \oplus {\mathbf C}^{S_1,1},          \notag    \\
{\mathcal C}^5 & = & {\mathbf C}^{S_0,5} \oplus {\mathbf C}^{S_1,3} \oplus
                    {\mathbf C}^{S_0,3} \oplus {\mathbf C}^{S_1,1},          \notag    \\
 ......       &   &                                                          \notag
\end{eqnarray}

\begin{figure}[h]

\begin{picture}(200,40)(0,-40)

\put(-65,-20){$S_0=\cdot$}

\put(-25,-20){$S_1=$}
\put(10,-10){\circle*{2}}
\put(10,-10){\line(-1,-2){10}}
\put(5,-20){1}
\put(0,-30){\circle*{2}}
\put(10,-10){\line(1,-2){10}}
\put(15,-20){2}
\put(20,-30){\circle*{2}}

\put(30,-20){$S_2=$}
\put(65,0){\circle*{2}}
\put(65,0){\line(-1,-2){10}}
\put(60,-10){1}
\put(55,-20){\circle*{2}}
\put(65,0){\line(1,-2){10}}
\put(70,-10){2}
\put(75,-20){\circle*{2}}

\put(75,-20){\circle*{2}}
\put(75,-20){\line(-1,-2){10}}
\put(70,-30){1}
\put(65,-40){\circle*{2}}
\put(75,-20){\line(1,-2){10}}
\put(80,-30){2}
\put(85,-40){\circle*{2}}

\put(95,-20){$S_3=$}
\put(140,0){\circle*{2}}
\put(140,0){\line(-1,-2){10}}
\put(135,-10){1}
\put(130,-20){\circle*{2}}
\put(140,0){\line(1,-2){10}}
\put(145,-10){2}
\put(150,-20){\circle*{2}}

\put(130,-20){\circle*{2}}
\put(130,-20){\line(-1,-2){10}}
\put(125,-30){1}
\put(120,-40){\circle*{2}}
\put(130,-20){\line(1,-2){10}}
\put(135,-30){2}
\put(140,-40){\circle*{2}}

\end{picture}

\caption{}              \label{Fig:44}

\end{figure}

\begin{df}
For a given seedling $S$, when $i$ is the number of edges of $S$,
$\{{\mathbf C}^{S,j},d^{i+j}\}_{j \geq 0}$ is called a {\em thread}
of $S$.
\end{df}

So the cohomology groups of $\{{\mathcal C}^i,d^i\}$ are the direct sum
of cohomology groups of threads $\{{\mathbf C}^{S,j},d^{i+j}\}_{j \geq 0}$.

Let us look into each of these threads.
For ${\mathbf C}^{S_0,0}$, where $S_0$ is a vertex,
${\mathbf C}^{S_0,i}$ is the module with the basis $\{T \otimes det(T)\}$,
where $T$ is a ladder with $i$ edges and the boundary maps extend the ladders
by adding an edge $e$ and replacing $det(T)$ with $e \wedge det(T)$.
Let us consider a chain complex which is isomorphic to the thread
${\mathbf C}^{S_0,0}$ of the ladder $S_0$.
For any $m$-bonsai, consider a vector space $V$
which has a basis $\{v_1,...,v_m\}$,
and let $V_n = V^{\otimes n}(n \geq 1)$.
Then we define a map $\delta ^n : V^n \to V^{n+1}$ as

\begin{eqnarray}
v_{i_1} \otimes v_{i_2} \otimes ... \otimes v_{i_n} \mapsto
& &\sum_{k=1}^{m} v_k \otimes v_{i_1} \otimes v_{i_2}
   \otimes ... \otimes v_{i_n} \\  \notag
&+&\sum_{k=1}^{m} (-1)^1 v_{i_1} \otimes v_k \otimes v_{i_2}
   \otimes ... \otimes v_{i_n} \\ \notag
&+&... \\ \notag
&+&\sum_{k=1}^{m} (-1)^n v_{i_1} \otimes ... \otimes v_{i_{n-1}}
   \otimes v_{i_n} \otimes v_k, \notag
\end{eqnarray}
and it is easily seen that this $\delta ^n$ is a boundary map, so we
have made $\{V^n,\delta^n\}$ a cochain complex. By the cochain map
$f$ as in Figure \ref{Fig:45}, the cochain complexes $\{{\mathbf
C}^{S_0,n}, d^n\}$ and $\{V^n,\delta^n\}$ are isomorphic, since
$d^n$ acts as in Figure \ref{Fig:46}, that is, we have $f \circ
\delta^n = d^n \circ f$.

\begin{figure}[h]

\begin{picture}(200,60)(-30,-60)

\put(0,0){\circle*{2}}
\put(0,0){\line(0,-1){15}}
\put(0,-10){$i_1$}
\put(0,-15){\circle*{2}}
\put(0,-15){\line(0,-1){15}}
\put(0,-25){$i_2$}
\put(0,-30){\circle*{2}}
\put(0,-35){.}
\put(0,-40){.}
\put(0,-45){.}
\put(0,-45){\circle*{2}}
\put(0,-45){\line(0,-1){15}}
\put(0,-55){$i_n$}
\put(0,-60){\circle*{2}}

\put(10,-25){$\otimes det($}

\put(37,0){\circle*{2}}
\put(37,0){\line(0,-1){15}}
\put(37,-10){$i_1$}
\put(37,-15){\circle*{2}}
\put(37,-15){\line(0,-1){15}}
\put(37,-25){$i_2$}
\put(37,-30){\circle*{2}}
\put(37,-35){.}
\put(37,-40){.}
\put(37,-45){.}
\put(37,-45){\circle*{2}}
\put(37,-45){\line(0,-1){15}}
\put(37,-55){$i_n$}
\put(37,-60){\circle*{2}}

\put(47,-25){$) \quad \mapsto \quad v_{i_1} \otimes v_{i_2} \otimes ... \otimes v_{i_n}$}

\end{picture}

\caption{}              \label{Fig:45}

\end{figure}

\begin{figure}[h]

\begin{picture}(300,190)(0,-165)

\put(-30,0){\circle*{2}}
\put(-30,0){\line(0,-1){15}}
\put(-30,-10){$i_1$}
\put(-30,-15){\circle*{2}}
\put(-30,-15){\line(0,-1){15}}
\put(-30,-25){$i_2$}
\put(-30,-30){\circle*{2}}
\put(-30,-35){.}
\put(-30,-40){.}
\put(-30,-45){.}
\put(-30,-45){\circle*{2}}
\put(-30,-45){\line(0,-1){15}}
\put(-30,-55){$i_n$}
\put(-30,-60){\circle*{2}}

\put(-20,-25){$\otimes det($}

\put(7,0){\circle*{2}}
\put(7,0){\line(0,-1){15}}
\put(7,-10){$i_1$}
\put(7,-15){\circle*{2}}
\put(7,-15){\line(0,-1){15}}
\put(7,-25){$i_2$}
\put(7,-30){\circle*{2}}
\put(7,-35){.}
\put(7,-40){.}
\put(7,-45){.}
\put(7,-45){\circle*{2}}
\put(7,-45){\line(0,-1){15}}
\put(7,-55){$i_n$}
\put(7,-60){\circle*{2}}

\put(17,-25){$) \mapsto \sum_{k=1}^{n} [$}

\put(70,0){\line(0,1){15}}
\put(70,15){\circle*{2}}
\put(70,5){$k$}

\put(70,0){\circle*{2}}
\put(70,0){\line(0,-1){15}}
\put(70,-10){$i_1$}
\put(70,-15){\circle*{2}}
\put(70,-15){\line(0,-1){15}}
\put(70,-25){$i_2$}
\put(70,-30){\circle*{2}}
\put(70,-35){.}
\put(70,-40){.}
\put(70,-45){.}
\put(70,-45){\circle*{2}}
\put(70,-45){\line(0,-1){15}}
\put(70,-55){$i_n$}
\put(70,-60){\circle*{2}}

\put(80,-25){$\otimes \quad \wedge det($}

\put(90,-15){\circle*{2}}
\put(90,-15){\line(0,-1){15}}
\put(92,-25){$k$}
\put(90,-30){\circle*{2}}

\put(127,0){\circle*{2}}
\put(127,0){\line(0,-1){15}}
\put(127,-10){$i_1$}
\put(127,-15){\circle*{2}}
\put(127,-15){\line(0,-1){15}}
\put(127,-25){$i_2$}
\put(127,-30){\circle*{2}}
\put(127,-35){.}
\put(127,-40){.}
\put(127,-45){.}
\put(127,-45){\circle*{2}}
\put(127,-45){\line(0,-1){15}}
\put(127,-55){$i_n$}
\put(127,-60){\circle*{2}}

\put(135,-25){$)+......+$}

\put(180,0){\circle*{2}}
\put(180,0){\line(0,-1){15}}
\put(180,-10){$i_1$}
\put(180,-15){\circle*{2}}
\put(180,-15){\line(0,-1){15}}
\put(180,-25){$i_2$}
\put(180,-30){\circle*{2}}
\put(180,-35){.}
\put(180,-40){.}
\put(180,-45){.}
\put(180,-45){\circle*{2}}
\put(180,-45){\line(0,-1){15}}
\put(180,-55){$i_n$}
\put(180,-60){\circle*{2}}

\put(180,-60){\line(0,-1){15}}
\put(180,-75){\circle*{2}}
\put(180,-70){$k$}

\put(190,-25){$\otimes \quad \wedge det($}

\put(200,-15){\circle*{2}}
\put(200,-15){\line(0,-1){15}}
\put(202,-25){$k$}
\put(200,-30){\circle*{2}}

\put(237,0){\circle*{2}}
\put(237,0){\line(0,-1){15}}
\put(237,-10){$i_1$}
\put(237,-15){\circle*{2}}
\put(237,-15){\line(0,-1){15}}
\put(237,-25){$i_2$}
\put(237,-30){\circle*{2}}
\put(237,-35){.}
\put(237,-40){.}
\put(237,-45){.}
\put(237,-45){\circle*{2}}
\put(237,-45){\line(0,-1){15}}
\put(237,-55){$i_n$}
\put(237,-60){\circle*{2}}

\put(245,-25){$)]$}

\put(17,-125){$=  \sum_{k=1}^{n} [$}

\put(65,-100){\line(0,1){15}}
\put(65,-85){\circle*{2}}
\put(65,-95){$k$}

\put(65,-100){\circle*{2}}
\put(65,-100){\line(0,-1){15}}
\put(65,-110){$i_1$}
\put(65,-115){\circle*{2}}
\put(65,-115){\line(0,-1){15}}
\put(65,-125){$i_2$}
\put(65,-130){\circle*{2}}
\put(65,-135){.}
\put(65,-140){.}
\put(65,-145){.}
\put(65,-145){\circle*{2}}
\put(65,-145){\line(0,-1){15}}
\put(65,-155){$i_n$}
\put(65,-160){\circle*{2}}

\put(75,-125){$\otimes (-1)^0 det($}

\put(127,-100){\line(0,1){15}}
\put(127,-85){\circle*{2}}
\put(127,-95){$k$}

\put(127,-100){\circle*{2}}
\put(127,-100){\line(0,-1){15}}
\put(127,-110){$i_1$}
\put(127,-115){\circle*{2}}
\put(127,-115){\line(0,-1){15}}
\put(127,-125){$i_2$}
\put(127,-130){\circle*{2}}
\put(127,-135){.}
\put(127,-140){.}
\put(127,-145){.}
\put(127,-145){\circle*{2}}
\put(127,-145){\line(0,-1){15}}
\put(127,-155){$i_n$}
\put(127,-160){\circle*{2}}

\put(135,-125){$)+......+$}

\put(180,-100){\circle*{2}}
\put(180,-100){\line(0,-1){15}}
\put(180,-110){$i_1$}
\put(180,-115){\circle*{2}}
\put(180,-115){\line(0,-1){15}}
\put(180,-125){$i_2$}
\put(180,-130){\circle*{2}}
\put(180,-135){.}
\put(180,-140){.}
\put(180,-145){.}
\put(180,-145){\circle*{2}}
\put(180,-145){\line(0,-1){15}}
\put(180,-155){$i_n$}
\put(180,-160){\circle*{2}}

\put(180,-160){\line(0,-1){15}}
\put(180,-175){\circle*{2}}
\put(180,-170){$k$}

\put(190,-125){$\otimes (-1)^n det($}

\put(242,-100){\circle*{2}}
\put(242,-100){\line(0,-1){15}}
\put(242,-110){$i_1$}
\put(242,-115){\circle*{2}}
\put(242,-115){\line(0,-1){15}}
\put(242,-125){$i_2$}
\put(242,-130){\circle*{2}}
\put(242,-135){.}
\put(242,-140){.}
\put(242,-145){.}
\put(242,-145){\circle*{2}}
\put(242,-145){\line(0,-1){15}}
\put(242,-155){$i_n$}
\put(242,-160){\circle*{2}}

\put(242,-160){\line(0,-1){15}}
\put(242,-175){\circle*{2}}
\put(242,-170){$k$}

\put(250,-125){$)]$}

\end{picture}

\caption{}              \label{Fig:46}

\end{figure}

In $\{V^n,\delta ^n\}_{n \geq 1}$,
by the definition of $\delta ^n$, inductively we have

\begin{eqnarray}
\\
\delta((v_1 \otimes ... \otimes v_i) \otimes v)
= \delta(v_1 \otimes ... \otimes v_i) \otimes v
+ (-1)^{i+1} (v_1 \otimes ... \otimes v_i) \otimes v \otimes \sum_{k=1}^{m} v_k
\notag
\end{eqnarray}
where $v_1 \otimes ... \otimes v_i \in V^i$ and $v \in V$.
Suppose  $\delta(\sum_{k=1}^m v'_k \otimes v_k) = 0$ where $v'_k \in V^i$,
then we have
\begin{eqnarray}
\sum_{k=1}^{m} \delta(v'_k) \otimes v_k +
 (-1)^{i+1} (\sum_{k=1}^m v'_k \otimes v_k) \otimes
\sum_{l=1}^{m} v_l \\ \notag
=\sum_{l=1}^{m} \{\delta(v'_l) - (-1)^k(\sum_{k=1}^m v'_k \otimes v_k)\}
\otimes v_l = 0.
\end{eqnarray}
Therefore, we have $(\sum_{k=1}^m v'_k \otimes v_k) = (-1)^k
\delta(v'_l)$ and it is a coboundary. So, $\{V^n, \delta ^n\}$ is
acyclic, and so is $\{{\mathbf C}^{S_0,n}, d^n\}$.

Here, we can directly calculate $H^0({\mathbf C}^{S_0,*})=0$,
since the boundary map image of a one-vertex bonsai is
the sum of one-edge bonsais over all labels 1,2,...,$n$.
So $\{{\mathbf C}^{S_0,n},d^n\}$ is acyclic with $H^0=0$.

Now for an arbitrary seedling $S$, when $S$ has $n$ edges,
there are $n+1$ vertices and
each vertex other than the root has one and only one edge
whose branch-end is that vertex.
When we order the edges of a bonsai $T$ with the shape $S$
as in Figure \ref{Fig:31}
and denote them as $e_l$'s ($l=1,2,...,n$),
we can denote the branch-end vertex of $e_l$
as $v_l$ and denote the root $v_0$.
Then the bonsais which appear in the basis of ${\mathbf C}^{S,j}$ are obtained
by extending the vertices of $T$ into upward-growing ladders,
and each ladder grown from $v_l$ is denoted as $L_l$,
as in the example of Figure \ref{Fig:47}.

\begin{figure}[h]

\begin{picture}(200,100)(-100,-100)

\put(6,4){\line(1,0){8}}
\put(6,4){\line(0,-1){8}}
\put(10,0){\circle*{2}}
\put(14,4){\line(0,-1){8}}
\put(6,-4){\line(1,0){8}}
\put(14,-4){$L_0$}

\put(10,0){\line(-1,-2){20}}
\put(-5,-20){$e_1$}

\put(-14,-36){\line(1,0){8}}
\put(-14,-36){\line(0,-1){8}}
\put(-10,-40){\circle*{2}}
\put(-6,-36){\line(0,-1){8}}
\put(-14,-44){\line(1,0){8}}
\put(-25,-40){$L_1$}

\put(10,0){\line(1,-2){20}}
\put(20,-20){$e_2$}

\put(26,-36){\line(1,0){8}}
\put(26,-36){\line(0,-1){8}}
\put(30,-40){\circle*{2}}
\put(34,-36){\line(0,-1){8}}
\put(26,-44){\line(1,0){8}}
\put(34,-40){$L_2$}

\put(30,-40){\line(-1,-2){20}}
\put(15,-60){$e_3$}

\put(6,-76){\line(1,0){8}}
\put(6,-76){\line(0,-1){8}}
\put(10,-80){\circle*{2}}
\put(14,-76){\line(0,-1){8}}
\put(6,-84){\line(1,0){8}}
\put(6,-94){$L_3$}

\put(30,-40){\line(1,-2){20}}
\put(40,-60){$e_4$}

\put(46,-76){\line(1,0){8}}
\put(46,-76){\line(0,-1){8}}
\put(50,-80){\circle*{2}}
\put(54,-76){\line(0,-1){8}}
\put(46,-84){\line(1,0){8}}
\put(46,-94){$L_4$}

\end{picture}

\caption{}              \label{Fig:47}

\end{figure}

To get the cohomology of ${\mathbf C}^{S,j}$,
let us consider the complexes ${\mathbf C}_k^{S_0,p}$,
where $k=0,...,n$ and each of ${\mathbf C}_k^{S_0,p}$
is a copy of ${\mathbf C}^{S_0,p}$,
i.e., each of ${\mathbf C}_k^{S_0,p}$
has the basis $\{L_k^p \otimes det(L_k^p)\}$,
where $L_k^p$ is a ladder with $p$ edges.
Then we have an isomorphism $F$ between
\begin{eqnarray}
{\mathbf D}^l = \bigoplus _{p_0+...+p_n=l} {\mathbf C}_0^{S_0,p_0}
                \otimes ... \otimes  {\mathbf C}_n^{S_0,p_n}
\end{eqnarray}
and ${\mathbf C}^{S,l}$ given by
\begin{eqnarray}
& L_0^{p_0} \otimes det(L_0^{p_0}) \otimes ... \otimes
  L_n^{p_n} \otimes det(L_n^{p_n}) \mapsto \\
& \Sigma (\mbox{The bonsai obtained by putting
                $L_i^{p_0}$ into the place of vertex $v_i$}) \notag \\
& \otimes det(L_0^{p_0}) \wedge e_1 \wedge
  det(L_1^{p_1}) \wedge ... \wedge e_n \wedge det(L_n^{p_n}) \notag
\end{eqnarray}
as in the example of Figure \ref{Fig:48},
for the seedling of Figure \ref{Fig:47}.

\begin{figure}[h]

\begin{picture}(300,200)(0,-150)

\put(0,30){\circle*{2}}
\put(0,30){\line(0,-1){10}}
\put(0,20){\circle*{2}}
\put(0,20){1}
\put(5,20){$\otimes det($}
\put(35,30){\circle*{2}}
\put(35,30){\line(0,-1){10}}
\put(35,20){\circle*{2}}
\put(35,20){1}
\put(40,20){$) \otimes$}
\put(55,20){\circle*{2}}
\put(60,20){$\otimes det($}
\put(87,20){\circle*{2}}
\put(90,20){$) \otimes$}
\put(105,35){\circle*{2}}
\put(105,35){\line(0,-1){10}}
\put(105,25){\circle*{2}}
\put(105,25){2}
\put(105,25){\line(0,-1){10}}
\put(105,15){\circle*{2}}
\put(105,15){1}
\put(110,20){$\otimes det($}
\put(135,35){\circle*{2}}
\put(135,35){\line(0,-1){10}}
\put(135,25){\circle*{2}}
\put(135,25){2}
\put(135,25){\line(0,-1){10}}
\put(135,15){\circle*{2}}
\put(135,15){1}
\put(140,20){$) \otimes$}
\put(155,20){\circle*{2}}
\put(160,20){$\otimes det($}
\put(187,20){\circle*{2}}
\put(190,20){$) \otimes$}
\put(202,30){\circle*{2}}
\put(202,30){\line(0,-1){10}}
\put(202,20){\circle*{2}}
\put(204,20){2}
\put(210,20){$\otimes det($}
\put(240,30){\circle*{2}}
\put(240,30){\line(0,-1){10}}
\put(240,20){\circle*{2}}
\put(240,20){2}
\put(245,20){$) \qquad \mapsto$}

\put(6,14){\line(1,0){8}}
\put(6,14){\line(0,-1){18}}
\put(10,0){\circle*{2}}
\put(10,0){1}
\put(10,0){\line(0,1){10}}
\put(10,10){\circle*{2}}
\put(14,14){\line(0,-1){18}}
\put(6,-4){\line(1,0){8}}
\put(14,-4){$L_0$}

\put(10,0){\line(-1,-2){10}}
\put(0,-10){$e_1$}

\put(-4,-16){\line(1,0){8}}
\put(-4,-16){\line(0,-1){8}}
\put(0,-20){\circle*{2}}
\put(4,-16){\line(0,-1){8}}
\put(-4,-24){\line(1,0){8}}
\put(-15,-20){$L_1$}

\put(10,0){\line(1,-2){10}}
\put(15,-10){$e_2$}

\put(16,-16){\line(1,0){10}}
\put(16,-16){\line(0,-1){28}}
\put(20,-20){\circle*{2}}
\put(20,-20){\line(0,-1){10}}
\put(20,-30){\circle*{2}}
\put(20,-30){2}
\put(20,-30){\line(0,-1){10}}
\put(20,-40){\circle*{2}}
\put(20,-40){1}
\put(26,-16){\line(0,-1){28}}
\put(16,-44){\line(1,0){10}}
\put(26,-20){$L_2$}

\put(20,-40){\line(-1,-2){10}}
\put(10,-50){$e_3$}

\put(6,-56){\line(1,0){8}}
\put(6,-56){\line(0,-1){8}}
\put(10,-60){\circle*{2}}
\put(14,-56){\line(0,-1){8}}
\put(6,-64){\line(1,0){8}}
\put(6,-74){$L_3$}

\put(20,-40){\line(1,-2){10}}
\put(25,-50){$e_4$}

\put(26,-56){\line(1,0){10}}
\put(26,-56){\line(0,-1){18}}
\put(30,-60){\circle*{2}}
\put(30,-60){\line(0,-1){10}}
\put(30,-70){\circle*{2}}
\put(30,-70){2}
\put(36,-56){\line(0,-1){18}}
\put(26,-74){\line(1,0){10}}
\put(36,-64){$L_4$}

\put(40,-30){$\otimes det($}
\put(70,-30){\circle*{2}}
\put(70,-30){\line(0,1){10}}
\put(70,-30){1}
\put(70,-20){\circle*{2}}
\put(75,-30){$)\wedge e_1 \wedge det($}
\put(130,-30){\circle*{2}}
\put(135,-30){$)\wedge e_2 \wedge det($}
\put(190,-35){\circle*{2}}
\put(190,-35){\line(0,1){10}}
\put(190,-35){1}
\put(190,-25){\circle*{2}}
\put(190,-25){\line(0,1){10}}
\put(190,-25){2}
\put(190,-15){\circle*{2}}
\put(195,-30){$)\wedge e_3 \wedge det($}
\put(250,-30){\circle*{2}}
\put(255,-30){$)\wedge e_4 \wedge det($}
\put(310,-30){\circle*{2}}
\put(310,-30){\line(0,1){10}}
\put(310,-30){2}
\put(310,-20){\circle*{2}}
\put(315,-30){$) \qquad =$}

\put(6,-86){\line(1,0){8}}
\put(6,-86){\line(0,-1){18}}
\put(10,-100){\circle*{2}}
\put(10,-100){1}
\put(10,-100){\line(0,1){10}}
\put(10,-90){\circle*{2}}
\put(14,-86){\line(0,-1){18}}
\put(6,-104){\line(1,0){8}}
\put(14,-104){$L_0$}

\put(10,-100){\line(-1,-2){10}}
\put(0,-110){$e_1$}

\put(-4,-116){\line(1,0){8}}
\put(-4,-116){\line(0,-1){8}}
\put(0,-120){\circle*{2}}
\put(4,-116){\line(0,-1){8}}
\put(-4,-124){\line(1,0){8}}
\put(-15,-120){$L_1$}

\put(10,-100){\line(1,-2){10}}
\put(15,-110){$e_2$}

\put(16,-116){\line(1,0){10}}
\put(16,-116){\line(0,-1){28}}
\put(20,-120){\circle*{2}}
\put(20,-120){\line(0,-1){10}}
\put(20,-130){\circle*{2}}
\put(20,-130){2}
\put(20,-130){\line(0,-1){10}}
\put(20,-140){\circle*{2}}
\put(20,-140){1}
\put(26,-116){\line(0,-1){28}}
\put(16,-144){\line(1,0){10}}
\put(26,-120){$L_2$}

\put(20,-140){\line(-1,-2){10}}
\put(10,-150){$e_3$}

\put(6,-156){\line(1,0){8}}
\put(6,-156){\line(0,-1){8}}
\put(10,-160){\circle*{2}}
\put(14,-156){\line(0,-1){8}}
\put(6,-164){\line(1,0){8}}
\put(6,-174){$L_3$}

\put(20,-140){\line(1,-2){10}}
\put(25,-150){$e_4$}

\put(26,-156){\line(1,0){10}}
\put(26,-156){\line(0,-1){18}}
\put(30,-160){\circle*{2}}
\put(30,-160){\line(0,-1){10}}
\put(30,-170){\circle*{2}}
\put(30,-170){2}
\put(36,-156){\line(0,-1){18}}
\put(26,-174){\line(1,0){10}}
\put(36,-164){$L_4$}

\put(40,-130){$\otimes det($}
\put(70,-130){\circle*{2}}
\put(70,-130){\line(0,1){10}}
\put(70,-130){1}
\put(70,-120){\circle*{2}}
\put(75,-130){$)\wedge e_1 \wedge e_2 \wedge det($}
\put(150,-135){\circle*{2}}
\put(150,-135){\line(0,1){10}}
\put(150,-135){1}
\put(150,-125){\circle*{2}}
\put(150,-125){\line(0,1){10}}
\put(150,-125){2}
\put(150,-115){\circle*{2}}
\put(155,-130){$)\wedge e_3 \wedge e_4 \wedge det($}
\put(230,-130){\circle*{2}}
\put(230,-130){\line(0,1){10}}
\put(230,-130){2}
\put(230,-120){\circle*{2}}
\put(235,-130){$)$}

\end{picture}

\caption{}              \label{Fig:48}

\end{figure}

From now on, we write $L_i^{p_i}$ just as $L_i$ for convenience of writing.

In $T \otimes det(T) \in {\mathbf C}^{S,l}$, $det(T)$ is

\begin{eqnarray} \label{eq:det}
det(L_0) \wedge e_{1,k_1} \wedge det(L_1) \wedge ... \wedge e_{n,k_n} \wedge det(L_n)
\end{eqnarray}
where $k_l$ is the label of the edge $e_l$,
and $d^n(T \otimes det(T))$ is

\begin{eqnarray}
&\Sigma
(\mbox{A bonsai $T'$ obtained by adding a new edge $f$ to one of the $L_i$})
  \otimes f \wedge det(T) \notag
\end{eqnarray}
and this is

\begin{eqnarray}
  & \Sigma(\mbox{A bonsai $T'$ obtained
               by adding a new edge $f$ to one of $L_i$})  \notag \\
  &  \otimes f \wedge det(L_0) \wedge e_{1,k_1} \wedge det(L_1)
     \wedge ... \wedge e_{n,k_n} \wedge det(L_n)  \notag \\
        \notag \\
= & \Sigma(\mbox{A bonsai $T'$ obtained
                 by adding a new edge $f$ to one of $L_i$})  \notag \\
  & \otimes  (-1)^{\beta}
     det(L_0) \wedge e_{1,k_1} \wedge det(L_1) \wedge ...
    \wedge e_{i,k_i} \wedge f \wedge det(L_i) \wedge ...
       \wedge e_{n,k_n} \wedge det(L_n)  \notag
\end{eqnarray}
where $\beta = (deg(L_0)+1)+(deg(L_1)+1)+...+(deg(L_{i-1})+1)$.

Here, $det(T')$ is obtained by replacing $det(L_i)$
with $f \wedge det(L_i)$ in  \eqref{eq:det},
and we get the sign $(-1)^{\beta}$
since when $f$ is added to $L_i$,
in the ordering of edges of $T'$,
the edges of $L_j(j=0,1,...,i-1)$ and edges
$e_1$, $e_2$,...,$e_i$ are prior to the edges of $L_i$.
So we can conclude that $d(T \otimes det(T))$ is the sum of
$(-1)^{deg(L_0)+1+...+deg(L_{i-1})+1} T' \otimes det(T')$,
where $T'$ is the bonsai obtained by adding a new edge in the subladder $L_i$.

Thus, when we construct a coboundary map $\partial ^l$ for $\{{\mathbf D}^l\}$,
acting on
${\mathbf C}_1 ^{S_0,p_1} \otimes ... \otimes {\mathbf C}_n ^{S_0,p_n}$ by

\begin{eqnarray}
& &\partial^l(L_0 \otimes det(L_0) \otimes ... \otimes L_n \otimes det(L_n))
   \notag\\
&=&d(L_0 \otimes det(L_0)) \otimes   L_1 \otimes det(L_1)
   \otimes ... \otimes L_n \otimes det(L_n)      \notag \\
&+& (-1)^{deg(L_0)+1}
    L_0 \otimes det(L_0)  \otimes d(L_1 \otimes det(L_1))
   \otimes ... \otimes L_n \otimes det(L_n)      \notag \\
&+& ...                                          \notag \\
&+& (-1)^{(deg(L_0)+1)+...+(deg(L_{n-1})+1)}
    L_0 \otimes det(L_0)  \otimes L_1 \otimes det(L_1)
   \otimes ... \otimes d(L_n \otimes det(L_n)),      \notag
\end{eqnarray}
the isomorphism $F$ becomes a cochain isomorphism
between $\{{\mathbf D}^l\}$ and $\{{\mathbf C}^{S,l}\}$,
and since each $\{{\mathbf C}_k ^{S_0,p_k}\}$ is acyclic and $H^0 =0$,
 by the K{\" u}nneth theorem,
$\{{\mathbf D}^l\}$ is acyclic and $H^0 =0$,
and so is $\{{\mathbf C}^{S,l}\}$.

Then, by a K{\" u}nneth argument,
every thread of $\{{\mathbf C}^{S,j},d^{i+j}\}$
($i$ is the number of edges of $S$) is acyclic with $H^0=0$,
and so we have the
\begin{thm} \label{thm:acyclic}
$\{{\mathcal C}^n,d^n\}$ is acyclic.
\end{thm}

\section{Vertex-appending Differential} \label{Section:vadiff}

In this section, we consider a new differential,
different from the ladder-extension.
Again, all bonsais of this section are $m$-bonsais.

\begin{df} \label{Def:vadiff}
We define the {\em vertex-appending differential} $\partial$ as follows;
Consider a determinanted bonsai $T \otimes det(T)$.
Then $\partial(T \otimes det(T))$ is the sum of $T' \otimes e \wedge det(T)$,
where $T'$ is a bonsai obtained by

i) appending a vertex to $T$

ii) except to tips of $T$,

and so, getting a new edge $e$.

If there is no available appending position an a bonsai, the map
$\partial$ assigns 0 to that bonsai.
\end{df}

For example, in 3-bonsai, we have an example like Figure \ref{Fig:49}
(in bonsais, determinanted terms are omitted.
Note that one vertex in the first example is also a tip
and the fourth bonsai is a cocycle).

\begin{figure}[h]

\begin{picture}(200,140)(0,-130)

\put(10,-10){\circle*{2}}

\put(25,-10){$\rightarrow \qquad 0$}

\put(10,-40){\circle*{2}}
\put(10,-40){\line(0,-1){20}}
\put(10,-50){2}
\put(10,-60){\circle*{2}}

\put(25,-50){$\rightarrow$}

\put(60,-40){\circle*{2}}
\put(60,-40){\line(-1,-2){10}}
\put(50,-60){\circle*{2}}
\put(55,-50){1}
\put(60,-40){\line(1,-2){10}}
\put(70,-60){\circle*{2}}
\put(65,-50){2}

\put(80,-50){$-$}

\put(100,-40){\circle*{2}}
\put(100,-40){\line(-1,-2){10}}
\put(90,-60){\circle*{2}}
\put(95,-50){2}
\put(100,-40){\line(1,-2){10}}
\put(110,-60){\circle*{2}}
\put(105,-50){3}

\put(10,-80){\circle*{2}}
\put(10,-80){\line(-1,-2){10}}
\put(0,-100){\circle*{2}}
\put(5,-90){2}
\put(10,-80){\line(1,-2){10}}
\put(20,-100){\circle*{2}}
\put(15,-90){3}

\put(25,-90){$\rightarrow$}

\put(90,-80){\circle*{2}}
\put(90,-80){\line(-2,-1){40}}
\put(50,-100){\circle*{2}}
\put(70,-90){1}
\put(90,-80){\line(-1,-2){10}}
\put(80,-100){\circle*{2}}
\put(85,-90){2}
\put(90,-80){\line(1,-2){10}}
\put(100,-100){\circle*{2}}
\put(95,-90){3}

\put(10,-120){\circle*{2}}
\put(10,-120){\line(-1,-2){10}}
\put(0,-140){\circle*{2}}
\put(5,-130){1}
\put(10,-120){\line(0,-1){20}}
\put(10,-140){\circle*{2}}
\put(10,-130){2}
\put(10,-120){\line(1,-2){10}}
\put(20,-140){\circle*{2}}
\put(15,-130){3}

\put(25,-130){$\rightarrow \qquad 0$}

\end{picture}

\caption{}                 \label{Fig:49}

\end{figure}

First, we have

\begin{thm}
${\partial}^{i+1} \circ {\partial}^i = 0$.
That is, $\partial$ is actually a differential.
\end{thm}

\begin{proof}
This proof is almost the same as that of Theorem \ref{Thm:lediff}.
Suppose $T''$ is obtained by appending a vertex to $T'$  with added
edge $e''$ so that i) and ii) of Definition \ref{Def:vadiff} are
satisfied, where $T'$ is obtained by appending a vertex to $T$ with
added edge $e'$ so that i) and ii) of Definition \ref{Def:vadiff}
are satisfied. Here, if $T'$ has no available position to append a
vertex, then $\partial(T')=0$ , so $T''=0$. Otherwise, when $e''$ is
$e''_{j,l_j}$ and $e'$ is $e''_{k,l_k}$ in the edge-ordering of
$T''$, in $\partial^{i+1} \circ \partial^i (T)$ hits the component
$T'' \otimes e''_{1,l_1} \wedge ...\wedge e''_{i+2,l_{i+2}}$ just by
adding the edge $e'$ and $e''$ to $T$, and it can be done only in
two ways;

i) adding $e'$ first to $T$: then the component of
$T'' \otimes e''_{1,l_1} \wedge ... \wedge e''_{i+2,l_{1+2}}$ is
$T'' \otimes e''_{j,l_j} \wedge e''_{k,l_k} \wedge e''_{1,l_1}
\wedge ... \wedge \widehat{e''_{j,l_j}} \wedge ... \wedge
\widehat{e''_{k,l_k}} \wedge ... \wedge e''_{i+2,l_{i+2}}$.

ii) adding $e''$ first to $T$: then the component of
$T'' \otimes e''_{1,l_1} \wedge ... \wedge e''_{i+2,l_{1+2}}$ is
$T'' \otimes e''_{k,l_k} \wedge e''_{j,l_j} \wedge e''_{1,l_1}
\wedge ... \wedge \widehat{e''_{j,l_j}} \wedge ... \wedge
\widehat{e''_{k,l_k}} \wedge ... \wedge e''_{i+2,l_{i+2}}$.

Since the order of $e''_{k,l_k}$ and $e''_{j,l_j}$ are different
in the wedge products,
the sum of the two terms in i) and ii) is 0.
This is true for all components of $\partial^{i+1} \circ \partial^i (T)$,
and so $\partial^{i+1} \circ \partial^i = 0$.
\end{proof}

\subsection{Definition of seedling.}
By the definition of $\partial^i$,
all terms in $\partial^i(T)$ are of the form $\pm T' \otimes e \wedge det(T)$,
where $T'$ runs over bonsais obtained by adding a new edge $e$ to $T$
so that i) and ii) of Definition \ref{Def:vadiff} hold.
So $T'$ has the form of appending a vertex to a vertex of T
which is not a tip.
Having this intuitive fact in mind,
let us present some new definitions and reorganize
the cochain complex of bonsais.

\begin{df}
For a bonsai $T$, an edge $e$ of $T$ is called {\em twiggy}
if it is at the end of a branch and the opposite end of the tip
is a branching vertex.
In Figure \ref{Fig:53}, $e$ is twiggy in $T$ and $e'$ is not,
and $f$ is not a twiggy edge of $T'$.

\begin{figure}[h]

\begin{picture}(200,50)(0,-40)

\put(20,-10){$T=$}

\put(50,0){\circle*{2}}
\put(50,0){\line(-1,-2){10}}
\put(40,-20){\circle*{2}}
\put(42,-10){\line(-2,-1){20}}
\put(-15,-25){$e$, twiggy}
\put(50,0){\line(1,-2){10}}
\put(60,-20){\circle*{2}}
\put(60,-20){\line(0,-1){20}}
\put(65,-35){$e'$, not twiggy}
\put(60,-40){\circle*{2}}

\put(115,-10){$T'=$}
\put(140,0){\circle*{2}}
\put(140,0){\line(0,-1){20}}
\put(140,-20){\circle*{2}}
\put(142,-15){$f$}

\end{picture}

\caption{}                 \label{Fig:53}

\end{figure}

\end{df}

\begin{df}
A bonsai which has no twiggy edge is called a {\em vertex-appending
seedling}. In this section, we will call this just a {\em seedling}.
The left bonsai in Figure \ref{Fig:52} is not a seedling, and the
ones in Figure \ref{Fig:54} are all seedlings in 2-bonsai. Note that
the one-vertex bonsai is a seedling. Intuitively, a seedling is a
bonsai which cannot be obtained by adding edges like i) and ii) of
Definition \ref{Def:vadiff}.

\begin{figure}[h]

\begin{picture}(200,50)(0,-40)

\put(50,0){\circle*{2}}
\put(50,0){\line(-1,-2){10}}
\put(40,-20){\circle*{2}}
\put(45,-15){$1$}
\put(40,-20){\line(0,-1){20}}
\put(40,-40){\circle*{2}}
\put(40,-35){$1$}
\put(50,0){\line(1,-2){10}}
\put(60,-20){\circle*{2}}
\put(55,-15){$2$}
\put(60,-20){\line(0,-1){20}}
\put(60,-35){$1$}
\put(60,-40){\circle*{2}}

\put(110,0){\circle*{2}}
\put(110,0){\line(0,-1){20}}
\put(110,-20){\circle*{2}}
\put(112,-15){$1$}
\put(110,-20){\line(0,-1){20}}
\put(110,-40){\circle*{2}}
\put(112,-35){$1$}

\put(170,0){\circle*{2}}
\put(170,0){\line(0,-1){20}}
\put(170,-10){2}
\put(170,-20){\circle*{2}}

\end{picture}

\caption{}                 \label{Fig:54}

\end{figure}

\end{df}

\begin{df}

For two seedlings $S$ and $S'$, we define an equivalence relation $S \sim S'$
if $S$ is obtained by changing labels of branch-end edges of $S'$.
For example, all four seedlings in Figure \ref{Fig:55} are equivalent,
and so are the first and second seedlings of Figure \ref{Fig:56},
but the first and third seedlings of Figure \ref{Fig:56} are not equivalent.

\begin{figure}[h]

\begin{picture}(200,50)(0,-40)

\put(10,0){\circle*{2}}
\put(10,0){\line(-1,-2){10}}
\put(0,-20){\circle*{2}}
\put(5,-15){$1$}
\put(0,-20){\line(0,-1){20}}
\put(0,-40){\circle*{2}}
\put(0,-35){$1$}
\put(10,0){\line(1,-2){10}}
\put(20,-20){\circle*{2}}
\put(15,-15){$2$}
\put(20,-20){\line(0,-1){20}}
\put(20,-35){$1$}
\put(20,-40){\circle*{2}}

\put(50,0){\circle*{2}}
\put(50,0){\line(-1,-2){10}}
\put(40,-20){\circle*{2}}
\put(45,-15){$1$}
\put(40,-20){\line(0,-1){20}}
\put(40,-40){\circle*{2}}
\put(40,-35){$2$}
\put(50,0){\line(1,-2){10}}
\put(60,-20){\circle*{2}}
\put(55,-15){$2$}
\put(60,-20){\line(0,-1){20}}
\put(60,-35){$1$}
\put(60,-40){\circle*{2}}

\put(90,0){\circle*{2}}
\put(90,0){\line(-1,-2){10}}
\put(80,-20){\circle*{2}}
\put(85,-15){$1$}
\put(80,-20){\line(0,-1){20}}
\put(80,-40){\circle*{2}}
\put(80,-35){$1$}
\put(90,0){\line(1,-2){10}}
\put(100,-20){\circle*{2}}
\put(95,-15){$2$}
\put(100,-20){\line(0,-1){20}}
\put(100,-35){$2$}
\put(100,-40){\circle*{2}}

\put(130,0){\circle*{2}}
\put(130,0){\line(-1,-2){10}}
\put(120,-20){\circle*{2}}
\put(125,-15){$1$}
\put(120,-20){\line(0,-1){20}}
\put(120,-40){\circle*{2}}
\put(120,-35){$2$}
\put(130,0){\line(1,-2){10}}
\put(140,-20){\circle*{2}}
\put(135,-15){$2$}
\put(140,-20){\line(0,-1){20}}
\put(140,-35){$2$}
\put(140,-40){\circle*{2}}

\end{picture}

\caption{}                 \label{Fig:55}

\end{figure}

\begin{figure}[h]

\begin{picture}(200,50)(10,-40)

\put(70,0){\circle*{2}}
\put(70,0){\line(0,-1){20}}
\put(70,-20){\circle*{2}}
\put(72,-15){$1$}
\put(70,-20){\line(0,-1){20}}
\put(70,-40){\circle*{2}}
\put(72,-35){$1$}

\put(110,0){\circle*{2}}
\put(110,0){\line(0,-1){20}}
\put(110,-20){\circle*{2}}
\put(112,-15){$1$}
\put(110,-20){\line(0,-1){20}}
\put(110,-40){\circle*{2}}
\put(112,-35){$2$}

\put(150,0){\circle*{2}}
\put(150,0){\line(0,-1){20}}
\put(150,-20){\circle*{2}}
\put(152,-15){$2$}
\put(150,-20){\line(0,-1){20}}
\put(150,-40){\circle*{2}}
\put(152,-35){$1$}

\end{picture}

\caption{}                 \label{Fig:56}

\end{figure}

\end{df}

Let us try the same trick as in the proof of acyclicity of the
branch-fixed differential. When $S$ is a seedling, let ${\mathbf
C}^{[S],0}$ be the subspace of the determinanted $m$-bonsai space
having $\{T \otimes det(T)\}$ as the basis, where $T$ is in the
equivalence class $[S]$ of $S$ by $\sim$. And let ${\mathbf
C}^{[S],i+1}$ be the space with the basis $\{T' \otimes det(T')\}$,
where $T'$ is obtained by adding an edge to $T$, where $\{T \otimes
det(T)\}$ is the basis of ${\mathbf C}^{S,i}$, as i) and ii) of
Definition \ref{Def:vadiff}. Every bonsai is obtained by adding some
edges to a seedling as given in Definition \ref{Def:vadiff} and if
$S$ and $S'$ are not equivalent seedlings, then the bonsais obtained
by adding edges to $S$ and $S'$ as given in Definition
\ref{Def:vadiff} are not equivalent; the space of determinanted
bonsais is the direct sum of ${\mathbf C}^{[S],i}$. Then, since
$\partial({\mathbf C}^{[S],i}) \subset {\mathbf C}^{[S],i+1}$, the
cohomology groups of determinanted bonsais by the differential
$\partial$ is the direct sum of cohomology groups of the threads
$\{{\mathbf C}^{[S],i}\}$.

\subsection{The cohomology groups of the vertex-appending differential}
First let us consider a coboundary complex $\{{\mathbf D}^{m,i}\}$
consisting of corollas, such that each corolla is an $m$-bonsai, and
the boundary map is the vertex-appending differential, but in this
complex, appending to the one-vertex bonsai is allowed, so we have
to be careful not to be confused with the definition of the above
vertex-appending differential. By the definition of $\partial
^i:{\mathbf D}^{m,i} \to {\mathbf D}^{m,i+1}$, all terms in
$\partial ^i (T)$ are of the form $\pm T' \otimes e \wedge det(T)$,
where $T'$ runs over bonsais obtained by adding one vertex as
defined in Definition \ref{Def:vadiff}. Since appending to a tip is
forbidden, every $\partial^i$ is appending a vertex to the root of a
corolla. Now let us show some boundary map sequences of the thread
starting from one vertex. For one vertex, $\partial ^0$ acts as in
Figure \ref{Fig:50}.

\begin{figure}[h]

\begin{picture}(200,30)(0,-20)

\put(10,-10){\circle*{2}}

\put(25,-10){$\rightarrow$}

\put(50,0){\circle*{2}} \put(50,0){\line(0,-1){20}}
\put(50,-20){\circle*{2}} \put(52,-10){1}

\put(60,-10){+}

\put(70,0){\circle*{2}} \put(70,0){\line(0,-1){20}}
\put(70,-20){\circle*{2}} \put(72,-10){2}

\put(80,-10){+...+}

\put(110,0){\circle*{2}} \put(110,0){\line(0,-1){20}}
\put(110,-20){\circle*{2}} \put(112,-10){$m$}

\end{picture}

\caption{}                 \label{Fig:50}

\end{figure}

For the one-edge corolla, $\partial^1$ acts as in Figure
\ref{Fig:51}, where $1 \leq n \leq m$, and for the two-edge corolla,
$\partial^2$ acts as in Figure \ref{Fig:52}, where $1 \leq n_1 < n_2
\leq m$, and so on.

\begin{figure}[h]

\begin{picture}(200,30)(0,-20)

\put(10,0){\circle*{2}} \put(10,0){\line(0,-1){20}}
\put(10,-20){\circle*{2}} \put(10,-10){$n$}

\put(25,-10){$\rightarrow$}

\put(50,0){\circle*{2}} \put(50,0){\line(-1,-2){10}}
\put(40,-20){\circle*{2}} \put(45,-15){1}
\put(50,0){\line(1,-2){10}} \put(60,-20){\circle*{2}}
\put(55,-15){$n$}

\put(60,-10){+...+}

\put(110,0){\circle*{2}} \put(110,0){\line(-1,-2){10}}
\put(100,-20){\circle*{2}} \put(90,-15){$n-1$}
\put(110,0){\line(1,-2){10}} \put(120,-20){\circle*{2}}
\put(115,-15){$n$}

\put(130,-10){$-$}

\put(150,0){\circle*{2}} \put(150,0){\line(-1,-2){10}}
\put(140,-20){\circle*{2}} \put(140,-15){$n$}
\put(150,0){\line(1,-2){10}} \put(160,-20){\circle*{2}}
\put(150,-15){$n+1$}

\put(180,-10){$-...-$}

\put(220,0){\circle*{2}} \put(220,0){\line(-1,-2){10}}
\put(210,-20){\circle*{2}} \put(215,-15){$n$}
\put(220,0){\line(1,-2){10}} \put(230,-20){\circle*{2}}
\put(225,-15){$m$}

\end{picture}

\caption{}                 \label{Fig:51}

\end{figure}

\begin{figure}[h]

\begin{picture}(200,80)(0,-70)

\put(10,0){\circle*{2}} \put(10,0){\line(-1,-2){10}}
\put(0,-20){\circle*{2}} \put(0,-15){$n_1$}
\put(10,0){\line(1,-2){10}} \put(20,-20){\circle*{2}}
\put(15,-15){$n_2$}

\put(25,-10){$\rightarrow$}

\put(60,0){\circle*{2}} \put(60,0){\line(-1,-1){20}}
\put(40,-20){\circle*{2}} \put(45,-15){1}
\put(60,0){\line(0,-1){20}} \put(60,-20){\circle*{2}}
\put(60,-15){$n_1$} \put(60,0){\line(1,-1){20}}
\put(80,-20){\circle*{2}} \put(75,-15){$n_2$}

\put(105,-10){+...+}

\put(160,0){\circle*{2}} \put(160,0){\line(-1,-1){20}}
\put(140,-20){\circle*{2}} \put(130,-15){$n_1-1$}
\put(160,0){\line(0,-1){20}} \put(160,-20){\circle*{2}}
\put(160,-15){$n_1$} \put(160,0){\line(1,-1){20}}
\put(180,-20){\circle*{2}} \put(175,-15){$n_2$}

\put(25,-40){$-$}

\put(60,-30){\circle*{2}} \put(60,-30){\line(-1,-1){20}}
\put(40,-50){\circle*{2}} \put(35,-45){$n_1$}
\put(60,-30){\line(0,-1){20}} \put(60,-50){\circle*{2}}
\put(47,-45){$n_1+1$} \put(60,-30){\line(1,-1){20}}
\put(80,-50){\circle*{2}} \put(75,-45){$n_2$}

\put(105,-40){$-...-$}

\put(160,-30){\circle*{2}} \put(160,-30){\line(-1,-1){20}}
\put(140,-50){\circle*{2}} \put(135,-45){$n_1$}
\put(160,-30){\line(0,-1){20}} \put(160,-50){\circle*{2}}
\put(145,-45){$n_2-1$} \put(160,-30){\line(1,-1){20}}
\put(180,-50){\circle*{2}} \put(175,-45){$n_2$}

\put(25,-70){+}

\put(60,-60){\circle*{2}} \put(60,-60){\line(-1,-1){20}}
\put(40,-80){\circle*{2}} \put(35,-75){$n_1$}
\put(60,-60){\line(0,-1){20}} \put(60,-80){\circle*{2}}
\put(50,-75){$n_2$} \put(60,-60){\line(1,-1){20}}
\put(80,-80){\circle*{2}} \put(65,-75){$n_2+1$}

\put(105,-70){+...+}

\put(160,-60){\circle*{2}} \put(160,-60){\line(-1,-1){20}}
\put(140,-80){\circle*{2}} \put(135,-75){$n_1$}
\put(160,-60){\line(0,-1){20}} \put(160,-80){\circle*{2}}
\put(155,-75){$n_2$} \put(160,-60){\line(1,-1){20}}
\put(180,-80){\circle*{2}} \put(175,-75){$m$}

\end{picture}

\caption{}                 \label{Fig:52}

\end{figure}

This sequence of coboundary maps is the same as that of the reduced
cohomology of the $(m-1)$-simplex with vertices $v_1, v_2,...,v_m$,
once we identify the corolla with labels $i_1,i_2,...,i_k$ with the
simplex generated by vertices $v_{i_1},v_{i_2},...,v_{i_k}$. So the
cohomology groups of this thread of boundary maps is acyclic, and
the lowest degree group is trivial. Let us denote the module having
the basis consisting of one vertex having $m$ available positions of
vertex appending as ${\mathbf D}^{m,0}$, and the module having the
basis consisting of corollae with $n$ edges as shown above ${\mathbf
D}^{m,n}$. Also, let $\{{\mathbf B}^{m,n}\}$ be a cochain complex
defined by ${\mathbf B}^{m,n}={\mathbf D}^{m,n+1}$ for later
convenience. Then $\{{\mathbf B}^{m,n}\}$ is acyclic and $H^0=k$,
where $k$ is the base field, since the cohomology of $\{{\mathbf
B}^{m,n}\}$ is isomorphic to the cohomology (not the reduced
cohomology) of the $(m-1)$-simplex.

\subsection{The case of general seedlings}

Now let us show through an example
\begin{lem}
Any complex ${\{{\mathbf C}^{[S],i}\}}$ is isomorphic to it, which
is represented as a direct sum of tensor products of $\{{\mathbf
D}^{m,i}\}$'s  and $\{{\mathbf B}^{m,i}\}$'s (as in the proof of
acyclicity of the branch-fixed differential).
\end{lem}
In 5-bonsai, the seedling $S$ in Figure \ref{Fig:57}
can get twiggy edges at the positions of the twigs shown in the picture
which are grouped as surrounded by squares.

\begin{figure}[h]

\begin{picture}(200,120)(-100,-110)

\put(0,0){\circle*{2}}
\put(0,0){\line(0,-1){40}}

\put(-25,0){${\mathbf D}_1^{1,*}$}
\put(-25,0){\line(1,0){20}}
\put(-25,0){\line(0,-1){25}}
\put(-5,0){\line(0,-1){25}}
\put(-25,-25){\line(1,0){20}}
\put(0,0){\line(-1,-1){20}}

\put(5,0){${\mathbf D}_6^{3,*}$}
\put(0,0){\line(1,-2){10}}
\put(0,0){\line(1,-1){20}}
\put(0,0){\line(3,-2){30}}
\put(5,0){\line(1,0){30}}
\put(5,0){\line(0,-1){25}}
\put(35,0){\line(0,-1){25}}
\put(5,-25){\line(1,0){30}}

\put(0,-35){$e_1$}
\put(0,-40){\circle*{2}}
\put(0,-40){\line(-1,-1){40}}
\put(0,-40){\line(1,-1){40}}

\put(-35,-40){${\mathbf D}_2^{1,*}$}
\put(0,-40){\line(-3,-2){30}}
\put(-35,-40){\line(1,0){10}}
\put(-35,-40){\line(0,-1){25}}
\put(-25,-40){\line(0,-1){25}}
\put(-35,-65){\line(1,0){10}}

\put(-5,-75){${\mathbf D}_4^{2,*}$}
\put(0,-40){\line(-1,-2){10}}
\put(0,-40){\line(1,-2){10}}
\put(-15,-50){\line(1,0){30}}
\put(-15,-50){\line(0,-1){15}}
\put(15,-50){\line(0,-1){15}}
\put(-15,-65){\line(1,0){30}}

\put(-35,-75){$e_2$}
\put(-40,-80){\circle*{2}}
\put(-40,-80){\line(0,-1){30}}
\put(-40,-80){\line(-1,-1){20}}
\put(-40,-80){\line(-1,-2){10}}
\put(-40,-80){\line(1,-2){10}}
\put(-40,-80){\line(1,-1){20}}
\put(-40,-110){\circle*{2}}

\put(-65,-80){${\mathbf B}_3^{5,*}$}
\put(-65,-80){\line(1,0){50}}
\put(-65,-80){\line(0,-1){35}}
\put(-15,-80){\line(0,-1){35}}
\put(-65,-115){\line(1,0){50}}

\put(25,-65){$e_3$}
\put(40,-80){\circle*{2}}
\put(40,-80){\line(0,-1){30}}
\put(40,-80){\line(-1,-2){10}}
\put(40,-80){\line(1,-2){10}}
\put(40,-80){\line(1,-1){20}}
\put(40,-80){\line(3,-2){30}}
\put(40,-110){\circle*{2}}

\put(25,-80){${\mathbf B}_5^{5,*}$}
\put(25,-80){\line(1,0){50}}
\put(25,-80){\line(0,-1){35}}
\put(75,-80){\line(0,-1){35}}
\put(25,-115){\line(1,0){50}}

\end{picture}

\caption{}                 \label{Fig:57}

\end{figure}
Note that, in Figure \ref{Fig:57} adding an edge to each square is
the same as attaching an edge to the corolla at the vertex at which
the square is appended, each corolla corresponds to the module that
is written on each square (In Figure \ref{Fig:57}, ${\mathbf
D}_i^{m,*}$ is isomorphic to ${\mathbf D}^{m,*}$ and
 ${\mathbf B}_i^{m,*}$ is isomorphic to ${\mathbf B}^{m,*}$).

Keeping this in mind, we can define new modules $\{{\mathbf D}^l\}$
and an isomorphism $F$ of them with $\{{\mathbf C}^{[S],i}\}$
like the following;

\begin{eqnarray} \label{eq:tprep}
{\mathbf D}^l = \bigoplus _{p_1+...+p_6=l}
                {\mathbf D}_1^{1,p_1}
        \otimes {\mathbf D}_2^{1,p_2}
        \otimes {\mathbf B}_3^{5,p_3}
        \otimes {\mathbf D}_4^{2,p_4}
        \otimes {\mathbf B}_5^{5,p_5}
        \otimes {\mathbf D}_6^{3,p_6}
\end{eqnarray}
and when ${\mathbf M}_1 = {\mathbf D}_1^{1,p_1}$,
${\mathbf M}_2 = {\mathbf D}_2^{1,p_2}$,
${\mathbf M}_3 = {\mathbf B}_3^{5,p_3}$,
${\mathbf M}_4 = {\mathbf D}_4^{2,p_4}$,
${\mathbf M}_5 = {\mathbf B}_5^{5,p_5}$
and ${\mathbf M}_6 = {\mathbf D}_6^{3,p_6}$,
the map $F:{\mathbf D}^l \to {\mathbf C}^{[S],l}$ is defined as,
when $c_i \in {\mathbf M}_i$ is a corolla,

\begin{eqnarray} \label{eq:F}
& c_1 \otimes det(c_1) \otimes ... \otimes c_6 \otimes det(c_6) \mapsto \\
& \Sigma (\mbox{The bonsai obtained by attaching $c_i$
                to the square corresponding to ${\mathbf M}_i$}) \notag \\
& \otimes det(c_1) \wedge e_1 \wedge det(c_2) \wedge e_2 \wedge det(c_3)
  \wedge det(c_4) \wedge e_3 \wedge det(c_5) \wedge det(c_6) \notag
\end{eqnarray}
as in the example of Figure \ref{Fig:58}.

\begin{figure}[h]

\begin{picture}(200,185)(0,-110)

\put(-65,45){\circle*{2}}
\put(-65,45){\line(0,1){20}}
\put(-65,65){\circle*{2}}
\put(-65,50){1}
\put(-60,50){$\otimes det($}
\put(-30,45){\circle*{2}}
\put(-30,45){\line(0,1){20}}
\put(-30,65){\circle*{2}}
\put(-30,50){1}
\put(-25,50){$) \otimes$}
\put(-10,53){\circle*{2}}
\put(-5,50){$\otimes det($}
\put(25,53){\circle*{2}}
\put(30,50){$) \otimes$}
\put(45,45){\circle*{2}}
\put(45,45){\line(0,1){20}}
\put(45,65){\circle*{2}}
\put(45,50){2}
\put(50,50){$\otimes det($}
\put(80,45){\circle*{2}}
\put(80,45){\line(0,1){20}}
\put(80,65){\circle*{2}}
\put(80,50){2}
\put(85,50){$) \otimes$}
\put(-65,20){\circle*{2}}
\put(-65,20){\line(0,1){20}}
\put(-65,40){\circle*{2}}
\put(-65,25){2}
\put(-60,25){$\otimes det($}
\put(-30,20){\circle*{2}}
\put(-30,20){\line(0,1){20}}
\put(-30,40){\circle*{2}}
\put(-30,25){2}
\put(-25,25){$) \otimes$}
\put(-10,20){\circle*{2}}
\put(0,40){\circle*{2}}
\put(0,40){\line(-1,-2){10}}
\put(-10,25){3}
\put(0,40){\line(1,-2){10}}
\put(5,25){4}
\put(10,20){\circle*{2}}
\put(10,25){$\otimes det($}
\put(40,20){\circle*{2}}
\put(50,40){\circle*{2}}
\put(50,40){\line(-1,-2){10}}
\put(40,25){3}
\put(50,40){\line(1,-2){10}}
\put(55,25){4}
\put(60,20){\circle*{2}}
\put(60,25){$) \otimes$}
\put(75,28){\circle*{2}}
\put(80,25){$\otimes det($}
\put(110,25){\circle*{2}}
\put(115,25){$) \qquad \mapsto$}

\put(0,0){\circle*{2}}
\put(0,0){\line(0,-1){40}}

\put(-25,0){${\mathbf D}_1^{1,*}$}
\put(-25,0){\line(1,0){20}}
\put(-25,0){\line(0,-1){25}}
\put(-5,0){\line(0,-1){25}}
\put(-25,-25){\line(1,0){20}}
\put(0,0){\line(-1,-1){20}}
\put(-20,-20){\circle*{2}}

\put(5,0){${\mathbf D}_6^{3,*}$}
\put(0,0){\line(1,-2){7}}
\put(0,0){\line(1,-1){14}}
\put(0,0){\line(3,-2){21}}
\put(5,0){\line(1,0){30}}
\put(5,0){\line(0,-1){25}}
\put(35,0){\line(0,-1){25}}
\put(5,-25){\line(1,0){30}}

\put(0,-35){$e_1$}
\put(0,-40){\circle*{2}}
\put(0,-40){\line(-1,-1){40}}
\put(0,-40){\line(1,-1){40}}

\put(-35,-40){${\mathbf D}_2^{1,*}$}
\put(0,-40){\line(-3,-2){30}}
\put(-35,-40){\line(1,0){10}}
\put(-35,-40){\line(0,-1){25}}
\put(-25,-40){\line(0,-1){25}}
\put(-35,-65){\line(1,0){10}}

\put(-5,-75){${\mathbf D}_4^{2,*}$}
\put(0,-40){\line(-1,-2){7}}
\put(0,-40){\line(1,-2){10}}
\put(10,-60){\circle*{2}}
\put(-15,-50){\line(1,0){30}}
\put(-15,-50){\line(0,-1){15}}
\put(15,-50){\line(0,-1){15}}
\put(-15,-65){\line(1,0){30}}

\put(-35,-75){$e_2$}
\put(-40,-80){\circle*{2}}
\put(-40,-80){\line(0,-1){20}}
\put(-40,-80){\line(-1,-1){20}}
\put(-40,-80){\line(-1,-2){15}}
\put(-40,-80){\line(1,-2){10}}
\put(-40,-80){\line(1,-1){20}}
\put(-55,-110){\circle*{2}}

\put(-65,-80){${\mathbf B}_3^{5,*}$}
\put(-65,-80){\line(1,0){50}}
\put(-65,-80){\line(0,-1){35}}
\put(-15,-80){\line(0,-1){35}}
\put(-65,-115){\line(1,0){50}}

\put(25,-65){$e_3$}
\put(40,-80){\circle*{2}}
\put(40,-80){\line(0,-1){20}}
\put(40,-80){\line(-1,-2){10}}
\put(40,-80){\line(1,-2){15}}
\put(40,-80){\line(1,-1){30}}
\put(40,-80){\line(3,-2){30}}
\put(55,-110){\circle*{2}}
\put(70,-110){\circle*{2}}

\put(25,-80){${\mathbf B}_5^{5,*}$}
\put(25,-80){\line(1,0){50}}
\put(25,-80){\line(0,-1){35}}
\put(75,-80){\line(0,-1){35}}
\put(25,-115){\line(1,0){50}}

\put(80,-40){$\otimes det($}
\put(110,-45){\circle*{2}}
\put(110,-45){\line(0,1){20}}
\put(110,-25){\circle*{2}}
\put(110,-40){1}
\put(115,-40){$)\wedge e_1 \wedge det($}
\put(170,-37){\circle*{2}}
\put(175,-40){$)\wedge e_2 \wedge det($}
\put(230,-45){\circle*{2}}
\put(230,-45){\line(0,1){20}}
\put(230,-25){\circle*{2}}
\put(230,-40){2}
\put(235,-40){$)$}

\put(90,-65){$\wedge det($}
\put(120,-70){\circle*{2}}
\put(120,-70){\line(0,1){20}}
\put(120,-50){\circle*{2}}
\put(120,-65){2}
\put(125,-65){$)\wedge e_3 \wedge det($}
\put(185,-70){\circle*{2}}
\put(195,-50){\circle*{2}}
\put(195,-50){\line(-1,-2){10}}
\put(185,-65){3}
\put(195,-50){\line(1,-2){10}}
\put(200,-65){4}
\put(205,-70){\circle*{2}}
\put(205,-65){$)\wedge det($}
\put(240,-65){\circle*{2}}
\put(245,-65){$)$}

\end{picture}

\caption{}                 \label{Fig:58}

\end{figure}

Then in $T \otimes det(T) \in {\mathbf C}^{S,l}$, $det(T)$ is

\begin{eqnarray}
\\
det(c_1) \wedge e_{1,k_1} \wedge det(c_2) \wedge e_{2,k_2} \wedge det(c_3)
\wedge det(c_4) \wedge e_{3,k_3} \wedge det(c_5) \wedge det(c_6) \notag
\end{eqnarray}
where $k_l$ is the label of the edge $e_l$,
and $\partial ^n(T \otimes det(T))$ is

\begin{eqnarray}
\\
\Sigma
(\mbox{A bonsai $T'$ obtained by adding a new edge $f$ to one of $c_i$})
   \otimes f \wedge det(T) \notag
\end{eqnarray}

\begin{eqnarray}
=& \Sigma(\mbox{A bonsai $T'$ obtained
                by adding a new edge $f$ to one of $c_i$})  \notag \\
 & \otimes f \wedge det(c_1) \wedge e_{1,k_1} \wedge det(c_2)
   \wedge e_{2,k_2} \wedge det(c_3) \wedge det(c_4)
   \wedge e_{3,k_3} \wedge det(c_5) \wedge det(c_6) \notag
\end{eqnarray}

\begin{eqnarray}
=& \Sigma(\mbox{A bonsai $T'$ obtained
                by adding a new edge $f$ to  $c_1$})  \notag \\
 & \otimes  f \wedge det(c_1) \wedge e_{1,k_1}
   \wedge det(c_2) \wedge e_{2,k_2} \wedge det(c_3)
   \wedge det(c_4) \wedge e_{3,k_3} \wedge det(c_5) \wedge det(c_6) \notag
\end{eqnarray}

\begin{eqnarray}
+& \Sigma(\mbox{A bonsai $T'$ obtained
              by adding a new edge $f$ to  $c_2$})  \notag \\
 & \otimes  (-1)^{deg(c_1)+1}   \notag \\
 & det(c_1) \wedge e_{1,k_1} \wedge f \wedge det(c_2)
   \wedge e_{2,k_2} \wedge det(c_3)
   \wedge det(c_4) \wedge e_{3,k_3}
   \wedge det(c_5) \wedge det(c_6) \notag
\end{eqnarray}

\begin{eqnarray}
+& \Sigma(\mbox{A bonsai $T'$ obtained
                by adding a new edge $f$ to  $c_3$})  \notag \\
 & \otimes  (-1)^{(deg(c_1)+1)+(deg(c_2)+1)}   \notag \\
 & det(c_1) \wedge e_{1,k_1} \wedge det(c_2)
   \wedge e_{2,k_2} \wedge f \wedge det(c_3)
   \wedge det(c_4) \wedge e_{3,k_3} \wedge det(c_5) \wedge det(c_6) \notag
\end{eqnarray}

\begin{eqnarray}
+& \Sigma(\mbox{A bonsai $T'$ obtained
                by adding a new edge $f$ to  $c_4$})  \notag \\
 & \otimes  (-1)^{(deg(c_1)+1)+(deg(c_2)+1)+(deg(c_3)+0)}   \notag \\
 & det(c_1) \wedge e_{1,k_1} \wedge det(c_2) \wedge e_{2,k_2}
   \wedge  det(c_3) \wedge f \wedge det(c_4) \wedge e_{3,k_3}
   \wedge det(c_5) \wedge det(c_6) \notag
\end{eqnarray}

\begin{eqnarray}
+& \Sigma(\mbox{A bonsai $T'$ obtained
                by adding a new edge $f$ to  $c_5$})  \notag \\
 & \otimes
   (-1)^{(deg(c_1)+1)+(deg(c_2)+1)+(deg(c_3)+0)+(deg(c_4)+1)}   \notag \\
 & det(c_1) \wedge e_{1,k_1} \wedge det(c_2) \wedge e_{2,k_2}
   \wedge  det(c_3) \wedge det(c_4) \wedge e_{3,k_3}
   \wedge f \wedge det(c_5) \wedge det(c_6) \notag
\end{eqnarray}

\begin{eqnarray}
+& \Sigma(\mbox{A bonsai $T'$ obtained
                by adding a new edge $f$ to  $c_6$})  \notag \\
 & \otimes
   (-1)^{(deg(c_1)+1)+(deg(c_2)+1)+(deg(c_3)+0)+(deg(c_4)+1)+(deg(c_5)+0)}
   \notag \\
 & det(c_1) \wedge e_{1,k_1} \wedge det(c_2) \wedge e_{2,k_2}
   \wedge  det(c_3) \wedge det(c_4) \wedge e_{3,k_3}
   \wedge det(c_5) \wedge f \wedge det(c_6) \notag  \\
 & \notag
\end{eqnarray}

So, when $m_1=deg(c_1)+1$, $m_2=deg(c_2)+1$, $m_3=deg(c_3)+0$,
$m_4=deg(c_4)+1$ and $m_5=deg(c_5)+0$,
we can write $\partial(T \otimes det(T))$ as

\begin{eqnarray}
+& \Sigma(\mbox{A bonsai $T'$ obtained
                by adding a new edge $f$ to  $c_i$})  \notag \\
 & \otimes  (-1)^{m_1+...+m_{i-1}} det(c_1) \wedge e_{1,k_1} \wedge ... \wedge
   (f \wedge det(c_i)) \wedge ...  \wedge det(c_6). \notag
\end{eqnarray}

Hence, when we define the coboundary map on $\{{\mathbf D}^l\}$ as

\begin{eqnarray}
c_1 \otimes ... \otimes c_6 \mapsto
\Sigma (-1)^{{\beta}_i}
c_1 \otimes ... \otimes \partial(c_i)\otimes ... \otimes c_6
\end{eqnarray}
where ${\beta}_i = m_1 +...+m_{i-1}$ and ${\beta}_1 = 0$,
the map $F$ defined in \eqref{eq:F}
becomes a cochain isomorphism of $\{{\mathbf D}^l\}$
and $\{{\mathbf C}^{[S],l}\}$.
Then by the K{\" u}nneth theorem,
the cohomology of the cochain complex $\{{\mathbf D}^l\}$ is
expressed as the sum of
$H^{q_1}({\mathbf M}_1) \otimes ... \otimes H^{q_6}({\mathbf M}_6)$
for some $q_i$'s, and since
$H^i({\mathbf M}_1) = H^i({\mathbf D}^{1,*}) = 0$ for any $i$,
the cohomology of ${\mathbf C}^{[S],l}$
for $S$ of Figure \ref{Fig:57} is acyclic.

\begin{df}
As shown for the example of Figure \ref{Fig:57}, for any seedling
$S$, we have a cochain complex as in \eqref{eq:tprep} and an
isomorphism $F$ of it with ${\mathbf C}^{[S],l}$ as in \eqref{eq:F}.
We call this cochain complex as in \eqref{eq:tprep} the {\em tensor
product representation } of ${\mathbf C}^{[S],l}$.
\end{df}
Then, whether $\{{\mathbf C}^{[S],l}\}$ is acyclic or not depends on
whether its tensor product representation contains ${\mathbf
D}^{m,*}$. As shown in Figure \ref{Fig:57}, ${\mathbf B}^{m,*}$
appears only at the branch-end edges of a seedling, and ${\mathbf
D}^{m,*}$ appears only at the available positions of
vertex-appending other than at branch-end edges. So, the only case
where $\{{\mathbf C}^{[S],l}\}$ is not acyclic is that the bonsai
obtained by deleting all branch-end edges of $S$ is a cocycle, i.e.,
that bonsai has no available position of vertex appending and so
there is no room for ${\mathbf D}^{m,*}$ on $S$, like the bonsais of
\ref{Fig:59}, in 2-bonsai. The only nontrivial cohomology group of
$\{{\mathbf C}^{[S],l}\}$ is $H^0=k$ by K{\"u}nneth Theorem, where
$k$ is the base field, since all $H^0({\mathbf B}^{m,i})=k$

\begin{figure}[h]

\begin{picture}(200,70)(0,-60)

\put(10,0){\circle*{2}}
\put(10,0){\line(-1,-2){10}}
\put(0,-20){\circle*{2}}
\put(5,-15){$1$}
\put(0,-20){\line(0,-1){20}}
\put(0,-40){\circle*{2}}
\put(0,-35){$1$}
\put(10,0){\line(1,-2){10}}
\put(20,-20){\circle*{2}}
\put(15,-15){$2$}
\put(20,-20){\line(0,-1){20}}
\put(20,-35){$1$}
\put(20,-40){\circle*{2}}

\put(50,0){\circle*{2}}
\put(50,0){\line(-1,-2){10}}
\put(40,-20){\circle*{2}}
\put(45,-15){$1$}
\put(40,-20){\line(0,-1){20}}
\put(40,-40){\circle*{2}}
\put(40,-35){$2$}
\put(50,0){\line(1,-2){10}}
\put(60,-20){\circle*{2}}
\put(55,-15){$2$}
\put(60,-20){\line(-1,-2){10}}
\put(50,-40){\circle*{2}}
\put(55,-35){$1$}
\put(50,-40){\line(0,-1){20}}
\put(50,-60){\circle*{2}}
\put(50,-55){$1$}
\put(60,-20){\line(1,-2){10}}
\put(70,-40){\circle*{2}}
\put(65,-35){$2$}
\put(70,-40){\line(0,-1){20}}
\put(70,-60){\circle*{2}}
\put(70,-55){$1$}

\put(110,0){\circle*{2}}
\put(110,0){\line(-1,-2){10}}
\put(100,-20){\circle*{2}}
\put(105,-15){$1$}
\put(100,-20){\line(-1,-2){10}}
\put(90,-40){\circle*{2}}
\put(95,-35){$1$}
\put(90,-40){\line(0,-1){20}}
\put(90,-60){\circle*{2}}
\put(90,-55){$1$}
\put(100,-20){\line(1,-2){10}}
\put(110,-40){\circle*{2}}
\put(105,-35){$2$}
\put(110,-40){\line(0,-1){20}}
\put(110,-60){\circle*{2}}
\put(110,-55){$1$}
\put(110,0){\line(1,-2){10}}
\put(120,-20){\circle*{2}}
\put(115,-15){$2$}
\put(120,-20){\line(0,-1){20}}
\put(120,-40){\circle*{2}}
\put(120,-35){$2$}

\end{picture}

\caption{}                 \label{Fig:59}

\end{figure}

In $m$-bonsai, every cocycle $C$ is a planar tree, all of whose
vertices which are not branch-ends have ramification number $m$, so
if $C$ contains $n$ corollas with $m$ edges, it has $mn+1$ vertices,
$n$ of those vertices have $m$ successors and $(mn+1)-n$ of those
vertices have 0 successors, i.e., are the endpoints of edges, in the
language of \cite{S}. Then, by Theorem 5.3.10 of \cite{S}, the
number of such $C$ is
\begin{eqnarray}
\frac{1}{mn+1}
\left(
\begin{array}{c}
mn+1\\
(mn+1)-n,...,n
\end{array}
\right)
\end{eqnarray}
which is $\frac{(mn)!}{((m-1)n+1)!n!}$. A seedling $S$ is obtained
by adding one edge to every branch-end vertex of $C$ and $C$ has
$(mn+1)-n$ branch-end vertices. So $S$ has $mn+((mn+1)-n)=(2m-1)n+1$
vertices. Thus we have

\begin{thm}
The cohomology groups $H^i$ of $m$-bonsai Hopf algebra by the
vertex-appending differential is,
\begin{eqnarray}
H^i= \left\{ \begin{array}{ll}
k^{\frac{(mn)!}{((m-1)n+1)!n!}} & \mbox{if } i=(2m-1)n+1, n \geq 0 \\
0                               & \mbox{otherwise.}
\end{array} \right. \notag
\end{eqnarray}
\end{thm}

In 2-bonsai,
the representatives of $H^i$ are as in Figure \ref{Fig:6'}.

\begin{figure}[h]

\begin{picture}(60,120)(0,-110)

\put(-160,-25){The representative of $H^4$:}

\put(-40,-25){$\sum$}

\put(0,0){\circle*{2}}
\put(-7,-10){1}
\put(5,-10){2}
\put(0,0){\line(1,-2){10}}
\put(10,-20){\circle*{2}}
\put(0,0){\line(-1,-2){10}}
\put(-10,-20){\circle*{2}}

\put(-10,-20){\line(0,-1){20}}
\put(-10,-40){\circle*{2}}
\put(-10,-30){$i$}
\put(10,-20){\line(0,-1){20}}
\put(10,-40){\circle*{2}}
\put(10,-30){$j$}

\put(-160,-75){The representatives of $H^4$:}

\put(-40,-75){$\sum$}

\put(10,-50){\circle*{2}}
\put(3,-60){1}
\put(15,-60){2}
\put(10,-50){\line(1,-2){10}}
\put(20,-70){\circle*{2}}
\put(10,-50){\line(-1,-2){10}}
\put(0,-70){\circle*{2}}

\put(0,-70){\line(-1,-2){10}}
\put(-10,-90){\circle*{2}}
\put(-10,-80){1}
\put(-10,-90){\line(0,-1){20}}
\put(-10,-110){\circle*{2}}
\put(-10,-100){$i$}
\put(0,-70){\line(1,-2){10}}
\put(10,-90){\circle*{2}}
\put(5,-80){2}
\put(10,-90){\line(0,-1){20}}
\put(10,-110){\circle*{2}}
\put(10,-100){$j$}
\put(20,-70){\line(0,-1){20}}
\put(20,-90){\circle*{2}}
\put(20,-80){$k$}

\put(60,-75){$\sum$}

\put(110,-50){\circle*{2}}
\put(103,-60){1}
\put(115,-60){2}
\put(110,-50){\line(1,-2){10}}
\put(120,-70){\circle*{2}}
\put(110,-50){\line(-1,-2){10}}
\put(100,-70){\circle*{2}}

\put(100,-70){\line(0,-1){20}}
\put(100,-90){\circle*{2}}
\put(100,-80){$i$}

\put(120,-70){\line(-1,-2){10}}
\put(110,-90){\circle*{2}}
\put(110,-80){1}
\put(110,-90){\line(0,-1){20}}
\put(110,-110){\circle*{2}}
\put(110,-100){$j$}
\put(120,-70){\line(1,-2){10}}
\put(130,-90){\circle*{2}}
\put(125,-80){2}
\put(130,-90){\line(0,-1){20}}
\put(130,-110){\circle*{2}}
\put(130,-100){$k$}

\end{picture}

\caption{}             \label{Fig:6'}

\end{figure}

\section{Appending Operation $*$ and Its Deviation }

Over a general base field $k$ for a bonsai Hopf algebra, it is not
so easy to find a good algebraic relationship between $\partial(T_1
* T_2)$ and $(\partial T_1) * T_2 \pm T_1 * (\partial T_2)$ where
$\partial$ is the vertex-appending differential, mainly because of
the signs of determinanted bonsais. But if  $k$ has characteristic
2, we don't need to consider signs. Moreover, the vertex-appending
differential becomes just appending of a vertex, taking no
consideration of determinanted terms, but it is still a boundary map
by the same argument as in the previous section. Then the relation
of $\partial(T_1 * T_2)$ and $(\partial T_1) * T_2 + T_1 * (\partial
T_2)$ becomes much simpler. In this section, we consider only the
case where the characteristic of $k$ is 2.

Let us define the binary operation $* _1$ by

\begin{eqnarray}
\partial(T_1 * T_2) =
(\partial T_1) * T_2 + T_1 * (\partial T_2) + T_1 * _1 T_2
\end{eqnarray}
and call this operation $* _1$ the {\em first deviation} of the
operation $*$. We let $* _2$ be defined by
\begin{eqnarray}
\partial(T_1 * _1 T_2) =
(\partial T_1) * _1 T_2
+ T_1 * _1 (\partial T_2)
+ T_1 * _2 T_2
\end{eqnarray}
and call this operation the {\em second deviation} of the operation $*$.
We define 3rd, 4th,... deviations iteratively.
Note that $* _2$ is the first deviation of $* _1$.

First, for the appending operation $*$, we have

\begin{thm}
In bonsai Hopf algebra, with its base field of characteristic 2,
$T_1 *_1 T_2$ is the sum of all $T'$'s,
where $T'$ is any bonsai obtained by connecting
a tip $v$ of $T_2$ and the root of $T_1$
with one edge and attatch another edge to that vetrex of $T_2$,
or by connecting a non-tip of $T_2$
and the root of $T_1$ with a length-2 ladder.
In both cases, the edges added to $T_1$ and $T_2$
are allowed to have all possible labels,
as in the example in 3-bonsai of Figure \ref{Fig:60}.
\end{thm}

\begin{figure}[h]

\begin{picture}(200,120)(60,-90)

\put(0,0){\circle*{2}}
\put(0,0){\line(-1,-3){5}}
\put(0,0){\line(0,-1){15}}
\put(0,0){\line(1,-3){5}}
\put(-5,-15){\circle*{2}}
\put(0,-15){\circle*{2}}
\put(5,-15){\circle*{2}}
\put(5,-10){$*_1$}
\put(15,0){\circle*{2}}
\put(15,0){\line(0,-1){15}}
\put(15,-15){\circle*{2}}
\put(15,-10){1}

\put(25,-10){=}

\put(35,15){\circle*{2}}
\put(35,15){\line(0,-1){15}}
\put(35,5){1}
\put(35,0){\circle*{2}}
\put(35,0){\line(0,-1){15}}
\put(35,-10){1}
\put(35,0){\line(1,-1){15}}
\put(50,-15){\circle*{2}}
\put(45,-10){2}
\put(35,-15){\circle*{2}}
\put(35,-15){\line(-1,-3){5}}
\put(35,-15){\line(0,-1){15}}
\put(35,-15){\line(1,-3){5}}
\put(30,-30){\circle*{2}}
\put(35,-30){\circle*{2}}
\put(40,-30){\circle*{2}}

\put(55,-10){+}

\put(65,15){\circle*{2}}
\put(65,15){\line(0,-1){15}}
\put(65,5){1}
\put(65,0){\circle*{2}}
\put(65,0){\line(0,-1){15}}
\put(65,-10){1}
\put(65,0){\line(1,-1){15}}
\put(80,-15){\circle*{2}}
\put(75,-10){3}
\put(65,-15){\circle*{2}}
\put(65,-15){\line(-1,-3){5}}
\put(65,-15){\line(0,-1){15}}
\put(65,-15){\line(1,-3){5}}
\put(60,-30){\circle*{2}}
\put(65,-30){\circle*{2}}
\put(70,-30){\circle*{2}}

\put(85,-10){+}

\put(95,15){\circle*{2}}
\put(95,15){\line(0,-1){15}}
\put(95,5){1}
\put(95,0){\circle*{2}}
\put(95,0){\line(0,-1){15}}
\put(95,-10){2}
\put(95,0){\line(1,-1){15}}
\put(110,-15){\circle*{2}}
\put(105,-10){1}
\put(95,-15){\circle*{2}}
\put(95,-15){\line(-1,-3){5}}
\put(95,-15){\line(0,-1){15}}
\put(95,-15){\line(1,-3){5}}
\put(90,-30){\circle*{2}}
\put(95,-30){\circle*{2}}
\put(100,-30){\circle*{2}}

\put(115,-10){+}

\put(125,15){\circle*{2}}
\put(125,15){\line(0,-1){15}}
\put(125,5){1}
\put(125,0){\circle*{2}}
\put(125,0){\line(0,-1){15}}
\put(125,-10){2}
\put(125,0){\line(1,-1){15}}
\put(140,-15){\circle*{2}}
\put(135,-10){3}
\put(125,-15){\circle*{2}}
\put(125,-15){\line(-1,-3){5}}
\put(125,-15){\line(0,-1){15}}
\put(125,-15){\line(1,-3){5}}
\put(120,-30){\circle*{2}}
\put(125,-30){\circle*{2}}
\put(130,-30){\circle*{2}}

\put(145,-10){+}

\put(155,15){\circle*{2}}
\put(155,15){\line(0,-1){15}}
\put(155,5){1}
\put(155,0){\circle*{2}}
\put(155,0){\line(0,-1){15}}
\put(155,-10){3}
\put(155,0){\line(1,-1){15}}
\put(170,-15){\circle*{2}}
\put(165,-10){1}
\put(155,-15){\circle*{2}}
\put(155,-15){\line(-1,-3){5}}
\put(155,-15){\line(0,-1){15}}
\put(155,-15){\line(1,-3){5}}
\put(150,-30){\circle*{2}}
\put(155,-30){\circle*{2}}
\put(160,-30){\circle*{2}}

\put(175,-10){+}

\put(185,15){\circle*{2}}
\put(185,15){\line(0,-1){15}}
\put(185,5){1}
\put(185,0){\circle*{2}}
\put(185,0){\line(0,-1){15}}
\put(185,-10){3}
\put(185,0){\line(1,-1){15}}
\put(200,-15){\circle*{2}}
\put(195,-10){2}
\put(185,-15){\circle*{2}}
\put(185,-15){\line(-1,-3){5}}
\put(185,-15){\line(0,-1){15}}
\put(185,-15){\line(1,-3){5}}
\put(180,-30){\circle*{2}}
\put(185,-30){\circle*{2}}
\put(190,-30){\circle*{2}}

\put(25,-70){+}

\put(50,-45){\circle*{2}}
\put(50,-45){\line(-1,-2){10}}
\put(40,-65){\circle*{2}}
\put(45,-55){1}
\put(50,-45){\line(1,-2){10}}
\put(60,-65){\circle*{2}}
\put(55,-55){2}
\put(60,-65){\line(0,-1){20}}
\put(60,-85){\circle*{2}}
\put(62,-81){1}
\put(60,-85){\line(-1,-3){5}}
\put(60,-85){\line(0,-1){15}}
\put(60,-85){\line(1,-3){5}}
\put(55,-100){\circle*{2}}
\put(60,-100){\circle*{2}}
\put(65,-100){\circle*{2}}

\put(65,-70){+}

\put(90,-45){\circle*{2}}
\put(90,-45){\line(-1,-2){10}}
\put(80,-65){\circle*{2}}
\put(85,-55){1}
\put(90,-45){\line(1,-2){10}}
\put(100,-65){\circle*{2}}
\put(95,-55){2}
\put(100,-65){\line(0,-1){20}}
\put(100,-85){\circle*{2}}
\put(102,-81){2}
\put(100,-85){\line(-1,-3){5}}
\put(100,-85){\line(0,-1){15}}
\put(100,-85){\line(1,-3){5}}
\put(95,-100){\circle*{2}}
\put(100,-100){\circle*{2}}
\put(105,-100){\circle*{2}}

\put(105,-70){+}

\put(130,-45){\circle*{2}}
\put(130,-45){\line(-1,-2){10}}
\put(120,-65){\circle*{2}}
\put(125,-55){1}
\put(130,-45){\line(1,-2){10}}
\put(140,-65){\circle*{2}}
\put(135,-55){2}
\put(140,-65){\line(0,-1){20}}
\put(140,-85){\circle*{2}}
\put(142,-81){3}
\put(140,-85){\line(-1,-3){5}}
\put(140,-85){\line(0,-1){15}}
\put(140,-85){\line(1,-3){5}}
\put(135,-100){\circle*{2}}
\put(140,-100){\circle*{2}}
\put(145,-100){\circle*{2}}

\put(145,-70){+}

\put(170,-45){\circle*{2}}
\put(170,-45){\line(-1,-2){10}}
\put(160,-65){\circle*{2}}
\put(165,-55){1}
\put(170,-45){\line(1,-2){10}}
\put(180,-65){\circle*{2}}
\put(175,-55){3}
\put(180,-65){\line(0,-1){20}}
\put(180,-85){\circle*{2}}
\put(182,-81){1}
\put(180,-85){\line(-1,-3){5}}
\put(180,-85){\line(0,-1){15}}
\put(180,-85){\line(1,-3){5}}
\put(175,-100){\circle*{2}}
\put(180,-100){\circle*{2}}
\put(185,-100){\circle*{2}}

\put(185,-70){+}

\put(210,-45){\circle*{2}}
\put(210,-45){\line(-1,-2){10}}
\put(200,-65){\circle*{2}}
\put(205,-55){1}
\put(210,-45){\line(1,-2){10}}
\put(220,-65){\circle*{2}}
\put(215,-55){3}
\put(220,-65){\line(0,-1){20}}
\put(220,-85){\circle*{2}}
\put(222,-81){2}
\put(220,-85){\line(-1,-3){5}}
\put(220,-85){\line(0,-1){15}}
\put(220,-85){\line(1,-3){5}}
\put(215,-100){\circle*{2}}
\put(220,-100){\circle*{2}}
\put(225,-100){\circle*{2}}

\put(225,-70){+}

\put(250,-45){\circle*{2}}
\put(250,-45){\line(-1,-2){10}}
\put(240,-65){\circle*{2}}
\put(245,-55){1}
\put(250,-45){\line(1,-2){10}}
\put(260,-65){\circle*{2}}
\put(255,-55){3}
\put(260,-65){\line(0,-1){20}}
\put(260,-85){\circle*{2}}
\put(262,-81){3}
\put(260,-85){\line(-1,-3){5}}
\put(260,-85){\line(0,-1){15}}
\put(260,-85){\line(1,-3){5}}
\put(255,-100){\circle*{2}}
\put(260,-100){\circle*{2}}
\put(265,-100){\circle*{2}}

\end{picture}

\caption{}                 \label{Fig:60}

\end{figure}

\begin{proof}
Let us use a graphical illustration. Bonsais which are summands of
$\partial(T_1*T_2)$ are as in Figure \ref{Fig:61}, where $T_1$ and
$T_2$ and the appended vertex are drawn as broomsticks, and $i$,$j$
and $k$ are indices of twigs and $T_1 * T_2$ is the sum of these two
kinds of bonsai over $i$, $j$ and $k$. In this proof, the black
circles in the pictures represent the twigs at non-tips of bonsais,
and the white circles represent the twigs at tips of bonsais. Here,
$i$ is on the vertices of $T_2$ which are not tips, and $k$ is on
the vertices of $T_1$ which are not tips and $i'$ is on the vertices
of $T_2$ which are tips, but not tips in $\partial(T_1 * T_2)$ since
the connecting edge is attached.

\begin{figure}[h]

\begin{picture}(160,150)(40,-140)

\put(0,0){\line(0,-1){20}}

\put(0,-35){$T_2$}
\put(0,-20){\line(2,-1){40}}
\put(0,-20){\line(-2,-1){40}}
\put(-40,-40){\line(1,0){80}}

\put(16,-40){\line(0,-1){80}}

\put(19,-50){$j$}

\put(-27,-50){$i$}
\put(-33,-75){$T$}
\put(-30,-40){\line(0,-1){20}}
\put(-30,-40){\circle*{6}}
\put(-30,-60){\line(-1,-2){10}}
\put(-30,-60){\line(1,-2){10}}
\put(-40,-80){\line(1,0){20}}

\put(16,-135){$T_1$}
\put(16,-120){\line(2,-1){40}}
\put(16,-120){\line(-2,-1){40}}
\put(-24,-140){\line(1,0){80}}

\put(70,-90){+}

\put(120,0){\line(0,-1){20}}

\put(120,-35){$T_2$}
\put(120,-20){\line(2,-1){40}}
\put(120,-20){\line(-2,-1){40}}
\put(80,-40){\line(1,0){80}}

\put(138,-40){\line(0,-1){30}}
\put(141,-60){$j$}

\put(138,-85){$T_1$}
\put(138,-70){\line(2,-1){40}}
\put(138,-70){\line(-2,-1){40}}
\put(98,-90){\line(1,0){80}}

\put(156,-90){\line(0,-1){30}}
\put(156,-90){\circle*{6}}
\put(159,-110){$k$}

\put(153,-135){$T$}
\put(156,-120){\line(-1,-2){10}}
\put(156,-120){\line(1,-2){10}}
\put(146,-140){\line(1,0){20}}

\put(190,-90){+}

\put(240,0){\line(0,-1){20}}

\put(240,-35){$T_2$}
\put(240,-20){\line(2,-1){40}}
\put(240,-20){\line(-2,-1){40}}
\put(200,-40){\line(1,0){80}}

\put(256,-40){\line(0,-1){80}}

\put(259,-50){$j$}

\put(246,-50){$i'$}
\put(231,-75){$T$}
\put(256,-40){\line(-1,-1){20}}
\put(256,-40){\circle{6}}
\put(236,-60){\line(-1,-2){10}}
\put(236,-60){\line(1,-2){10}}
\put(226,-80){\line(1,0){20}}

\put(256,-135){$T_1$}
\put(256,-120){\line(2,-1){40}}
\put(256,-120){\line(-2,-1){40}}
\put(216,-140){\line(1,0){80}}

\end{picture}

\caption{}              \label{Fig:61}

\end{figure}

The bonsais which are summands of $T_1 * \partial T_2$
are as in Figure \ref{Fig:62}.
In the right bonsai of Figure \ref{Fig:62},
the label of the edge connecting $T$ and $T_1$
can be anything out of $1,2,...m$.

\begin{figure}[h]

\begin{picture}(160,150)(40,-140)

\put(0,0){\line(0,-1){20}}

\put(0,-35){$T_2$}
\put(0,-20){\line(2,-1){40}}
\put(0,-20){\line(-2,-1){40}}
\put(-40,-40){\line(1,0){80}}

\put(26,-40){\line(0,-1){80}}

\put(29,-50){$j$}

\put(-17,-50){$i$}
\put(-23,-75){$T$}
\put(-20,-40){\line(0,-1){20}}
\put(-20,-40){\circle*{6}}
\put(-20,-60){\line(-1,-2){10}}
\put(-20,-60){\line(1,-2){10}}
\put(-30,-80){\line(1,0){20}}

\put(26,-135){$T_1$}
\put(26,-120){\line(2,-1){40}}
\put(26,-120){\line(-2,-1){40}}
\put(-14,-140){\line(1,0){80}}

\put(90,-90){+}

\put(150,0){\line(0,-1){20}}

\put(150,-35){$T_2$}
\put(150,-20){\line(2,-1){40}}
\put(150,-20){\line(-2,-1){40}}
\put(110,-40){\line(1,0){80}}

\put(168,-40){\line(0,-1){30}}
\put(171,-60){$j$}

\put(163,-85){$T$}
\put(168,-70){\line(-1,-2){10}}
\put(168,-70){\line(1,-2){10}}
\put(158,-90){\line(1,0){20}}

\put(166,-90){\line(0,-1){30}}

\put(168,-135){$T_1$}
\put(166,-120){\line(2,-1){40}}
\put(166,-120){\line(-2,-1){40}}
\put(126,-140){\line(1,0){80}}

\end{picture}

\caption{}              \label{Fig:62}

\end{figure}

The bonsais which are summands of $\partial T_1 * T_2$ are as in
Figure \ref{Fig:63}.
\begin{figure}[h]

\begin{picture}(100,150)(-50,-140)

\put(0,0){\line(0,-1){20}}

\put(0,-35){$T_2$}
\put(0,-20){\line(2,-1){40}}
\put(0,-20){\line(-2,-1){40}}
\put(-40,-40){\line(1,0){80}}

\put(18,-40){\line(0,-1){30}}
\put(21,-60){$j$}

\put(18,-85){$T_1$}
\put(18,-70){\line(2,-1){40}}
\put(18,-70){\line(-2,-1){40}}
\put(-22,-90){\line(1,0){80}}

\put(36,-90){\line(0,-1){30}}
\put(39,-110){$k$}

\put(33,-135){$T$}
\put(36,-120){\line(-1,-2){10}}
\put(36,-120){\line(1,-2){10}}
\put(26,-140){\line(1,0){20}}

\end{picture}

\caption{}              \label{Fig:63}

\end{figure}
Then the discrepancy between $\partial(T_1 * T_2)$ and $\partial T_1
* T_2 + T_1 * \partial T_2$ is the sum of the third bonsai of Figure
\ref{Fig:61} and the second bonsai of Figure \ref{Fig:62}, which
gives the wanted formula.

\end{proof}

\begin{thm}
Mod 2, for any $m$-bonsais $T_1$ and $T_2$ , we have $T_1 *_2 T_2 =
0$. In other words, $\partial(T_1 *_1 T_2) = \partial T_1 *_1 T_2 +
T_1 *_1 \partial T_2$.
\end{thm}

\begin{proof}
Bonsais which are summands of $T_1 *_1 T_2$
are those in Figure \ref{Fig:64},
which represent the bonsais obtained by connecting $T_1$ and $T_2$
with one edge and attatching another edge
and obtained by connecting $T_1$ and $T_2$ with a length-2 ladder.
As in the proof of the previous theorem,
a black circle represents a non-tip of a bonsai and
a white circle represents a tip of a bonsai.

\begin{figure}[h]

\begin{picture}(160,150)(-40,-140)

\put(0,0){\line(0,-1){20}}

\put(0,-35){$T_2$}
\put(0,-20){\line(2,-1){40}}
\put(0,-20){\line(-2,-1){40}}
\put(-40,-40){\line(1,0){80}}

\put(16,-40){\circle{6}}
\put(16,-40){\line(0,-1){80}}
\put(19,-70){$j$}
\put(16,-40){\line(1,-1){30}}
\put(41,-65){$j'$}
\put(46,-70){\circle*{2}}

\put(16,-135){$T_1$}
\put(16,-120){\line(2,-1){40}}
\put(16,-120){\line(-2,-1){40}}
\put(-24,-140){\line(1,0){80}}

\put(60,-90){+}

\put(110,0){\line(0,-1){20}}

\put(110,-35){$T_2$}
\put(110,-20){\line(2,-1){40}}
\put(110,-20){\line(-2,-1){40}}
\put(70,-40){\line(1,0){80}}

\put(126,-40){\circle*{6}}
\put(126,-40){\line(0,-1){80}}
\put(129,-70){$j$}
\put(126,-80){\circle*{3}}

\put(126,-135){$T_1$}
\put(126,-120){\line(2,-1){40}}
\put(126,-120){\line(-2,-1){40}}
\put(86,-140){\line(1,0){80}}

\end{picture}

\caption{}              \label{Fig:64}

\end{figure}

The bonsais in $\partial(T_1 *_1 T_2)$ obtained by $\partial$ acting
on the left bonsai in Figure \ref{Fig:64} are those in Figure
\ref{Fig:65}, and the bonsais obtained by $\partial$ acting on the
right bonsai in Figure \ref{Fig:64} are those in Figure
\ref{Fig:66}. In both pictures, the broomstick lettered $T$ is a
one-vertex bonsai which is appended by the vertex-appending
differential $\partial$.

The bonsais in $\partial(T_1 *_1 T_2)$ corresponding to the left
bonsai in Figure \ref{Fig:64} are like Figure \ref{Fig:65}, and the
ones corresponding to the right one are like Figure \ref{Fig:66}.
For later use, we denote those bonsais A, B,...,G as assigned in the
Figures. In both pictures, the broomstick lettered $T$ is a
one-vertex bonsai which is appended by the vertex-appending
differential $\partial$.

\begin{figure}[h]

\begin{picture}(160,150)(60,-140)

\put(-40,0){A}

\put(0,0){\line(0,-1){20}}

\put(0,-35){$T_2$}
\put(0,-20){\line(2,-1){40}}
\put(0,-20){\line(-2,-1){40}}
\put(-40,-40){\line(1,0){80}}

\put(16,-40){\circle*{6}}
\put(16,-40){\line(0,-1){80}}
\put(19,-65){$j$}
\put(16,-80){\circle*{3}}

\put(-1,-55){$i$}
\put(-17,-85){$T$}

\put(16,-40){\line(-1,-1){30}}
\put(-14,-70){\line(-1,-2){10}}
\put(-14,-70){\line(1,-2){10}}
\put(-24,-90){\line(1,0){20}}

\put(16,-135){$T_1$}
\put(16,-120){\line(2,-1){40}}
\put(16,-120){\line(-2,-1){40}}
\put(-27,-140){\line(1,0){80}}

\put(40,-90){+}

\put(50,0){B}

\put(90,0){\line(0,-1){20}}

\put(90,-35){$T_2$}
\put(90,-20){\line(2,-1){40}}
\put(90,-20){\line(-2,-1){40}}
\put(50,-40){\line(1,0){80}}

\put(106,-40){\circle*{6}}
\put(106,-40){\line(0,-1){80}}
\put(109,-65){$j$}
\put(106,-80){\circle*{3}}

\put(63,-55){$i$}
\put(57,-75){$T$}
\put(60,-40){\circle*{6}}
\put(60,-40){\line(0,-1){20}}
\put(60,-60){\line(-1,-2){10}}
\put(60,-60){\line(1,-2){10}}
\put(50,-80){\line(1,0){20}}

\put(106,-135){$T_1$}
\put(106,-120){\line(2,-1){40}}
\put(106,-120){\line(-2,-1){40}}
\put(66,-140){\line(1,0){80}}

\put(130,-90){+}

\put(140,0){C}

\put(180,0){\line(0,-1){20}}

\put(180,-35){$T_2$}
\put(180,-20){\line(2,-1){40}}
\put(180,-20){\line(-2,-1){40}}
\put(140,-40){\line(1,0){80}}

\put(196,-40){\circle*{6}}
\put(196,-40){\line(0,-1){80}}
\put(199,-65){$j$}
\put(196,-75){\circle*{3}}
\put(196,-75){\line(1,-1){20}}
\put(201,-80){$j'$}

\put(212,-110){$T$}
\put(216,-95){\line(-1,-2){10}}
\put(216,-95){\line(1,-2){10}}
\put(206,-115){\line(1,0){20}}

\put(196,-135){$T_1$}
\put(196,-120){\line(2,-1){40}}
\put(196,-120){\line(-2,-1){40}}
\put(156,-140){\line(1,0){80}}

\put(220,-90){+}

\put(230,0){D}

\put(270,0){\line(0,-1){20}}

\put(270,-35){$T_2$}
\put(270,-20){\line(2,-1){40}}
\put(270,-20){\line(-2,-1){40}}
\put(230,-40){\line(1,0){80}}

\put(288,-55){\circle*{3}}
\put(288,-40){\circle*{6}}
\put(288,-40){\circle*{6}}
\put(288,-40){\line(0,-1){30}}
\put(291,-50){$j$}

\put(288,-85){$T_1$}
\put(288,-70){\line(2,-1){40}}
\put(288,-70){\line(-2,-1){40}}
\put(248,-90){\line(1,0){80}}

\put(306,-90){\circle*{6}}
\put(306,-90){\line(0,-1){30}}
\put(309,-110){$k$}

\put(303,-135){$T$}
\put(306,-120){\line(-1,-2){10}}
\put(306,-120){\line(1,-2){10}}
\put(296,-140){\line(1,0){20}}

\end{picture}

\caption{}              \label{Fig:65}

\end{figure}

\begin{figure}[h]

\begin{picture}(160,150)(40,-140)

\put(-40,0){E}

\put(0,0){\line(0,-1){20}}

\put(0,-35){$T_2$}
\put(0,-20){\line(2,-1){40}}
\put(0,-20){\line(-2,-1){40}}
\put(-40,-40){\line(1,0){80}}

\put(16,-40){\circle{6}}
\put(16,-40){\line(0,-1){80}}
\put(19,-65){$j$}
\put(16,-40){\line(1,-1){30}}
\put(41,-65){$j'$}
\put(46,-70){\circle*{2}}

\put(-27,-50){$i$}
\put(-33,-75){$T$}
\put(-30,-40){\circle*{6}}
\put(-30,-40){\line(0,-1){20}}
\put(-30,-60){\line(-1,-2){10}}
\put(-30,-60){\line(1,-2){10}}
\put(-40,-80){\line(1,0){20}}

\put(16,-135){$T_1$}
\put(16,-120){\line(2,-1){40}}
\put(16,-120){\line(-2,-1){40}}
\put(-24,-140){\line(1,0){80}}

\put(60,-90){+}

\put(70,0){F}

\put(110,0){\line(0,-1){20}}

\put(110,-35){$T_2$}
\put(110,-20){\line(2,-1){40}}
\put(110,-20){\line(-2,-1){40}}
\put(70,-40){\line(1,0){80}}

\put(126,-40){\circle{6}}
\put(126,-40){\line(0,-1){80}}
\put(129,-65){$j$}
\put(126,-40){\line(1,-1){30}}
\put(151,-65){$j'$}
\put(156,-70){\circle*{2}}
\put(126,-40){\line(-1,-1){30}}
\put(106,-65){$j''$}

\put(92,-85){$T$}
\put(96,-70){\line(-1,-2){10}}
\put(96,-70){\line(1,-2){10}}
\put(86,-90){\line(1,0){20}}

\put(126,-135){$T_1$}
\put(126,-120){\line(2,-1){40}}
\put(126,-120){\line(-2,-1){40}}
\put(86,-140){\line(1,0){80}}

\put(170,-90){+}

\put(180,0){G}

\put(220,0){\line(0,-1){20}}

\put(220,-35){$T_2$}
\put(220,-20){\line(2,-1){40}}
\put(220,-20){\line(-2,-1){40}}
\put(180,-40){\line(1,0){80}}

\put(238,-40){\circle{6}}
\put(238,-40){\line(0,-1){30}}
\put(241,-60){$j$}
\put(238,-40){\line(1,-1){30}}
\put(263,-65){$j'$}
\put(268,-70){\circle*{2}}

\put(238,-85){$T_1$}
\put(238,-70){\line(2,-1){40}}
\put(238,-70){\line(-2,-1){40}}
\put(198,-90){\line(1,0){80}}

\put(256,-90){\circle*{6}}
\put(256,-90){\line(0,-1){30}}
\put(259,-110){$k$}

\put(253,-135){$T$}
\put(256,-120){\line(-1,-2){10}}
\put(256,-120){\line(1,-2){10}}
\put(246,-140){\line(1,0){20}}

\end{picture}

\caption{}              \label{Fig:66}

\end{figure}

For the bonsai F, actually we have a term like \ref{Fig:67} also in
$\partial(T_1 *_1 T_2)$, and since we are working mod 2, the bonsais
looking like F are all canceled. So we have F=0.

\begin{figure}[h]

\begin{picture}(160,150)(40,-140)

\put(110,0){\line(0,-1){20}}

\put(110,-35){$T_2$}
\put(110,-20){\line(2,-1){40}}
\put(110,-20){\line(-2,-1){40}}
\put(70,-40){\line(1,0){80}}

\put(126,-40){\circle{6}}
\put(126,-40){\line(0,-1){80}}
\put(129,-60){$j$}
\put(126,-40){\line(1,-1){30}}
\put(151,-65){$j'$}
\put(96,-70){\circle*{2}}
\put(126,-40){\line(-1,-1){30}}
\put(111,-65){$j''$}

\put(152,-85){$T$}
\put(156,-70){\line(-1,-2){10}}
\put(156,-70){\line(1,-2){10}}
\put(146,-90){\line(1,0){20}}

\put(126,-135){$T_1$}
\put(126,-120){\line(2,-1){40}}
\put(126,-120){\line(-2,-1){40}}
\put(86,-140){\line(1,0){80}}

\end{picture}

\caption{}              \label{Fig:67}

\end{figure}

Now, the bonsais in $\partial (T_1) *_1 T_2$  are as in Figure \ref{Fig:68}.
Let us denote them as a and b.\

\begin{figure}[h]

\begin{picture}(160,150)(0,-140)

\put(-40,0){a}

\put(0,0){\line(0,-1){20}}

\put(0,-35){$T_2$}
\put(0,-20){\line(2,-1){40}}
\put(0,-20){\line(-2,-1){40}}
\put(-40,-40){\line(1,0){80}}

\put(18,-40){\circle{6}}
\put(18,-40){\line(0,-1){30}}
\put(21,-60){$j$}
\put(18,-40){\line(1,-1){30}}
\put(43,-65){$j'$}
\put(48,-70){\circle*{2}}

\put(18,-85){$T_1$}
\put(18,-70){\line(2,-1){40}}
\put(18,-70){\line(-2,-1){40}}
\put(-22,-90){\line(1,0){80}}

\put(36,-90){\circle*{6}}
\put(36,-90){\line(0,-1){30}}
\put(39,-110){$k$}

\put(33,-135){$T$}
\put(36,-120){\line(-1,-2){10}}
\put(36,-120){\line(1,-2){10}}
\put(26,-140){\line(1,0){20}}

\put(60,-90){+}

\put(70,0){b}

\put(110,0){\line(0,-1){20}}

\put(110,-35){$T_2$}
\put(110,-20){\line(2,-1){40}}
\put(110,-20){\line(-2,-1){40}}
\put(70,-40){\line(1,0){80}}

\put(128,-40){\circle*{6}}
\put(128,-40){\line(0,-1){30}}
\put(128,-55){\circle*{3}}
\put(131,-50){$j$}

\put(128,-85){$T_1$}
\put(128,-70){\line(2,-1){40}}
\put(128,-70){\line(-2,-1){40}}
\put(88,-90){\line(1,0){80}}

\put(146,-90){\circle*{6}}
\put(146,-90){\line(0,-1){30}}
\put(149,-110){$k$}

\put(143,-135){$T$}
\put(146,-120){\line(-1,-2){10}}
\put(146,-120){\line(1,-2){10}}
\put(136,-140){\line(1,0){20}}

\end{picture}

\caption{}              \label{Fig:68}

\end{figure}

The bonsais in $T_1 *_1 \partial(T_2)$ are like those in Figure
\ref{Fig:69}, where we denote the bonsais as c, d, e and f.

\begin{figure}[h]

\begin{picture}(160,150)(40,-140)

\put(-40,0){c}

\put(0,0){\line(0,-1){20}}

\put(0,-35){$T_2$}
\put(0,-20){\line(2,-1){40}}
\put(0,-20){\line(-2,-1){40}}
\put(-40,-40){\line(1,0){80}}

\put(16,-40){\circle{6}}
\put(16,-40){\line(0,-1){80}}
\put(19,-65){$j$}
\put(16,-40){\line(1,-1){15}}
\put(26,-50){$j'$}
\put(31,-55){\circle*{2}}

\put(-27,-50){$i$}
\put(-33,-75){$T$}
\put(-30,-40){\circle*{6}}
\put(-30,-40){\line(0,-1){20}}
\put(-30,-60){\line(-1,-2){10}}
\put(-30,-60){\line(1,-2){10}}
\put(-40,-80){\line(1,0){20}}

\put(16,-135){$T_1$}
\put(16,-120){\line(2,-1){40}}
\put(16,-120){\line(-2,-1){40}}
\put(-24,-140){\line(1,0){80}}

\put(40,-70){+}

\put(50,0){d}

\put(90,0){\line(0,-1){20}}

\put(90,-35){$T_2$}
\put(90,-20){\line(2,-1){40}}
\put(90,-20){\line(-2,-1){40}}
\put(50,-40){\line(1,0){80}}

\put(106,-40){\line(0,-1){20}}

\put(102,-75){$T$}
\put(106,-60){\line(-1,-2){10}}
\put(106,-60){\line(1,-2){10}}
\put(96,-80){\line(1,0){20}}

\put(106,-80){\circle{6}}
\put(106,-80){\line(0,-1){40}}
\put(106,-95){$k'$}
\put(106,-80){\line(1,-1){20}}
\put(121,-95){$k''$}
\put(126,-100){\circle*{2}}

\put(106,-135){$T_1$}
\put(106,-120){\line(2,-1){40}}
\put(106,-120){\line(-2,-1){40}}
\put(66,-140){\line(1,0){80}}

\put(130,-70){+}

\put(140,0){e}

\put(180,0){\line(0,-1){20}}

\put(180,-35){$T_2$}
\put(180,-20){\line(2,-1){40}}
\put(180,-20){\line(-2,-1){40}}
\put(140,-40){\line(1,0){80}}

\put(196,-40){\circle*{6}}
\put(196,-40){\line(0,-1){80}}
\put(199,-65){$j$}
\put(196,-80){\circle*{3}}

\put(153,-55){$i$}
\put(147,-75){$T$}
\put(150,-40){\circle*{6}}
\put(150,-40){\line(0,-1){20}}
\put(150,-60){\line(-1,-2){10}}
\put(150,-60){\line(1,-2){10}}
\put(140,-80){\line(1,0){20}}

\put(196,-135){$T_1$}
\put(196,-120){\line(2,-1){40}}
\put(196,-120){\line(-2,-1){40}}
\put(156,-140){\line(1,0){80}}

\put(220,-70){+}

\put(230,0){f}

\put(270,0){\line(0,-1){20}}

\put(270,-35){$T_2$}
\put(270,-20){\line(2,-1){40}}
\put(270,-20){\line(-2,-1){40}}
\put(230,-40){\line(1,0){80}}

\put(250,-40){\line(1,-1){40}}
\put(268,-55){$j$}
\put(290,-80){\circle*{3}}
\put(290,-80){\line(0,-1){40}}

\put(253,-55){$i$}
\put(247,-75){$T$}
\put(250,-40){\circle*{6}}
\put(250,-40){\line(0,-1){20}}
\put(250,-60){\line(-1,-2){10}}
\put(250,-60){\line(1,-2){10}}
\put(240,-80){\line(1,0){20}}

\put(290,-135){$T_1$}
\put(290,-120){\line(2,-1){40}}
\put(290,-120){\line(-2,-1){40}}
\put(250,-140){\line(1,0){80}}

\end{picture}

\caption{}              \label{Fig:69}

\end{figure}

In Figures \ref{Fig:65}-\ref{Fig:69}, we have a=G, b=D, c=E, d=C, e=B and f=A.
We already showed that F=0, so we have
$\partial(T_1 *_1 T_2) = \partial(T_1) *_1 T_2 + T_1 *_1 \partial(T_2)$.

\end{proof}

\section{Planar Clear-edged $m$-bonsai and its Derivative}

In planar clear-edged $m$-bonsai, we can define a differential and
calculate some cohomologies as for $m$-bonsai.

\begin{df}   \label{Def:vadiff2}
For the planar clear-edged $m$-bonsai, we define the
{\em vertex-appending differential} $\partial$ as follows;
Consider a determinated planar clear-edged $m$-bonsai
$T \otimes det(T)$. Then $\partial(T \otimes det(T))$
is the sum of $T' \otimes \wedge det(T)$, where $T'$ is a
planar clear-edged $m$-bonsai obtained by

i) appending a vertex to $T$

ii) except to tips of $T$,

and so, getting a new edge $e$.

If there is no available appending position an a bonsai, the map
assigns 0 to that bonsai.
\end{df}

For example, in planar clear-edged 3-bonsai,
we can get an example like Figure \ref{Fig:113}
(in bonsais of Figure \ref{Fig:113}, determinanted terms are omitted.
Note that one vertex in the first example is also a tip
and the third bonsai is a cocycle.
In the picture, newly appended vertices are drawn as open vertices,
not indicating colors).

\begin{figure}[h]

\begin{picture}(200,140)(0,-130)

\put(10,-10){\circle*{4}}

\put(25,-10){$\rightarrow \qquad 0$}

\put(10,-40){\circle*{4}}
\put(10,-40){\line(0,-1){20}}

\put(10,-60){\circle*{4}}

\put(25,-50){$\rightarrow$}

\put(60,-40){\circle*{4}}
\put(60,-40){\line(-1,-2){10}}
\put(50,-60){\circle{4}}

\put(60,-40){\line(1,-2){10}}
\put(70,-60){\circle*{4}}

\put(80,-50){$-$}

\put(100,-40){\circle*{4}}
\put(100,-40){\line(-1,-2){10}}
\put(90,-60){\circle*{4}}

\put(100,-40){\line(1,-2){10}}
\put(110,-60){\circle{4}}

\put(115,-50){$= \quad 0$}

\put(10,-80){\circle*{4}}
\put(10,-80){\line(-1,-2){10}}
\put(0,-100){\circle*{4}}

\put(10,-80){\line(1,-2){10}}
\put(20,-100){\circle*{4}}

\put(25,-90){$\rightarrow$}

\put(60,-80){\circle*{4}}
\put(60,-80){\line(-1,-2){10}}
\put(50,-100){\circle{4}}
\put(60,-80){\line(0,-1){20}}
\put(60,-100){\circle*{4}}
\put(60,-80){\line(1,-2){10}}
\put(70,-100){\circle*{4}}

\put(80,-90){$-$}

\put(100,-80){\circle*{4}}
\put(100,-80){\line(-1,-2){10}}
\put(90,-100){\circle*{4}}
\put(100,-80){\line(0,-1){20}}
\put(100,-100){\circle{4}}
\put(100,-80){\line(1,-2){10}}
\put(110,-100){\circle*{4}}

\put(120,-90){$+$}

\put(140,-80){\circle*{4}}
\put(140,-80){\line(-1,-2){10}}
\put(130,-100){\circle*{4}}
\put(140,-80){\line(0,-1){20}}
\put(140,-100){\circle*{4}}
\put(140,-80){\line(1,-2){10}}
\put(150,-100){\circle{4}}

\put(160,-90){$=$}

\put(180,-80){\circle*{4}}
\put(180,-80){\line(-1,-2){10}}
\put(170,-100){\circle*{4}}
\put(180,-80){\line(0,-1){20}}
\put(180,-100){\circle*{4}}
\put(180,-80){\line(1,-2){10}}
\put(190,-100){\circle*{4}}

\put(10,-120){\circle*{4}}
\put(10,-120){\line(-1,-2){10}}
\put(0,-140){\circle*{4}}

\put(10,-120){\line(0,-1){20}}
\put(10,-140){\circle*{4}}

\put(10,-120){\line(1,-2){10}}
\put(20,-140){\circle*{4}}

\put(25,-130){$\rightarrow \qquad 0$}

\end{picture}

\caption{}                 \label{Fig:113}

\end{figure}

And by the totally same argument as for edge-numbered bonsai, we get

\begin{thm}
${\partial}^{i+1} \circ {\partial}^i = 0$.
That is, $\partial$ is a differential.
\end{thm}

Now let us consider the cohomology groups of this differential.
First let us consider a cochain complex $\{{\mathbf D}^i\}$
consisting of corollae,
with the boundary map ${\bar \partial}$
being the vertex-appending differential,
but in this complex, appending to the one-vertex bonsai is allowed,
Here, let us denote  as ${\mathbf D}^0$ the module
having the basis consisting of one vertex,
and as ${\mathbf D}^n$ the one-dimensional module
having the basis consisting of the corolla
with $n$ edges as shown above.
By the definition of $\partial ^i$,
all terms in $\partial ^i (T)$ are of the from
$\pm T' \otimes e \wedge det(T)$,
where $T'$ runs over bonsais obtained
by adding one vertex as defined in Definition \ref{Def:vadiff2}.
Since appending to an edge-end is forbidden,
every $\partial^i$ is appending a vertex to the root of a corolla.
Now let us show some boundary map sequences of the thread starting
from one vertex.
For one vertex, $\partial ^0$ acts as in Figure \ref{Fig:114}.

\begin{figure}[h]

\begin{picture}(200,30)(-50,-20)

\put(10,-10){\circle*{2}}

\put(25,-10){$\rightarrow$}

\put(50,0){\circle*{2}}
\put(50,0){\line(0,-1){20}}
\put(50,-20){\circle*{2}}

\end{picture}

\caption{}                 \label{Fig:114}

\end{figure}

For the one-edge corolla, ${\partial}^1$ acts as in the second map of
Figure \ref{Fig:113},
for the two-edge corolla, ${\partial}^2$ acts as in the third map of
Figure \ref{Fig:113},
and so on.

This sequence of coboundary maps is

\begin{eqnarray}
k     \overset{id}{\rightarrow}
k     \overset{0 }{\rightarrow}
k     \overset{id}{\rightarrow}
k     \overset{0 }{\rightarrow}
...
\end{eqnarray}
where $k$ is the base field.

So the cohomology groups of this thread of boundary maps is acyclic,
and the lowest degree group is trivial.

Also, let $\{{\mathbf B}^n\}$ be a cochain complex defined by
${\mathbf B}^n = {\mathbf D}^{n+1}$ for later convenience.
Then $\{{\mathbf B}^{n}\}$ is acyclic and $H^0=k$,
where $k$ is the base field.

Now let us consider the general case.
By the definition of $\partial^i$,
all terms in $\partial^i(T)$ are of the form $\pm T' \otimes e \wedge det(T)$,
where $T'$ runs over bonsais obtained by adding a new edge $e$ to $T$
so that i) and ii) of Definition \ref{Def:vadiff2} hold.
So $T'$ has the form of appending a vertex to a vertex of T
other than a tip.
Having this intuitive fact in mind,
let us present some new definitions and reorganize
the cochain complex of bonsais.

\begin{df}
For a bonsai $T$, an edge $e$ of $T$ is called {\em twiggy}
if it is at the end of a branch and the opposite end of the tip
is a branching vertex.
In Figure \ref{Fig:115}, $e$ is twiggy in $T$ and $e'$ is not,
and $f$ is not a twiggy edge of $T'$.

\begin{figure}[h]

\begin{picture}(200,50)(0,-40)

\put(20,-10){$T=$}

\put(50,0){\circle*{2}}
\put(50,0){\line(-1,-2){10}}
\put(40,-20){\circle*{2}}
\put(42,-10){\line(-2,-1){20}}
\put(-15,-25){$e$, twiggy}
\put(50,0){\line(1,-2){10}}
\put(60,-20){\circle*{2}}
\put(60,-20){\line(0,-1){20}}
\put(65,-35){$e'$, not twiggy}
\put(60,-40){\circle*{2}}

\put(115,-10){$T'=$}
\put(140,0){\circle*{2}}
\put(140,0){\line(0,-1){20}}
\put(140,-20){\circle*{2}}
\put(142,-15){$f$}

\end{picture}

\caption{}                 \label{Fig:115}

\end{figure}

\end{df}

\begin{df}  \label{Def:clearseedling}
A bonsai which has no twiggy edge is called a
{\em vertex-appending seedling}.
In this section, we will just call this {\em seedling}.
The bonsais in Figure \ref{Fig:116} are all seedlings.
Note that the one-vertex bonsai is a seedling.
Intuitively, a seedling is a bonsai which cannot be obtained by
adding edges like i) and ii) of Definition \ref{Def:vadiff}.

\begin{figure}[h]

\begin{picture}(200,50)(0,-40)

\put(50,0){\circle*{2}}
\put(50,0){\line(-1,-2){10}}
\put(40,-20){\circle*{2}}

\put(40,-20){\line(0,-1){20}}
\put(40,-40){\circle*{2}}

\put(50,0){\line(1,-2){10}}
\put(60,-20){\circle*{2}}

\put(60,-20){\line(0,-1){20}}

\put(60,-40){\circle*{2}}

\put(110,0){\circle*{2}}
\put(110,0){\line(0,-1){20}}
\put(110,-20){\circle*{2}}

\put(110,-20){\line(0,-1){20}}
\put(110,-40){\circle*{2}}

\put(170,0){\circle*{2}}
\put(170,0){\line(0,-1){20}}

\put(170,-20){\circle*{2}}

\end{picture}

\caption{}                 \label{Fig:116}

\end{figure}

\end{df}

\subsection{Cohomology Groups of $\partial$ when $m=\infty$}

Let us try the same trick as in the proof of acyclicity
of the branch-fixed differential.

\begin{df}
When $S$ is a seedling, let ${\mathbf C}^{S,0}$
be the subspace of the determinanted planar clear-edged
$\infty$-bonsai space having
$\{S \otimes det(S)\}$ as the basis.
And let ${\mathbf C}^{S,i+1}$ be the space with the basis
$\{T' \otimes det(T')\}$,
where $T'$ is obtained by adding an edge to $T$,
where $\{T \otimes det(T)\}$ is the basis of ${\mathbf C}^{S,i}$,
as i) and ii) of Definition \ref{Def:vadiff2}.
Then since every bonsai is obtained by adding some edges
to a seedling as given in Definition \ref{Def:vadiff2}
and if $S$ and $S'$ are different seedlings,
then the bonsais obtained by adding edges to $S$ and $S'$
as given in Definition \ref{Def:vadiff2} are different,
the space of determinanted bonsais is the direct sum of
the ${\mathbf C}^{S,i}$.
We call this complex $\{{\mathbf C}^{S,i}\}$
a {\em thread} starting from $S$.
\end{df}
Then, since $\partial({\mathbf C}^{S,i}) \subset {\mathbf
C}^{S,i+1}$, the cohomology groups of determinanted bonsais by the
differential $\partial$ are the direct sum of cohomology groups of
the threads $\{{\mathbf C}^{S,i}\}$.

Now let us show through an example that,
for any thread ${\{{\mathbf C}^{S,i}\}}$,
we can get a cochain complex which is isomorphic to it,
obtained from the direct sums of tensor products of
$\{{\mathbf D}^{i}\}$'s  and $\{{\mathbf B}^{i}\}$'s
as in the proof of acyclicity of branch-fixed differential.
In 5-bonsai, the seedling $S$ in Figure \ref{Fig:117}
can have twiggy edges at the positions of the twigs shown in the picture,
and those twigs are grouped as surrounded by squares.

\begin{figure}[h]

\begin{picture}(200,120)(-100,-110)

\put(0,0){\circle*{2}}
\put(0,0){\line(0,-1){40}}

\put(-25,0){${\mathbf D}_1^{*}$}
\put(-25,0){\line(1,0){20}}
\put(-25,0){\line(0,-1){25}}
\put(-5,0){\line(0,-1){25}}
\put(-25,-25){\line(1,0){20}}
\put(0,0){\line(-1,-1){20}}
\put(0,0){\line(-1,-2){10}}

\put(5,0){${\mathbf D}_7^{*}$}
\put(0,0){\line(1,-2){10}}
\put(0,0){\line(1,-1){20}}
\put(0,0){\line(3,-2){30}}
\put(5,0){\line(1,0){30}}
\put(5,0){\line(0,-1){25}}
\put(35,0){\line(0,-1){25}}
\put(5,-25){\line(1,0){30}}

\put(0,-35){$e_1$}
\put(0,-40){\circle*{2}}
\put(0,-40){\line(-1,-1){40}}
\put(0,-40){\line(1,-1){40}}

\put(-35,-40){${\mathbf D}_2^{*}$}
\put(0,-40){\line(-3,-2){30}}
\put(0,-40){\line(-2,-1){30}}
\put(-35,-40){\line(1,0){10}}
\put(-35,-40){\line(0,-1){25}}
\put(-25,-40){\line(0,-1){25}}
\put(-35,-65){\line(1,0){10}}

\put(20,-40){${\mathbf D}_6^{*}$}
\put(0,-40){\line(3,-2){30}}
\put(0,-40){\line(2,-1){30}}
\put(25,-40){\line(1,0){10}}
\put(35,-40){\line(0,-1){25}}
\put(25,-40){\line(0,-1){25}}
\put(25,-65){\line(1,0){10}}

\put(-5,-75){${\mathbf D}_4^{*}$}
\put(0,-40){\line(-1,-2){10}}
\put(0,-40){\line(1,-2){10}}
\put(-15,-50){\line(1,0){30}}
\put(-15,-50){\line(0,-1){15}}
\put(15,-50){\line(0,-1){15}}
\put(-15,-65){\line(1,0){30}}

\put(-35,-75){$e_2$}
\put(-40,-80){\circle*{2}}
\put(-40,-80){\line(0,-1){30}}
\put(-40,-80){\line(-1,-1){20}}
\put(-40,-80){\line(-1,-2){10}}
\put(-40,-80){\line(1,-2){10}}
\put(-40,-80){\line(1,-1){20}}
\put(-40,-110){\circle*{2}}

\put(-65,-80){${\mathbf B}_3^{*}$}
\put(-65,-80){\line(1,0){50}}
\put(-65,-80){\line(0,-1){35}}
\put(-15,-80){\line(0,-1){35}}
\put(-65,-115){\line(1,0){50}}

\put(25,-70){$e_3$}
\put(40,-80){\circle*{2}}
\put(40,-80){\line(0,-1){30}}
\put(40,-80){\line(-1,-2){10}}
\put(40,-80){\line(1,-2){10}}
\put(40,-80){\line(1,-1){20}}
\put(40,-80){\line(3,-2){30}}
\put(40,-110){\circle*{2}}

\put(25,-80){${\mathbf B}_5^{*}$}
\put(25,-80){\line(1,0){50}}
\put(25,-80){\line(0,-1){35}}
\put(75,-80){\line(0,-1){35}}
\put(25,-115){\line(1,0){50}}

\end{picture}

\caption{}                 \label{Fig:117}

\end{figure}
Note that, in Figure \ref{Fig:117}, adding edges to the corolla in
each square is the same as attaching edges to the vertex at which
the square is appended, each corolla corresponds to the module that
is written on each square (In Figure \ref{Fig:117}, ${\mathbf
D}_i^{m,*}$ is isomorphic to ${\mathbf D}^{m,*}$ and
 ${\mathbf B}_i^{m,*}$ is isomorphic to ${\mathbf B}^{m,*}$).

Then, as in Section 11, we can get the cochain isomorphism of
$\{{\mathbf C}^{S,i}\}$ and
${\mathbf D}_1^* \otimes {\mathbf D}_2^* \otimes {\mathbf B}_3^* \otimes
 {\mathbf D}_4^* \otimes {\mathbf B}_5^* \otimes {\mathbf D}_6^* \otimes
 {\mathbf D}_7^*$.
and use K{\" u}nneth's theorem. Since the cohomology of $\{{\mathbf
D}^*\}$ is acyclic with trivial base degree cohomology, it is clear
that any $\{{\mathbf C}^{S,*}\}$ whose tensor product representation
has $\{{\mathbf D}^*\}$ is acyclic with trivial base degree
cohomology. The only seedlings having no $\{{\mathbf D}^*\}$ are the
first two bonsais of Figure \ref{Fig:113}, i.e., the one-vertex
bonsai $v$ and the one-edge bonsai $e$. Since $v$ is a cocycle, we
have $H^0 = k$ and since ${\mathbf C}^{e,*} = {\mathbf B}^*$, we
have $H^1 = H^0({\mathbf B}^*) =  k$. This illustrates a general
argument. Thus we have

\begin{thm}
The cohomology groups $H^i$ of the planar clear-edged $m$-bonsai
Hopf algebra by vertex-appending differential are,
\begin{eqnarray}
H^i= \left\{ \begin{array}{ll}
k & \mbox{if } i=0 \mbox{ or } 1 \\
0 & \mbox{otherwise.}
\end{array} \right. \notag
\end{eqnarray}
\end{thm}

\subsection{Cohomology Groups of $\partial$ when $m < \infty$: Terminology}
\label{subsec:cohomology}

In this subsection, let us consider the cohomology groups in case
$m < \infty$.
It is not as easy as in the previous subsection
to get a simple tensor product representation of a thread
$\{{\mathbf C}^{S,i}\}$ when each vertex has the upper bound $m$ of
ramification number.
So we need to change our strategy for the case of $m< \infty$.
Since every $\{{\mathbf C}^{S,i}\}$ is finite-dimensional,
we can calculate the cohomology groups
by considering a finite number of bonsais.
So from now on, we develop  an ``inductive strategy'' for calculating
the cohomology groups of the thread $\{{\mathbf C}^{S,i}\}$
for any given $S$.

Let us illustrate the basic idea of the `` inductive strategy''
using an example. In 2-bonsai, let $S_i$ be the ladder of length
$i$. Then the cohomology $H^j(S_1)$ of $\{{\mathbf C}^{S_1,j}\}$ is,
by a sequence as in Figure \ref{Fig:118}, $H^1(S_1)=H^2(S_2)=k$, the
base field.

\begin{figure}[h]

\begin{picture}(100,30)(0,0)

\put(0,10){$0 \overset{}{\rightarrow}$}
\put(20,20){\line(0,-1){20}}
\put(20,20){\circle*{3}}
\put(20,0){\circle*{3}}
\put(25,10){$\overset{0}{\rightarrow}$}
\put(50,20){\line(-1,-2){10}}
\put(50,20){\line(1,-2){10}}
\put(50,20){\circle*{3}}
\put(40,0){\circle*{3}}
\put(60,0){\circle*{3}}
\put(60,10){$\overset{0}{\rightarrow} 0$}

\end{picture}

\caption{}                 \label{Fig:118}

\end{figure}

Let us find an inductive step to get
$H^i(S_{j+1})$'s from $H^i(S_j)$'s,
so that we can get the cohomology group of every $\{{\mathbf C}^{S_i,j}\}$.

As in Figure \ref{Fig:119},
when $T$ is a linear combination of $m$-bonsais,
the other two expressions of Figure \ref{Fig:119}
represent linear combinations
of bonsais obtained by attaching a bonsai
to the roots of bonsais which are
the components of $T$.

\begin{figure}[h]

\begin{picture}(100,80)(0,-80)

\put(0,-15){$T=$}
\put(25,-5){\circle*{3}}
\put(25,-5){\line(0,-1){20}}
\put(25,-25){\circle*{3}}
\put(30,-15){$+ \quad 2$}
\put(65,-5){\circle*{3}}
\put(65,-5){\line(-1,-2){10}}
\put(65,-5){\line(1,-2){10}}
\put(55,-25){\circle*{3}}
\put(75,-25){\circle*{3}}

\put(-100,-60){\circle{15}}
\put(-103,-63){$T$}
\put(-100,-52){\line(0,1){20}}
\put(-100,-32){\circle*{3}}
\put(-88,-55){$=$}
\put(-75,-35){\circle*{3}}
\put(-75,-35){\line(0,-1){20}}
\put(-75,-55){\circle*{3}}
\put(-75,-55){\line(0,-1){20}}
\put(-75,-75){\circle*{3}}
\put(-70,-55){$+ \quad 2$}
\put(-35,-35){\circle*{3}}
\put(-35,-35){\line(0,-1){20}}
\put(-35,-55){\circle*{3}}
\put(-35,-55){\line(-1,-2){10}}
\put(-35,-55){\line(1,-2){10}}
\put(-45,-75){\circle*{3}}
\put(-25,-75){\circle*{3}}

\put(20,-60){\circle{15}}
\put(17,-63){$T$}
\put(20,-52){\line(0,1){20}}
\put(20,-32){\circle*{3}}
\put(20,-32){\line(-1,-1){15}}
\put(5,-47){\circle*{3}}
\put(32,-55){$=$}
\put(65,-35){\circle*{3}}
\put(65,-35){\line(0,-1){20}}
\put(65,-35){\line(-1,-1){15}}
\put(50,-50){\circle*{3}}
\put(65,-55){\circle*{3}}
\put(65,-55){\line(0,-1){20}}
\put(65,-75){\circle*{3}}
\put(70,-55){$+ \quad 2$}
\put(115,-35){\circle*{3}}
\put(115,-35){\line(0,-1){20}}
\put(115,-35){\line(-1,-1){15}}
\put(100,-50){\circle*{3}}
\put(115,-55){\circle*{3}}
\put(115,-55){\line(-1,-2){10}}
\put(115,-55){\line(1,-2){10}}
\put(105,-75){\circle*{3}}
\put(125,-75){\circle*{3}}

\end{picture}

\caption{}                 \label{Fig:119}

\end{figure}

Then, when $T$ is a linear combination of bonsais in
$\{{\mathbf C}^{S_i,j}\}$ in 2-bonsai, the map $\partial$ on
$\{{\mathbf C}^{S_{i+1},j}\}$ is expressed in Figure \ref{Fig:120}.

\begin{figure}[h]

\begin{picture}(100,80)(0,-110)

\put(0,-60){\circle{15}}
\put(-3,-63){$T$}
\put(0,-52){\line(0,1){20}}
\put(0,-32){\circle*{3}}
\put(12,-55){$\to$}
\put(40,-60){\circle{15}}
\put(37,-63){$T$}
\put(40,-52){\line(0,1){20}}
\put(40,-32){\circle*{3}}
\put(40,-32){\line(-1,-1){15}}
\put(25,-47){\circle*{3}}
\put(52,-55){$+ \quad (-1)^{degT+1}$}
\put(130,-60){\circle{15}}
\put(127,-63){$T$}
\put(130,-52){\line(0,1){20}}
\put(130,-32){\circle*{3}}
\put(130,-32){\line(1,-1){15}}
\put(145,-47){\circle*{3}}
\put(150,-55){$-$}
\put(170,-60){\circle{15}}
\put(165,-63){$\partial T$}
\put(170,-52){\line(0,1){20}}
\put(170,-32){\circle*{3}}

\put(0,-110){\circle{15}}
\put(0,-80){\line(-1,-1){15}}
\put(-15,-95){\circle*{3}}
\put(-3,-113){$T$}
\put(0,-102){\line(0,1){20}}
\put(0,-82){\circle*{3}}
\put(12,-105){$\to$}
\put(40,-110){\circle{15}}
\put(40,-82){\line(-1,-1){15}}
\put(25,-97){\circle*{3}}
\put(35,-113){$\partial T$}
\put(40,-102){\line(0,1){20}}
\put(40,-82){\circle*{3}}

\put(100,-110){\circle{15}}
\put(100,-80){\line(1,-1){15}}
\put(115,-95){\circle*{3}}
\put(97,-113){$T$}
\put(100,-102){\line(0,1){20}}
\put(100,-82){\circle*{3}}
\put(122,-105){$\to -$}
\put(155,-110){\circle{15}}
\put(155,-80){\line(1,-1){15}}
\put(170,-95){\circle*{3}}
\put(150,-113){$\partial T$}
\put(155,-102){\line(0,1){20}}
\put(155,-82){\circle*{3}}

\end{picture}

\caption{}                 \label{Fig:120}

\end{figure}

From Figure \ref{Fig:120}, we can find that the kernel of $\partial$
on $\{{\mathbf C}^{S_{i+1},j}\}$ is generated by the linear combinations
of bonsais shown in Figure \ref{Fig:121}
in which $T \in ker \partial$ and $\partial T' = \partial T'' = T$,
and that the image of $\partial$ on $\{{\mathcal C}^{S^{i+1},j}\}$
is generated by the linear combinations of bonsais shown
in Figure \ref{Fig:122}.

\begin{figure}[h]

\begin{picture}(100,80)(60,-110)

\put(40,-60){\circle{15}}
\put(37,-63){$T'$}
\put(40,-52){\line(0,1){20}}
\put(40,-32){\circle*{3}}
\put(40,-32){\line(-1,-1){15}}
\put(25,-47){\circle*{3}}
\put(52,-55){$+ \quad (-1)^{degT''+1}$}
\put(130,-60){\circle{15}}
\put(127,-63){$T''$}
\put(130,-52){\line(0,1){20}}
\put(130,-32){\circle*{3}}
\put(130,-32){\line(1,-1){15}}
\put(145,-47){\circle*{3}}
\put(150,-55){$-$}
\put(170,-60){\circle{15}}
\put(165,-63){$T$}
\put(170,-52){\line(0,1){20}}
\put(170,-32){\circle*{3}}
\put(190,-55){= A}

\put(40,-110){\circle{15}}
\put(40,-82){\line(-1,-1){15}}
\put(25,-97){\circle*{3}}
\put(35,-113){$T$}
\put(40,-102){\line(0,1){20}}
\put(40,-82){\circle*{3}}
\put(70,-105){= B}

\put(128,-105){$-$}
\put(155,-110){\circle{15}}
\put(155,-80){\line(1,-1){15}}
\put(170,-95){\circle*{3}}
\put(150,-113){$T$}
\put(155,-102){\line(0,1){20}}
\put(155,-82){\circle*{3}}
\put(185,-105){= C}

\end{picture}

\caption{}                 \label{Fig:121}

\end{figure}

\begin{figure}[h]

\begin{picture}(100,80)(60,-110)

\put(40,-60){\circle{15}}
\put(37,-63){$T'$}
\put(40,-52){\line(0,1){20}}
\put(40,-32){\circle*{3}}
\put(40,-32){\line(-1,-1){15}}
\put(25,-47){\circle*{3}}
\put(52,-55){$+ \quad (-1)^{degT'+1}$}
\put(130,-60){\circle{15}}
\put(127,-63){$T'$}
\put(130,-52){\line(0,1){20}}
\put(130,-32){\circle*{3}}
\put(130,-32){\line(1,-1){15}}
\put(145,-47){\circle*{3}}
\put(150,-55){$-$}
\put(170,-60){\circle{15}}
\put(165,-63){$\partial T'$}
\put(170,-52){\line(0,1){20}}
\put(170,-32){\circle*{3}}
\put(190,-55){= a}

\put(40,-110){\circle{15}}
\put(40,-82){\line(-1,-1){15}}
\put(25,-97){\circle*{3}}
\put(35,-113){$\partial T$}
\put(40,-102){\line(0,1){20}}
\put(40,-82){\circle*{3}}
\put(70,-105){= b}

\put(128,-105){$-$}
\put(155,-110){\circle{15}}
\put(155,-80){\line(1,-1){15}}
\put(170,-95){\circle*{3}}
\put(150,-113){$\partial T$}
\put(155,-102){\line(0,1){20}}
\put(155,-82){\circle*{3}}
\put(185,-105){= c}

\end{picture}

\caption{}                 \label{Fig:122}

\end{figure}

So we can write

\begin{eqnarray}
\frac{ker \partial}{im \partial}=
\frac{\langle A,B,C \rangle}{\langle a,b,c \rangle}.
\end{eqnarray}

In A, we have $\partial T = 0$, $\partial T' = T$ and $\partial T''
= T$, so we have A $-$ a = $(-1)^{deg T'' +1}$C. Hence we have C is
generated by A and a. Also, in c, when $T' =\partial T$, we have
$a-b=(-1)^{degT} c$. So c is generated by a and b. So we have

\begin{eqnarray}
\frac{ker \partial}{im \partial}=
\frac{\langle A,B \rangle}{\langle a,b \rangle}.
\end{eqnarray}

In Figure \ref{Fig:121}, $\partial T'' = T$. So A in Figure
\ref{Fig:121} can be redrawn as Figure \ref{Fig:123}.

\begin{figure}[h]

\begin{picture}(100,100)(60,-110)

\put(40,-60){\circle{15}}
\put(37,-63){$T'$}
\put(40,-52){\line(0,1){20}}
\put(40,-32){\circle*{3}}
\put(40,-32){\line(-1,-1){15}}
\put(25,-47){\circle*{3}}
\put(52,-55){$+ \quad (-1)^{degT''+1}$}
\put(130,-60){\circle{15}}
\put(127,-63){$T''$}
\put(130,-52){\line(0,1){20}}
\put(130,-32){\circle*{3}}
\put(130,-32){\line(1,-1){15}}
\put(145,-47){\circle*{3}}
\put(150,-55){$-$}
\put(170,-60){\circle{15}}
\put(165,-63){$\partial T''$}
\put(170,-52){\line(0,1){20}}
\put(170,-32){\circle*{3}}

\put(-40,-105){=}
\put(-10,-115){\circle{25}}
\put(-25,-118){$T'-T''$}
\put(-10,-102){\line(0,1){20}}
\put(-10,-82){\circle*{3}}
\put(-10,-82){\line(-1,-1){15}}
\put(-25,-97){\circle*{3}}
\put(12,-105){$+$}
\put(40,-110){\circle{15}}
\put(37,-113){$T''$}
\put(40,-102){\line(0,1){20}}
\put(40,-82){\circle*{3}}
\put(40,-82){\line(-1,-1){15}}
\put(25,-97){\circle*{3}}
\put(52,-105){$+ \quad (-1)^{degT''+1}$}
\put(130,-110){\circle{15}}
\put(127,-113){$T''$}
\put(130,-102){\line(0,1){20}}
\put(130,-82){\circle*{3}}
\put(130,-82){\line(1,-1){15}}
\put(145,-97){\circle*{3}}
\put(150,-105){$-$}
\put(170,-110){\circle{15}}
\put(165,-113){$\partial T''$}
\put(170,-102){\line(0,1){20}}
\put(170,-82){\circle*{3}}

\end{picture}

\caption{}                 \label{Fig:123}

\end{figure}

Since $\partial T' = \partial T'' =T$, we have $\partial (T'-T'') =0$
and so A can be rewritten as B+a. Hence we have

\begin{eqnarray}
\frac{ker \partial}{im \partial}=
\frac{\langle B+a,B \rangle}{\langle a,b \rangle}
=\frac{\langle a,B \rangle}{\langle a,b \rangle}=
\frac{\langle a \rangle \oplus \langle B \rangle}
     {\langle a \rangle \oplus \langle b \rangle}
=\frac{\langle B \rangle}{\langle b \rangle}.
\end{eqnarray}

Obviously, $\frac{\langle B \rangle}{\langle b \rangle}$
is isomorphic to the cohomology group of ${\mathbf C}^{S_i,j}$.
Since B and b have two more edges than bonsais in ${\mathbf C}^{S_i,j}$,
we can see

\begin{thm}
When $H^j(S_i)$ is the $j$-th cohomology group of the thread
${\mathbf C}^{S_i,j}$, $H^{j+2}(S_{i+1})=H^j(S_i)$.
\end{thm}

As in the previous theorem, we can generalize the process of getting
the cohomology of the thread starting from $S'$ which is obtained by
attaching the root of a seedling $S$ to a tip of a corolla, when the
cohomology of the thread starting from $S$ is already known. Let us
define some new terminology. We first define a new kind of
``seedling''.

\begin{df}
We define a {\em grafting seedling} as a bonsai defined as one of
the following;

1) a seedling

2) a bonsai obtained by attaching even-arity corollas to vertices of
a seedling $S$ which are more than two edges from any tip, and by
replacing branch-end edges of $S$ to even-arity corollas.

In Figure \ref{Fig:123'}, the first three bonsais are grafting
seedlings and the last is not.
\end{df}

\begin{figure}[h]

\begin{picture}(200,50)(0,-40)

\put(0,0){\circle*{2}} \put(0,0){\line(-1,-2){10}}
\put(-10,-20){\circle*{2}}

\put(0,0){\line(-1,-4){5}} \put(-5,-20){\circle*{2}}

\put(0,0){\line(1,-4){5}} \put(5,-20){\circle*{2}}

\put(-10,-20){\line(0,-1){20}} \put(-10,-40){\circle*{2}}

\put(0,0){\line(1,-2){10}} \put(10,-20){\circle*{2}}

\put(10,-20){\line(0,-1){20}}

\put(10,-40){\circle*{2}}

\put(50,0){\circle*{2}} \put(50,0){\line(0,-1){20}}
\put(50,-20){\circle*{2}}

\put(50,0){\line(1,-4){5}} \put(55,-20){\circle*{2}}

\put(50,0){\line(1,-2){10}} \put(60,-20){\circle*{2}}

\put(50,-20){\line(0,-1){20}} \put(50,-40){\circle*{2}}

\put(100,0){\circle*{2}} \put(100,0){\line(-1,-2){10}}
\put(90,-20){\circle*{2}}

\put(90,-20){\line(1,-4){5}} \put(95,-40){\circle*{2}}

\put(90,-20){\line(-1,-4){5}} \put(85,-40){\circle*{2}}

\put(100,0){\line(1,-2){10}} \put(110,-20){\circle*{2}}

\put(110,-20){\line(0,-1){20}}

\put(110,-40){\circle*{2}}

\put(180,0){\circle*{2}} \put(180,0){\line(-1,-2){10}}
\put(170,-20){\circle*{2}}

\put(180,0){\line(0,-1){20}} \put(180,-20){\circle*{2}}

\put(170,-20){\line(0,-1){20}} \put(170,-40){\circle*{2}}

\put(180,0){\line(1,-2){10}} \put(190,-20){\circle*{2}}

\put(190,-20){\line(0,-1){20}}

\put(190,-40){\circle*{2}}

\end{picture}

\caption{}                 \label{Fig:123'}

\end{figure}

\begin{df} \label{def:gs}
A grafting seedling $gs(n;T_1,T_2,...T_{n+1};S_1,S_2,...,S_n)$,
which we call a {\em grafting seedling} is constructed like the
following; In the corolla $C$ with arity $n$, the corollae $T_1$,
$T_2$, ... , $T_{n+1}$ of {\em even} arities are attached so that
the root of $T_1$ is attached to the root of $C$ on the left of the
leftmost edge of $C$, the root of $T_2$ is attached to the root of
$C$ between the first leftmost edge and the second leftmost edge of
$C$,..., and so on, and the grafting seedlings $S_1$, $S_2$, ...
,$S_n$ are attached so that the root of $S_1$ is attached to the tip
of the first leftmost edge of $C$, the root of $S_2$ is attached to
the tip of the second leftmost edge of $C$, ... , and so on, as in
Figure \ref{Fig:124} in 6-bonsai.

\begin{figure}[h]

\begin{picture}(100,160)(60,-160)

\put(0,-15){$gs(3;T_1,T_2,T_3,T_4;S_1,S_2,S_3)=$}

\put(40,-30){\line(-1,-2){10}}
\put(30,-50){\circle*{2}}
\put(40,-30){\line(-1,-4){5}}
\put(35,-50){\circle*{2}}

\put(40,-30){\line(1,-2){10}}
\put(50,-50){\circle*{2}}
\put(40,-30){\line(1,-4){5}}
\put(45,-50){\circle*{2}}

\put(40,-30){\line(0,-1){20}}
\put(40,-50){\circle*{3}}
\put(40,-30){\circle*{3}}
\put(40,-30){\line(-1,-1){20}}
\put(20,-50){\circle*{3}}
\put(40,-30){\line(1,-1){20}}
\put(60,-50){\circle*{3}}

\put(20,-50){\line(0,-1){30}}
\put(20,-65){\circle*{3}}
\put(20,-80){\circle*{3}}

\put(40,-50){\line(0,-1){15}}
\put(40,-65){\circle*{3}}

\put(60,-50){\line(0,-1){15}}
\put(60,-65){\circle*{3}}

\put(0,-100){$T_1=$}
\put(30,-96){\circle*{3}}
\put(40,-100){$T_2=$}
\put(75,-90){\circle*{3}}
\put(75,-90){\line(-1,-2){10}}
\put(65,-110){\circle*{3}}
\put(75,-90){\line(1,-2){10}}
\put(85,-110){\circle*{3}}
\put(100,-100){$T_3=$}

\put(135,-90){\circle*{3}}
\put(135,-90){\line(-1,-2){10}}
\put(125,-110){\circle*{3}}
\put(135,-90){\line(1,-2){10}}
\put(145,-110){\circle*{3}}

\put(150,-100){$T_4=$}
\put(180,-95){\circle*{3}}

\put(0,-130){$S_1=$}
\put(30,-120){\circle*{3}}
\put(30,-120){\line(0,-1){30}}
\put(30,-135){\circle*{3}}
\put(30,-150){\circle*{3}}
\put(55,-130){$S_2=$}
\put(85,-120){\circle*{3}}
\put(85,-120){\line(0,-1){20}}
\put(85,-140){\circle*{3}}
\put(105,-130){$S_3=$}
\put(135,-120){\circle*{3}}
\put(135,-120){\line(0,-1){20}}
\put(135,-140){\circle*{3}}

\end{picture}

\caption{}                 \label{Fig:124}

\end{figure}

\end{df}

Note that the grafting seedling defined here is different from
the seedling we used until now.

\begin{df}
We define the relation $T_1 \to T_2$ of clear-edged
$m$-bonsais $T_1$ and $T_2$ as follows;

i) $T_2$ is a nonzero component of $\partial T_1$ and $T_2$ is
obtained by attaching one edge to the {\em root} of $T_1$

or

ii) $T_2$ is obtained by attaching an edge to a non-root vertex of $T_1$.

In 3-bonsai, we have examples as in Figure \ref{Fig:125}.
\end{df}

\begin{figure}[h]

\begin{picture}(100,50)(90,-50)

\put(0,0){\circle*{3}}
\put(0,0){\line(0,-1){20}}
\put(0,-20){\circle*{3}}
\put(0,-20){\line(-1,-2){10}}
\put(0,-20){\line(1,-2){10}}
\put(-10,-40){\circle*{3}}
\put(10,-40){\circle*{3}}

\put(10,-20){$\to$}

\put(40,0){\circle*{3}}
\put(40,0){\line(0,-1){20}}
\put(40,0){\line(-1,-1){15}}
\put(25,-15){\circle*{3}}
\put(40,-20){\circle*{3}}
\put(40,-20){\line(-1,-2){10}}
\put(40,-20){\line(1,-2){10}}
\put(30,-40){\circle*{3}}
\put(50,-40){\circle*{3}}

\put(80,0){\circle*{3}}
\put(80,0){\line(0,-1){20}}
\put(80,-20){\circle*{3}}
\put(80,-20){\line(-1,-2){10}}
\put(80,-20){\line(1,-2){10}}
\put(70,-40){\circle*{3}}
\put(90,-40){\circle*{3}}

\put(90,-20){$\to$}

\put(110,0){\circle*{3}}
\put(110,0){\line(0,-1){20}}
\put(110,-20){\circle*{3}}
\put(110,-20){\line(-1,-2){10}}
\put(110,-20){\line(0,-1){20}}
\put(110,-20){\line(1,-2){10}}
\put(100,-40){\circle*{3}}
\put(110,-40){\circle*{3}}
\put(120,-40){\circle*{3}}

\put(150,0){\circle*{3}}
\put(150,0){\line(0,-1){20}}
\put(150,-20){\circle*{3}}
\put(150,-20){\line(0,-1){20}}
\put(150,-40){\circle*{3}}

\put(160,-20){$\to$}

\put(190,0){\circle*{3}}
\put(190,0){\line(0,-1){20}}
\put(190,-20){\circle*{3}}
\put(190,-20){\line(-1,-2){10}}
\put(190,-20){\line(1,-2){10}}
\put(180,-40){\circle*{3}}
\put(200,-40){\circle*{3}}

\put(240,0){\circle*{3}}
\put(240,0){\line(0,-1){20}}
\put(240,0){\line(-1,-1){15}}
\put(240,-20){\circle*{3}}
\put(225,-15){\circle*{3}}
\put(240,-20){\line(-1,-2){10}}
\put(240,-20){\line(1,-2){10}}
\put(230,-40){\circle*{3}}
\put(250,-40){\circle*{3}}

\put(250,-20){$\longrightarrow$}
\put(255,-20){/ }

\put(285,0){\circle*{3}}
\put(285,0){\line(0,-1){20}}
\put(285,0){\line(-1,-1){15}}
\put(285,0){\line(-1,-2){10}}
\put(285,-20){\circle*{3}}
\put(270,-15){\circle*{3}}
\put(275,-20){\circle*{3}}
\put(285,-20){\line(-1,-2){10}}
\put(285,-20){\line(1,-2){10}}
\put(275,-40){\circle*{3}}
\put(295,-40){\circle*{3}}

\end{picture}

\caption{}                 \label{Fig:125}

\end{figure}

\begin{df} \label{def:K}
We define the relation $T_1 \Rightarrow T_1'$
if there is a sequence of $m$-bonsais such that
$T_1 \to T_2 \to ... \to T_n = T_1'$ or $T_1 = T_1'$.

Let ${\mathbf K}(gs(n;T_1,...,T_{n+1};S_1,...,S_n))$ be the vector
space generated by the bonsais $T'$ such that
$gs(n;T_1,...,T_{n+1};S_1,...,S_n) \Rightarrow T'$. Also, when $S$
is a grafting seedling and ${\mathbf C}$ is the cochain complex of
$m$-bonsai, we define ${\mathbf K}^i(S) := {\mathbf C}^i \cap
{\mathbf K}(S)$.
\end{df}

\begin{thm}
When $S$ is a grafting seedling and $H^i(S)$ is the $i$-th
cohomology group of the thread $\{{\mathbf K}^i(S)\}$, the $i$-th
cohomology group $H^i$ of $m$-bonsai is $H^i = \underset{\mbox{$S$
is a grafting seedling}}{\bigoplus} H^i(S)$.

\end{thm}

\begin{proof}
We have
${\mathbf C} = \underset{\mbox{$S$ is a grafting seedling}}{\bigoplus}
{\mathbf K}(S)$
since i)there is no $T$ such that $T \Rightarrow T'$, $T \neq T'$ and
$T'$ is a grafting seedling and
ii)if two grafting seedlings $S$ and $S'$ are not equal,
then ${\mathbf K}(S) \cap {\mathbf K}(S')= \emptyset$.
Also we have
$\partial {\mathbf K}^i(S) \subset \partial {\mathbf K}^{i+1} (S)$.
So we get the wanted result.
\end{proof}

\subsection{ The Cohomology Groups for each
$\{ {\mathbf K}^i(gs(n;T_1,...,T_{n+1};S_1,...,S_n)) \}$}

Now we have to calculate the cohomology groups for $\{ {\mathbf
K}^i(gs(n;T_1,...,T_{n+1};S_1,...,S_n)) \}$. First let us define
some notation.

\begin{df}

For any integer $n \geq 0$, the cochain complex
$\{ {\mathbf D}_{2n}^j \}$ is defined as follows;
when $j=2n$ or $2n+1$, ${\mathbf D}_{2n}^j$ is a one-dimensional
vector space with the basis $\{C_j\}$, where $C_j$ is the corolla of arity $j$
(when $j=0$, $C_j$ is the one-vertex bonsai),
and otherwise, ${\mathbf D}_{2n}^j = 0$.
And when $j=2n$, the boundary map
$\partial:{\mathbf D}_{2n}^j \to{\mathbf D}_{2n}^{j+1}$ is given by
$C_j \mapsto C_{j+1}$ and otherwise, $\partial =0$.

\end{df}

To calculate the cohomology groups of $\{ {\mathbf
K}^i(gs(n;T_1,...,T_{n+1};S_1,...,S_n)) \}$, we use a similar type
of tensor product representation as in Section 11.

\begin{df}
Let $B(n;U_1,...,U_{n+1};V_1,...,V_n)$ be the bonsai obtained by
replacing $T_i$ by $U_i$ and $S_i$ by $V_i$ in
$gs(n;T_1,...,T_{n+1};S_1,...,S_n)$, where $U_i$ is the corolla
$C_k$ of arity $k=degT_i$ or $degT_i+1$ (i.e., $U_i$ is a basis
element of $\{{\mathbf D}_{degT_i}\}$) and $V_i$ is a bonsai in the
thread $\{{\mathbf C}^{S_i,i}\}$ starting from $S_i$ (see Subsection
\ref{subsec:cohomology}).
\end{df}

Let us define the cochain complexes $\{{\mathbf D}_i\}$ and
$\{{\mathbf E}_i\}$ as $\{{\mathbf D}_{degT_i}\}$ and $\{{\mathbf
C}(S_i)\}$. Then we can define an isomophism $P$ from ${\mathbf
K}^i(S)$ to ${\mathbf L}^i ={\mathbf D}_1^{i_1} \otimes {\mathbf
E}_1^{j_1} \otimes ... \otimes {\mathbf D}_n^{i_n} \otimes {\mathbf
E}_n^{j_n} \otimes {\mathbf D}_{n+1}^{i_{n+1}}$ where
$i_1+...+i_{n+1}+j_1+...+j_n+n =i$ and $i_1+...+i_{n+1}+n \leq m$,
by sending $B(n;U_1,...U_{n+1};V_1,...,V_n)$ to $U_1 \otimes V_1
\otimes ... \otimes U_n \otimes V_n \otimes U_{n+1}$. the
differential $d$ in ${\mathbf L}$ is defined as, when $degU_1
+...+degU_{n+1} < m-n$,
\begin{eqnarray}
 &d(U_1 \otimes V_1 \otimes U_2 \otimes V_2 \otimes ...
  \otimes U_n \otimes V_n \otimes U_{n+1}) \notag \\
=&\partial U_1 \otimes V_1 \otimes
  U_2 \otimes V_2 \otimes ...
  \otimes U_n \otimes V_n \otimes U_{n+1} \notag \\
+&U_1 \otimes (-1)^{p_1}\partial V_1 \otimes
  U_2 \otimes V_2 \otimes ...
  \otimes U_n \otimes V_n \otimes U_{n+1}) \notag \\
+&U_1 \otimes V_1 \otimes (-1)^{q_1}
  \partial U_2 \otimes V_2 \otimes ...
  \otimes U_n \otimes V_n \otimes U_{n+1}) \notag \\
+&U_1 \otimes V_1 \otimes
  U_2 \otimes (-1)^{p_2}\partial V_2 \otimes ...
  \otimes U_n \otimes V_n \otimes U_{n+1}) \notag \\
 &+...+ \notag \notag \\
+&U_1 \otimes V_1 \otimes
  U_2 \otimes V_2 \otimes...
  \otimes U_n \otimes V_n \otimes
  (-1)^{q_n} \partial U_{n+1}), \notag
\end{eqnarray}
where $p_i=degU_1+degV_1+...+degU_{i-1}+degV_{i-1}+degU_i+i$ and
$q_i=degU_1+degV_1+...+degU_i+degV_i+i$, and if
$degU_1+...+degU_{n+1} = m-n$,
\begin{eqnarray}
 &d(U_1 \otimes V_1 \otimes U_2 \otimes V_2 \otimes ...
  \otimes U_n \otimes V_n \otimes U_{n+1}) \notag \\
=&\partial U_1 \otimes V_1 \otimes
  U_2 \otimes V_2 \otimes ...
  \otimes U_n \otimes V_n \otimes U_{n+1} \notag \\
+&U_1 \otimes V_1 \otimes (-1)^{q_1}
  \partial U_2 \otimes V_2 \otimes ...
  \otimes U_n \otimes V_n \otimes U_{n+1}) \notag \\
 &+...+ \notag \notag \\
+&U_1 \otimes V_1 \otimes
  U_2 \otimes V_2 \otimes...
  \otimes U_n \otimes V_n \otimes
  (-1)^{q_n} \partial U_{n+1}). \notag
\end{eqnarray}

Then by the definition of $\partial$ in ${\mathbf C}$, the
isomorphism $P$ is a cochain complex isomorphism from $({\mathbf
K}^i(S),\partial)$ to $({\mathbf L}^i ,d)$.

Now let us define a double complex $\{{\mathbf M}^i \otimes {\mathbf
N}^j\}^{i,j}$ where ${\mathbf M}^i = \sum {\mathbf D}_1^{i_1}
\otimes ...\otimes {\mathbf D}_{n+1}^{i_{n+1}}$ where
$i_1,...,i_{n+1}$ satisfy $i_1+...+i_{n+1}+n=i$ and $i \leq m$, and
${\mathbf N}^j = \sum_{j_1+...+j_n=j} {\mathbf E}_1^{j_1} \otimes
...\otimes {\mathbf E}_n^{j_n}$ where $j_1,...,j_n$ satisfy
$j_1+...+j_n=j$, and differentials $\bar{\partial _1}$ and
$\bar{\partial _2}$ are

\begin{eqnarray}
 &\bar{\partial _1}(U_1 \otimes U_2 \otimes ... \otimes U_{n+1}
                    \otimes V_1 \otimes ...\otimes V_n) \notag \\
=&\partial U_1 \otimes U_2 \otimes ... \otimes U_{n+1}
  \otimes V_1 \otimes ...\otimes V_n \notag \\
+&U_1 \otimes (-1)^{q_1} \partial U_2 \otimes ...
  \otimes U_{n+1} \otimes V_1 \otimes ...\otimes V_n \notag \\
 &+...+ \notag \notag \\
+&U_1 \otimes U_2 \otimes ... \otimes
  (-1)^{q_n} \partial U_{n+1}
  \otimes V_1 \otimes ...\otimes V_n \notag
\end{eqnarray}

where $U_1 \otimes ... \otimes U_{n+1} \in {\mathbf M}^j(j < m)$
(if $j \geq m$, since ${\mathbf M}^j =0$, $\bar{\partial _1} = 0$),
and

\begin{eqnarray}
 &\bar{\partial _2}(U_1 \otimes ... \otimes U_{n+1} \otimes
                    V_1 \otimes V_2 \otimes ... \otimes V_n ) \notag \\
=&U_1 \otimes ... \otimes U_{n+1} \otimes
  (-1)^{p_1}\partial V_1 \otimes V_2 \otimes ... \otimes V_n \notag \\
+&U_1 \otimes ... \otimes U_{n+1} \otimes
  V_1 \otimes (-1)^{p_2} \partial V_2 \otimes ...
  \otimes V_n ) \notag \\
 &+...+ \notag \notag \\
+&U_1 \otimes ... \otimes U_{n+1} \otimes
  V_1 \otimes V_2 \otimes... \otimes
  (-1)^{p_n} \partial V_n). \notag
\end{eqnarray}

Then each of $\bar{\partial _1}$ and $\bar{\partial _2}$ is a
differential of K{\" u}nneth products of $\{{\mathbf D}_{2n}^i\}$'s
and $\{{\mathbf C}^{S,j}\}$'s, respectively. Let
$\bar{\partial}=\bar{\partial _1}+\bar{\partial _2}$. When the
bijection $Q:\{{{\mathbf L}^{i}}\} \to \{{\mathbf
M}^i\otimes{\mathbf N}^j\}$ is given by $U_1 \otimes V_1 \otimes
...\otimes U_n \otimes V_n \otimes U_{n+1} \mapsto
 U_1 \otimes ... \otimes U_{n+1} \otimes V_1 \otimes ...\otimes V_n$,
immediately by the definitions of $d$ and $\bar{\partial}$, we have
\begin{eqnarray} \label{E:partial}
\bar{\partial} \circ Q = Q \circ d.
\end{eqnarray}
Hence $\bar{\partial}$ satisfies $\bar{\partial} \circ
\bar{\partial}=0$. So we have $0=\bar{\partial} \circ \bar{\partial}
     =\bar{\partial_1} \circ \bar{\partial_1}
     +\bar{\partial_1} \circ \bar{\partial_2}
     +\bar{\partial_2} \circ \bar{\partial_1}
     +\bar{\partial_2} \circ \bar{\partial_2}
     =\bar{\partial_1} \circ \bar{\partial_2}
     +\bar{\partial_2} \circ \bar{\partial_1}$,
therefore $\{{\mathbf M}^i \otimes {\mathbf N}^j\}$ is a double
complex. Also, by \eqref{E:partial}, $Q$ induces the cochain complex
isomorphism $(\{{\mathbf L}^i\},d) \to (\{{\mathbf M}^i \otimes
{\mathbf N}^j\},\bar{\partial})$. So by the isomorphism $Q \circ P$,
$(\{{\mathbf K}(gs(n;T_1,...T_{n+1};S_1,...,S_n))\}, \partial)$ and
$(\{{\mathbf M} \otimes {\mathbf N}\}, \bar{\partial})$ are
isomorphic. In order to calculate the cohomology of $(\{{\mathbf
K}^i(gs(n;T_1,...,T_{n+1};S_1,...,S_n))\})$, we can use the
cohomology groups $(\{{\mathbf M} \otimes {\mathbf N}\},
\bar{\partial})$. Let us use the spectral sequence starting with
$\bar{\partial _1}$.

Since $\bar{\partial _1}$ acts only on $\{{\mathbf M}^i\}$, we can
write $E_1^{i,j} = H_{\bar{\partial _1}}(\{{\mathbf M}^i \otimes
{\mathbf N}^j\}) = H^i_{\bar{\partial _1}}(\{{\mathbf M}\}) \otimes
{\mathbf N}^j$, and similarly, since $\bar{\partial _2}$ acts only
on $\{{\mathbf N}^j\}$, we can write $E_2^{i,j} = H_{\bar{\partial
_2}}(E_1) = H^i_{\bar{\partial _1}}(\{{\mathbf M}\}) \otimes
H^j_{\bar{\partial _2}}(\{{\mathbf N}\})$. Since $\bar{\partial _2}$
is a K{\" u}nneth product of ${\mathbf E}_i$'s, $H_{\bar{\partial
_2}}(\{{\mathbf N}\})$ is the K{\" u}nneth product $H({\mathbf E}_1)
\otimes ... \otimes H({\mathbf E}_n) =H(S_1) \otimes ... \otimes
H(S_n)$. But as we can see in the definition of $\bar{\partial _1}$,
it is not exactly the canonical differential of K{\" u}nneth
product, so the calculation of $H(\{{\mathbf M}\})$ takes some more
consideration.

When $i<m$, ${\mathbf M}^i =
\bigoplus _{i_1+...+i_{n+1}=i-n}
{\mathbf D}_1^{i_1} \otimes ... \otimes {\mathbf D}_{n+1}^{i_{n+1}}$
and $\bar{\partial _1}$ is a K{\" u}nneth differential.
So when $i<m$,
$H_{\bar{\partial _1}}^i(\{{\mathbf M}\})$ is a K{\" u}nneth product of
$H(\{{\mathbf D}_j\})$'s, and so it is 0,
since each $\{{\mathbf D}_j\}$ is acyclic.

When $i > m$, ${\mathbf M}^j =0$.
So when $i \geq m$, $\bar{\partial _1}=0$ on ${\mathbf M}^i$
and ${H^i _{{\bar {\partial _1}} } (\{{\mathbf M}\})}=0(i>m)$.

Let us calculate $H_{\bar{\partial _1}}^m(\{{\mathbf M}\}) =ker
\bar{\partial _1}|_{{\mathbf M}^m} /im  \bar{\partial _1}|_{{\mathbf
M}^{m-1}}$.

First, if $(i_1+1)+..+(i_{n+1}+1)+n < m$, ${\mathbf M}^m=0$. So,
$H_{\bar{\partial _1}}^m(\{{\mathbf M}\})=0$.

Suppose that $(i_1+1)+..+(i_{n+1}+1)+n \geq m$. Since the
clear-edged $m$-bonsai is a vector space over the field $k$, we just
need to calculate the dimension of $H^m$. We have
\begin{eqnarray} \label{E:dim1}
dim(H_{\bar{\partial _1}}^m(\{{\mathbf M}\}))
=dim(ker \bar{\partial _1}|_{{\mathbf M}^m})
-dim(im  \bar{\partial _1}|_{{\mathbf M}^{m-1}}).
\end{eqnarray}
\begin{eqnarray} \label{E:dim2}
dim(ker \bar{\partial _1}|_{{\mathbf M}^m} )
=dim({\mathbf M}^m)
\end{eqnarray}
and
\begin{eqnarray} \label{E:dim3}
dim(im \bar{\partial _1}|_{{\mathbf M}^{m-1}})
=dim({\mathbf M}^{m-1}) - dim(ker \bar{\partial _1}|_{{\mathbf M}^{m-1}}).
\end{eqnarray}
Since
\begin{eqnarray} \label{E:dim4}
H_{\bar{\partial _1}}^j(\{{\mathbf M}\})=0 \mbox{ when } j < m,
\end{eqnarray}
we have
\begin{eqnarray} \label{E:dim5}
\\
dim(ker \bar{\partial _1}|_{{\mathbf M}^{m-1}})
=dim(im \bar{\partial _1}|_{{\mathbf M}^{m-2}})
=dim({\mathbf M}^{m-2}) - dim(ker \bar{\partial _1}|_{{\mathbf M}^{m-2}}),
\notag
\end{eqnarray}
and so, by \eqref{E:dim1}-\eqref{E:dim5},
\begin{eqnarray}
 &dim(H_{\bar{\partial _1}}^m(\{{\mathbf M}\}))  \\
=&dim( {\mathbf M}^m )
 -(dim({\mathbf M}^{m-1}) - dim(ker \bar{\partial _1}|_{{\mathbf M}^{m-1}}))
 \notag \\
=&dim({\mathbf M}^m )  -dim({\mathbf M}^{m-1})
  +dim({\mathbf M}^{m-2}) - dim(ker \bar{\partial _1}|_{{\mathbf M}^{m-2}}).
 \notag
\end{eqnarray}
Continuing like this, we have
\begin{eqnarray}
 &dim(H_{\bar{\partial _1}}^m(\{{\mathbf M}\}))  \\
=&dim({\mathbf M}^m )  -dim({\mathbf M}^{m-1})
  +dim({\mathbf M}^{m-2}) - dim({\mathbf M}^{m-3}) +...
 \notag
\end{eqnarray}
and since ${\mathbf M}^i=0$ if $i < n+P$ where
\begin{eqnarray}
P=degT_1+...+degT_{n+1},
\end{eqnarray}
we have
\begin{eqnarray}
 dim(H_{\bar{\partial _1}}^m(\{{\mathbf M}\}))
=dim({\mathbf M}^m )  -dim({\mathbf M}^{m-1}) \\
+...+(-1)^{m-(n+P)}
  dim({\mathbf M}^{n+P}).
 \notag
\end{eqnarray}
Let $N$ be this number and let us calculate it. By the definition of
${\mathbf M}^i$, its dimension is that of
\begin{eqnarray}
\bigoplus_{i_1+...+i_{n+1}+n=i}
{\mathbf D}_1^{i_1} \otimes ...\otimes {\mathbf D}_n^{i_n}.
\end{eqnarray}
Every ${\mathbf D}_k^{i_k}$ is one-dimensional when
$p_k=i_k-degT_k$ is 0 or 1, and 0 otherwise.
So the above direct sum is,
\begin{eqnarray}
\bigoplus_{p_1+...+p_{n+1}+P+n=i}
{\mathbf D}_1^{p_1+degT_1}
\otimes ...\otimes {\mathbf D}_n^{p_n+degT_n}.
\end{eqnarray}
Hence, $dim({\mathbf M}^i)$
is the number of $(p_1,...,p_{n+1})$'s satisfying
$p_1+...+p_{n+1}+P+n=i$ and each $p_k$ is 0 or 1. Hence
\begin{eqnarray}
dim({\mathbf M}^i)=
\begin{pmatrix}
n+1 \\
i-P-n
\end{pmatrix}
\end{eqnarray}
and
\begin{eqnarray}
N=
\begin{pmatrix}
n+1 \\
m-P-n
\end{pmatrix}
-
\begin{pmatrix}
n+1 \\
(m-1)-P-n
\end{pmatrix}
+...+(-1)^{m-P-n}
\begin{pmatrix}
n+1\\
0
\end{pmatrix}
. \notag
\end{eqnarray}

Now we have

\begin{eqnarray}
E^{i,j}_2= \left\{ \begin{array}{ll} \bigoplus_{j_1+...+j_n=j} k^N
\otimes H^{j_1}(S_1) \otimes ...\otimes H^{j_n}(S_n)
  & \mbox{if } i=m  \\
0 & \mbox{otherwise.}
\end{array} \right. \notag
\end{eqnarray}

Since $E_2^{i,j}=0$ except when $j=m$, every ``knight's move map''
on $E^{i,j}$'s is trivial. So the spectral sequence collapses and we
have $H_{\bar{\partial}}^n = \bigoplus_{i+j=n} E^{i,j} =\bigoplus
_{j_1+...+j_n=n-m} k^N \otimes H^{j_1}(S_1) \otimes ...\otimes
H^{j_n}(S_n)$, and since
\begin{eqnarray}
 k^N \otimes H^{j_1}(S_1) \otimes ...\otimes H^{j_n}(S_n)
= (H^{j_1}(S_1) \otimes ...\otimes H^{j_n}(S_n))^{\bigoplus N},
\end{eqnarray}
we finally have

\begin{thm}
When $H^i(gs(n;T_1,...,T_{n+1};S_1,...,S_n))$ is the $i$-th
cohomology group of the thread ${\mathbf
K}^i(gs(n;T_1,...,T_{n+1};S_1,...,S_n))$, the $i$-th cohomology
group $H^i$ of clear-edged $m$-bonsai is $H^i = \underset{\mbox{$S$
is a grafting seedling}}{\bigoplus} H^i(S)$.

And, if $P=degT_1+...+degT_n <m-2n+1$, then
$H^{i}(gr(n;T_1,...,T_{n+1};S_1,...,S_n))=0$. Otherwise,
\begin{eqnarray}
H^{i}(gr(n;T_1,...,T_{n+1};S_1,...,S_n))
=\bigoplus_{j_1+...+j_n=n-m}
[H^{j_1}(S_1)\otimes ... \otimes H^{j_n}(S_n)]^{\oplus N} \notag
\end{eqnarray}
where
\begin{eqnarray}
N=
\begin{pmatrix}
n+1 \\
m-P-n
\end{pmatrix}
-
\begin{pmatrix}
n+1 \\
(m-1)-P-n
\end{pmatrix}
+...+(-1)^{m-(n+P)}
\begin{pmatrix}
n+1\\
0
\end{pmatrix}
.
 \notag
\end{eqnarray}

\end{thm}

\section{Differential and Appending of Clear-edged Bonsai}

Let us consider the relationship between the appending operation $*$
on the clear-edged $m$-bonsai Hopf algebra ${\mathcal H}_{c,m}$ and
the vertex appending differential $\partial$. We work mod 2 again.
First, the operation $T_1*T_2$ is the sum of all $m$-bonsais
obtained by connecting the root of $T_1$ and a vertex of $T_2$ with
an edge, as illustrated for 3-bonsai in Figure \ref{Fig:126}.

\begin{figure}[h]

\begin{picture}(200,80)(60,-70)

\put(-10,0){\circle*{2}}
\put(-10,0){\line(1,-2){10}}
\put(0,-20){\circle*{2}}
\put(-10,0){\line(-1,-2){10}}
\put(-20,-20){\circle*{2}}

\put(5,-10){*}

\put(25,0){\circle*{2}}
\put(25,0){\line(1,-2){10}}
\put(35,-20){\circle*{2}}
\put(25,0){\line(-1,-2){10}}
\put(15,-20){\circle*{2}}

\put(40,-10){=}

\put(95,0){\circle*{2}}
\put(95,0){\line(1,-2){10}}
\put(105,-20){\circle*{2}}
\put(95,0){\line(-1,-2){10}}
\put(85,-20){\circle*{2}}
\put(95,0){\line(-1,-1){20}}
\put(75,-20){\circle*{2}}

\put(75,-20){\circle*{2}}
\put(75,-20){\line(1,-2){10}}
\put(85,-40){\circle*{2}}
\put(75,-20){\line(-1,-2){10}}
\put(65,-40){\circle*{2}}

\put(110,-20){+}

\put(135,0){\circle*{2}}
\put(135,0){\line(1,-2){10}}
\put(145,-20){\circle*{2}}
\put(135,0){\line(-1,-2){10}}
\put(125,-20){\circle*{2}}
\put(135,0){\line(0,-1){20}}

\put(135,-20){\circle*{2}}
\put(135,-20){\line(1,-2){10}}
\put(145,-40){\circle*{2}}
\put(135,-20){\line(-1,-2){10}}
\put(125,-40){\circle*{2}}

\put(150,-20){+}

\put(175,0){\circle*{2}}
\put(175,0){\line(1,-2){10}}
\put(185,-20){\circle*{2}}
\put(175,0){\line(-1,-2){10}}
\put(165,-20){\circle*{2}}
\put(175,0){\line(1,-1){20}}

\put(195,-20){\circle*{2}}
\put(195,-20){\line(1,-2){10}}
\put(205,-40){\circle*{2}}
\put(195,-20){\line(-1,-2){10}}
\put(185,-40){\circle*{2}}

\put(210,-20){+}

\put(235,0){\circle*{2}}
\put(235,0){\line(1,-2){10}}
\put(245,-20){\circle*{2}}
\put(235,0){\line(-1,-2){10}}
\put(225,-20){\circle*{2}}

\put(225,-20){\line(0,-1){20}}

\put(225,-40){\circle*{2}}
\put(225,-40){\line(1,-2){10}}
\put(235,-60){\circle*{2}}
\put(225,-40){\line(-1,-2){10}}
\put(215,-60){\circle*{2}}

\put(250,-20){+}

\put(275,0){\circle*{2}}
\put(275,0){\line(1,-2){10}}
\put(285,-20){\circle*{2}}
\put(275,0){\line(-1,-2){10}}
\put(265,-20){\circle*{2}}

\put(285,-20){\line(0,-1){20}}

\put(285,-40){\circle*{2}}
\put(285,-40){\line(1,-2){10}}
\put(295,-60){\circle*{2}}
\put(285,-40){\line(-1,-2){10}}
\put(275,-60){\circle*{2}}

\end{picture}

\caption{}                 \label{Fig:126}

\end{figure}

In this section, we will show that $T_1 *_2 T_2 =0$ for every $T_1$
and $T_2$ as in Section 10.

Temporarily in this section, we use a differential $\bar{\partial}$
of corollas which is the same as $\partial$ except
$\bar{\partial}(v) = e$, where $v$ is the one-vertex bonsai and $e$
is the one-edge bonsai.

\subsection{Brief Table of Contents}

In this section, first we will describe $T_1 *_1 T_2$ for each of
the following cases when $T_2$ is a corolla, by dividing the cases
as follows;

i) When $\partial T_2 \neq 0$ and $deg(T_2) \leq m-2$

ii) When $\partial T_2 \neq 0$ and $deg(T_2) = m-1$

iii) When $\partial T_2 = 0$ and $deg(T_2) \leq m-2$

iv) When $\partial T_2 = 0$ and $deg(T_2) = m-1$

v) When $deg(T_2) = m$

Second, we will show that $T_1 *_2 T_2$ for each of the following
cases when $T_2$ is a corolla.

i) When $\partial T_2 \neq 0$ and $deg(T_2) \leq m-3$

ii) When $\partial T_2 \neq 0$ and $deg(T_2) = m-2$

iii) When $\partial T_2 \neq 0$ and $deg(T_2) = m-1$

iv) When $\partial T_2 = 0$ and $deg(T_2) \leq m-2$

v) When $\partial T_2 = 0$ and $deg(T_2) = m-1$

vi) When $deg(T_2) = m$

Finally, we show that $T_1 *_2 T_2$ for a general $T_2$.

\subsection{
$T_1 *_1 T_2$ when $T_2$ is a corolla}

For clear-edged bonsai, since it is not an operad,
we cannot use broomstick diagrams for graphical proof.
Let us look into $T_1 *_1 T_2$
by dividing the cases of $\partial T_2$ and $deg(T_2)$.

\subsubsection{
When $\partial T_2 \neq 0$ and $deg(T_2) \leq m-2$ }

If $T_1$ is not the one-vertex bonsai, then $(\partial T_1) * T_2$
is the sum of terms in $\partial(T_1 * T_2)$ obtained by attaching
edges to $T_1$. So $T_1 *_1 T_2 = (\partial T_1)*T_2 + T_1*(\partial
T_2) -
\partial(T_1*T_2)$ is (``$-$'' in this equation is in fact ``+'',
since we are working mod 2),

$T_1 * (\partial T_2)$

$+\sum$(a term in $\partial(T_1 * T_2)$ which is obtained by
attaching an edge to a vertex of $T_1 * T_2$ not in $T_1$ so that i)
and ii) of Definition \ref{Def:vadiff2} is satisfied)

Then the first summand is the sum of i) bonsais $A_1$ obtained by
connecting a non-root vertex of $\partial T_2$ and the root of $T_1$
with an edge, which is depicted as in the first equation of Figure
\ref{Fig:128} in 3-bonsai and ii) bonsais $A_2$ obtained by
connecting the root of $\partial T_2$ and the root of $T_1$ with an
edge, which is depicted as in the second equation in Figure
\ref{Fig:128} in 3-bonsai.

\begin{figure}[h]

\begin{picture}(200,180)(60,-170)

\put(-10,-10){$T_1=$}

\put(25,0){\circle*{2}}
\put(25,0){\line(1,-2){10}}
\put(35,-20){\circle*{2}}
\put(25,0){\line(-1,-2){10}}
\put(15,-20){\circle*{2}}

\put(50,-10){$T_2=$}

\put(85,0){\circle*{2}}
\put(85,0){\line(1,-2){10}}
\put(95,-20){\circle*{2}}
\put(85,0){\line(-1,-2){10}}
\put(75,-20){\circle*{2}}

\put(30,-60){$A_1=$}

\put(95,-30){\circle*{2}}
\put(95,-30){\line(1,-2){10}}
\put(105,-50){\circle*{2}}
\put(95,-30){\line(-1,-2){10}}
\put(85,-50){\circle*{2}}
\put(95,-30){\line(-1,-1){20}}
\put(75,-50){\circle*{2}}

\put(75,-50){\circle*{2}}
\put(75,-50){\line(0,-1){20}}

\put(75,-70){\circle*{2}}
\put(75,-70){\line(1,-2){10}}
\put(85,-90){\circle*{2}}
\put(75,-70){\line(-1,-2){10}}
\put(65,-90){\circle*{2}}

\put(110,-60){+}

\put(135,-30){\circle*{2}}
\put(135,-30){\line(1,-2){10}}
\put(145,-50){\circle*{2}}
\put(135,-30){\line(-1,-2){10}}
\put(125,-50){\circle*{2}}
\put(135,-30){\line(0,-1){20}}

\put(135,-50){\circle*{2}}
\put(135,-50){\line(0,-1){20}}

\put(135,-70){\circle*{2}}
\put(135,-70){\line(1,-2){10}}
\put(145,-90){\circle*{2}}
\put(135,-70){\line(-1,-2){10}}
\put(125,-90){\circle*{2}}

\put(150,-60){+}

\put(175,-30){\circle*{2}}
\put(175,-30){\line(1,-2){10}}
\put(185,-50){\circle*{2}}
\put(175,-30){\line(-1,-2){10}}
\put(165,-50){\circle*{2}}
\put(175,-30){\line(1,-1){20}}

\put(195,-50){\circle*{2}}
\put(195,-50){\line(0,-1){20}}

\put(195,-70){\circle*{2}}
\put(195,-70){\line(1,-2){10}}
\put(205,-90){\circle*{2}}
\put(195,-70){\line(-1,-2){10}}
\put(185,-90){\circle*{2}}

\put(30,-120){$A_2=$}

\put(95,-100){\circle*{2}}
\put(95,-100){\line(1,-2){10}}
\put(105,-120){\circle*{2}}
\put(95,-100){\line(-1,-2){10}}
\put(85,-120){\circle*{2}}
\put(95,-100){\line(-1,-1){20}}
\put(75,-120){\circle*{2}}
\put(95,-100){\line(0,-1){20}}
\put(95,-120){\circle*{2}}

\put(75,-120){\circle*{2}}
\put(75,-120){\line(1,-2){10}}
\put(85,-140){\circle*{2}}
\put(75,-120){\line(-1,-2){10}}
\put(65,-140){\circle*{2}}

\put(110,-120){+}

\put(135,-100){\circle*{2}}
\put(135,-100){\line(1,-2){10}}
\put(145,-120){\circle*{2}}
\put(135,-100){\line(-1,-2){10}}
\put(125,-120){\circle*{2}}
\put(135,-100){\line(0,-1){20}}
\put(135,-100){\line(1,-1){20}}
\put(155,-120){\circle*{2}}

\put(135,-120){\circle*{2}}
\put(135,-120){\line(1,-2){10}}
\put(145,-140){\circle*{2}}
\put(135,-120){\line(-1,-2){10}}
\put(125,-140){\circle*{2}}

\put(160,-120){+}

\put(185,-100){\circle*{2}}
\put(185,-100){\line(1,-2){10}}
\put(195,-120){\circle*{2}}
\put(185,-100){\line(-1,-2){10}}
\put(175,-120){\circle*{2}}
\put(185,-100){\line(1,-1){20}}
\put(185,-100){\line(0,-1){20}}
\put(185,-120){\circle*{2}}
\put(205,-120){\circle*{2}}

\put(195,-120){\circle*{2}}
\put(195,-120){\line(1,-2){10}}
\put(205,-140){\circle*{2}}
\put(195,-120){\line(-1,-2){10}}
\put(185,-140){\circle*{2}}

\put(220,-120){+}

\put(245,-100){\circle*{2}}
\put(245,-100){\line(1,-2){10}}
\put(255,-120){\circle*{2}}
\put(245,-100){\line(-1,-2){10}}
\put(235,-120){\circle*{2}}
\put(245,-100){\line(1,-1){20}}
\put(245,-100){\line(0,-1){20}}
\put(245,-120){\circle*{2}}

\put(265,-120){\circle*{2}}
\put(265,-120){\line(1,-2){10}}
\put(275,-140){\circle*{2}}
\put(265,-120){\line(-1,-2){10}}
\put(255,-140){\circle*{2}}

\end{picture}

\caption{}                 \label{Fig:128}

\end{figure}

The second summand is the sum of $A_3$, $A_4$ and $A_5$, where
$A_3$, $A_4$ and $A_5$ are as follows:

$A_3$ is the sum of bonsais obtained by connecting a vertex $v$
of $T_2$ and the root of $T_1$ with one edge and attaching an edge to $v$,
as in Figure \ref{Fig:129} in 3-bonsai.

\begin{figure}[h]

\begin{picture}(200,100)(60,-90)

\put(-10,-10){$T_1=$}

\put(25,0){\circle*{2}}
\put(25,0){\line(1,-2){10}}
\put(35,-20){\circle*{2}}
\put(25,0){\line(-1,-2){10}}
\put(15,-20){\circle*{2}}

\put(50,-10){$T_2=$}

\put(85,0){\circle*{2}}
\put(85,0){\line(1,-2){10}}
\put(95,-20){\circle*{2}}
\put(85,0){\line(-1,-2){10}}
\put(75,-20){\circle*{2}}

\put(20,-60){         $A_3$  =}

\put(85,-30){\circle*{2}}
\put(85,-30){\line(1,-2){10}}
\put(95,-50){\circle*{2}}
\put(85,-30){\line(-1,-2){10}}
\put(75,-50){\circle*{2}}

\put(75,-50){\circle*{2}}
\put(75,-50){\line(0,-1){20}}
\put(75,-50){\line(-1,-1){20}}
\put(55,-70){\circle*{2}}

\put(75,-70){\circle*{2}}
\put(75,-70){\line(1,-2){10}}
\put(85,-90){\circle*{2}}
\put(75,-70){\line(-1,-2){10}}
\put(65,-90){\circle*{2}}

\put(100,-60){+}

\put(135,-30){\circle*{2}}
\put(135,-30){\line(1,-2){10}}
\put(145,-50){\circle*{2}}
\put(135,-30){\line(-1,-2){10}}
\put(125,-50){\circle*{2}}

\put(125,-50){\circle*{2}}
\put(125,-50){\line(0,-1){20}}
\put(125,-50){\line(1,-1){20}}
\put(145,-70){\circle*{2}}

\put(125,-70){\circle*{2}}
\put(125,-70){\line(1,-2){10}}
\put(135,-90){\circle*{2}}
\put(125,-70){\line(-1,-2){10}}
\put(115,-90){\circle*{2}}

\put(150,-60){+}

\put(175,-30){\circle*{2}}
\put(175,-30){\line(1,-2){10}}
\put(185,-50){\circle*{2}}
\put(175,-30){\line(-1,-2){10}}
\put(165,-50){\circle*{2}}

\put(185,-50){\circle*{2}}
\put(185,-50){\line(0,-1){20}}
\put(185,-50){\line(-1,-1){20}}
\put(165,-70){\circle*{2}}

\put(185,-70){\circle*{2}}
\put(185,-70){\line(1,-2){10}}
\put(195,-90){\circle*{2}}
\put(185,-70){\line(-1,-2){10}}
\put(175,-90){\circle*{2}}

\put(200,-60){+}

\put(225,-30){\circle*{2}}
\put(225,-30){\line(1,-2){10}}
\put(235,-50){\circle*{2}}
\put(225,-30){\line(-1,-2){10}}
\put(215,-50){\circle*{2}}

\put(235,-50){\circle*{2}}
\put(235,-50){\line(0,-1){20}}
\put(235,-50){\line(1,-1){20}}
\put(255,-70){\circle*{2}}

\put(235,-70){\circle*{2}}
\put(235,-70){\line(1,-2){10}}
\put(245,-90){\circle*{2}}
\put(235,-70){\line(-1,-2){10}}
\put(225,-90){\circle*{2}}

\end{picture}

\caption{}                 \label{Fig:129}

\end{figure}

$A_4$: bonsais $B$ obtained as follows; suppose that  $T_2$ is
constructed by attaching the roots of corollae $X_1$ and $X_2$ so
that $X_1$ is on the left and $X_2$ is on the right. Then $B$ is
obtained by attaching the roots of $\bar{\partial} X_1$, $V$ and
$X_2$ from the left or $X_1$, $V$ and $\bar{\partial} X_2$ from the
left, where $V$ is obtained by attaching the root of $T_1$ to the
lower vertex of the one-edge clear-edged bonsai, illustrated in
Figure \ref{Fig:130}. In Figure \ref{Fig:130}, the first term of
$A_4$ is constructed by $\bar{\partial} X_1$, $V$ and $X_2$, and the
second term is constructed by $X_1$, $V$ and $\bar{\partial} X_2$.

\begin{figure}[h]

\begin{picture}(200,100)(60,-90)

\put(-10,-10){$T_1=$}

\put(25,0){\circle*{2}}
\put(25,0){\line(1,-2){10}}
\put(35,-20){\circle*{2}}
\put(25,0){\line(-1,-2){10}}
\put(15,-20){\circle*{2}}

\put(50,-10){$T_2=$}

\put(85,0){\circle*{2}}
\put(85,0){\line(1,-2){10}}
\put(95,-20){\circle*{2}}
\put(85,0){\line(-1,-2){10}}
\put(75,-20){\circle*{2}}

\put(110,-10){$X_1=$}

\put(145,0){\circle*{2}}
\put(145,0){\line(1,-2){10}}
\put(155,-20){\circle*{2}}
\put(145,0){\line(-1,-2){10}}
\put(135,-20){\circle*{2}}

\put(170,-10){$X_2=$}

\put(200,-7){\circle*{2}}

\put(230,-10){$V=$}

\put(265,0){\circle*{2}}
\put(265,0){\line(0,-1){20}}
\put(265,-20){\circle*{2}}
\put(265,-20){\line(1,-2){10}}
\put(275,-40){\circle*{2}}
\put(265,-20){\line(-1,-2){10}}
\put(255,-40){\circle*{2}}

\put(-20,-65){Terms in $A_4$}
\put(50,-65){=}
\put(-20,-75){corresponding}
\put(-20,-85){to $X_1$ and $X_2$}

\put(85,-50){\circle*{2}}
\put(85,-50){\line(1,-2){10}}
\put(95,-70){\circle*{2}}
\put(85,-50){\line(-1,-2){10}}
\put(75,-70){\circle*{2}}
\put(85,-50){\line(0,-1){20}}
\put(85,-70){\circle*{2}}
\put(85,-50){\line(-1,-1){20}}
\put(65,-70){\circle*{2}}

\put(95,-70){\circle*{2}}
\put(95,-70){\line(1,-2){10}}
\put(105,-90){\circle*{2}}
\put(95,-70){\line(-1,-2){10}}
\put(85,-90){\circle*{2}}

\put(105,-65){+}

\put(125,-50){\circle*{2}}
\put(125,-50){\line(1,-2){10}}
\put(135,-70){\circle*{2}}
\put(125,-50){\line(-1,-2){10}}
\put(115,-70){\circle*{2}}
\put(125,-50){\line(0,-1){20}}
\put(125,-70){\circle*{2}}
\put(125,-50){\line(1,-1){20}}
\put(145,-70){\circle*{2}}

\put(135,-70){\circle*{2}}
\put(135,-70){\line(1,-2){10}}
\put(145,-90){\circle*{2}}
\put(135,-70){\line(-1,-2){10}}
\put(125,-90){\circle*{2}}

\end{picture}

\caption{}                 \label{Fig:130}

\end{figure}

$A_5$: bonsais $B$ obtained as follows; Suppose $T_2$ is constructed
by attaching the roots of bonsais $Y_1$, $E$ and $Y_2$ from the
left, where $Y_1$ and $Y_2$ are corollae and $E$ is the one-edge
bonsai. Then $B$ is constructed by attaching the roots of
$\bar{\partial} Y_1$, $V$ and $Y_2$ from the left or $Y_1$, $V$ and
$\bar{\partial} Y_2$ from the left, where $V$ is obtained by
connecting the root of $T_1$ to the lower vertex of $E$ with one
edge. This is illustrated in Figure \ref{Fig:131}.

\begin{figure}[h]

\begin{picture}(200,100)(60,-90)

\put(-10,-10){$T_1=$}

\put(25,0){\circle*{2}}
\put(25,0){\line(1,-2){10}}
\put(35,-20){\circle*{2}}
\put(25,0){\line(-1,-2){10}}
\put(15,-20){\circle*{2}}

\put(50,-10){$T_2=$}

\put(85,0){\circle*{2}}
\put(85,0){\line(1,-2){10}}
\put(95,-20){\circle*{2}}
\put(85,0){\line(-1,-2){10}}
\put(75,-20){\circle*{2}}

\put(110,-10){$Y_1=$}

\put(145,0){\circle*{2}}
\put(145,0){\line(-1,-2){10}}
\put(135,-20){\circle*{2}}

\put(170,-10){$E=$}

\put(205,0){\circle*{2}}
\put(205,0){\line(1,-2){10}}
\put(215,-20){\circle*{2}}

\put(230,-10){$Y_2=$}

\put(260,-7){\circle*{2}}

\put(-20,-45){Terms in $A_5$}
\put(50,-45){=}
\put(-20,-55){corresponding}
\put(-20,-65){to $Y_1$ and $Y_2$}

\put(85,-30){\circle*{2}}
\put(85,-30){\line(1,-2){10}}
\put(95,-50){\circle*{2}}
\put(85,-30){\line(-1,-2){10}}
\put(75,-50){\circle*{2}}
\put(85,-30){\line(1,-1){20}}
\put(105,-50){\circle*{2}}
\put(95,-50){\line(0,-1){20}}

\put(95,-70){\circle*{2}}
\put(95,-70){\line(1,-2){10}}
\put(105,-90){\circle*{2}}
\put(95,-70){\line(-1,-2){10}}
\put(85,-90){\circle*{2}}

\end{picture}

\caption{}                 \label{Fig:131}

\end{figure}

If $T_1$ is the one-vertex bonsai, $A_2$ becomes 0, since it is
equal to $\partial \partial T_2$. And $A_3 = 0$, since it is twice a
multiple of the bonsais obtained by attaching the root of the
two-edge corolla to a tip of $T_2$ (note that we are working mod 2).
Also, $A_4= \partial \partial T_2 =0$. So $T_1 *_1 T_2 = A_1 + A_5$.

\subsubsection{
When $\partial T_2 \neq 0$ and $deg(T_2) = m-1$}

When $T_1$ is not the one-vertex bonsai: This is almost the same as
the previous case, but we cannot add more than one edge to the root
of $T_2$. So $T_1 *_1 T_2$ is the sum of the terms $A_1$, $A_3$ and
$A_5$.

When $T_1$ is the one-vertex bonsai: $A_3 = 0$, since it is twice a
multiple of the bonsais obtained by attaching the root of the
two-edge corolla to a tip of $T_2$(note that we are working mod 2).
So $T_1 *_1 T_2 = A_1 + A_5$.

\subsubsection{
When $\partial T_2 = 0$ and $deg(T_2) \leq m-2$}

When $T_1$ is not the one-vertex bonsai:
As in the first case,
$T_1 *_1 T_2 = (\partial T_1)*T_2 + T_1*(\partial T_2) - \partial(T_1*T_2)$
is

$T_1 * (\partial T_2)$

$+\sum$(a term in $\partial(T_1 * T_2)$ that is obtained by
attaching an edge to a vertex of $T_1 * T_2$ not in $T_1$ so that i)
and ii) of Definition \ref{Def:vadiff2} is satisfied)

Here $\partial T_2 = 0$. So we just have the latter summand in $T_1
*_1 T_2$. As in the first case again, we have $T_1 *_1 T_2 = A_3 +
A_4 + A_5$.

When $T_1$ is the one-vertex bonsai: $A_3 = 0$, since it is a twice
multiple of the bonsais obtained by attaching the root of the
two-edge corolla to a tip of $T_2$ (note that we are working mod 2).
Also, $A_4= \partial \partial T_2 =0$. Hence, $T_1 *_1 T_2 = A_5$.

\subsubsection{
When $\partial T_2 = 0$ and $deg(T_2) = m-1$}

When $T_1$ is not the one-vertex bonsai: This is almost the same as
the previous case, but we cannot add more than one edge to the root
of $T_2$. So $T_1 *_1 T_2$ is the sum of the terms $A_3$ and $A_5$.

When $T_1$ is the one-vertex bonsai: $A_3 = 0$, since it is a twice
multiple of the bonsais obtained by attaching the root of the
two-edge corolla to a tip of $T_2$(note that we are working mod 2).
Hence, $T_1 *_1 T_2 = A_5$.

\subsubsection{When $deg(T_2) = m$}

When $T_1$ is not the one-vertex bonsai: This is almost the same as
the previous case, but we cannot add any more edges to the root of
$T_2$. So $T_1 *_1 T_2$ is $A_3$.

When $T_1$ is the one-vertex bonsai: $A_3 = 0$, since it is twice a
multiple of the bonsais obtained by attaching the root of the
two-edge corolla to a tip of $T_2$(note that we are working in mod
2). Hence, $T_1 *_1 T_2 = 0$.

\subsection{$T_1 *_2 T_2$ when $T_2$ is a corolla}

For each case in the last subsection, let us show that $T_1 *_2 T_2 = 0$.

\subsubsection{
When $\partial T_2 \neq 0$ and $deg(T_2) \leq m-3$ }

In this case, we have $deg(\partial T_2 \leq m-2)$ and
$\partial (\partial T_2) = 0$.
And as in the first case of the last subsection,
$T_1 *_2 T_2$ is

$T_1 *_1 (\partial T_2)$

$+\sum$(a term in $\partial(T_1 *_1 T_2)$ obtained by attaching an
edge to a vertex of $T_1 * T_2$ not in $T_1$, so that i) and ii) of
Definition \ref{Def:vadiff2} are satisfied)

When $T_1$ is not the one-vertex bonsai: By the third case of the
last subsection, $T_1 *_1 (\partial T_2)$ is $A_3+A_4+A_5$, and by
the first case of the last subsection, the sum of the terms in
$\partial(T_1 *_1 T_2)$ obtained by attaching an edge to a vertex
not in $T_1$ is ${\hat \partial}(T_1 *_1 T_2)= {\hat
\partial}A_1+{\hat \partial}A_2+{\hat \partial}A_3 +{\hat
\partial}A_4+{\hat \partial}A_5$, where ${\hat \partial}X$, when
$X$ is a sum of terms in $T_1 *_1 T_2$, is the sum of the bonsais in
$\partial X$ obtained by attaching an edge to a vertex which is not
originally in $T_1$.

In order to show that $T_1 *_2 T_2 = T_1 *_1 \partial T_2 +{\hat
\partial}(T_1 *_1 T_2) = 0$ pictorially, let us introduce a new
picture convention. In Figure \ref{Fig:132}, where $T_2$ is a
corolla with 4 edges, each triangle represents a corolla (including
the one-vertex bonsai).

\begin{figure}[h]

\begin{picture}(200,50)(60,-40)

\put(-10,-10){$U_1=$}

\put(25,0){\circle*{2}}
\put(25,0){\line(1,-2){10}}
\put(35,-20){\circle*{2}}
\put(25,0){\line(0,-1){20}}
\put(25,-20){\circle*{2}}
\put(25,0){\line(-1,-2){10}}
\put(15,-20){\circle*{2}}

\put(50,-10){$U_2=$}

\put(85,0){\circle*{2}}
\put(85,0){\line(1,-2){10}}
\put(95,-20){\circle*{2}}
\put(85,0){\line(0,-1){20}}
\put(85,-20){\circle*{2}}
\put(85,0){\line(-1,-2){10}}
\put(75,-20){\circle*{2}}

\put(110,-10){$V_1=$}

\put(145,0){\circle*{2}}
\put(145,0){\line(1,-2){10}}
\put(155,-20){\circle*{2}}
\put(145,0){\line(0,-1){20}}
\put(145,-20){\circle*{2}}
\put(145,0){\line(-1,-2){10}}
\put(135,-20){\circle*{2}}

\put(170,-10){$V_2=$}

\put(200,0){\circle*{2}}
\put(200,0){\line(0,-1){20}}
\put(200,-20){\circle*{2}}

\put(-20,-45){$\partial T_2 =$}

\put(35,-30){\circle*{2}}
\put(35,-30){\line(-1,-1){20}}
\put(15,-50){\circle*{2}}
\put(35,-30){\line(1,-2){10}}
\put(45,-50){\circle*{2}}
\put(35,-30){\line(-1,-2){10}}
\put(25,-50){\circle*{2}}

\put(35,-30){\line(0,-1){20}}
\put(35,-50){\circle*{2}}

\put(35,-30){\line(1,-1){20}}
\put(55,-50){\circle*{2}}

\put(55,-45){=}

\put(85,-30){\circle*{2}}
\put(85,-30){\line(1,-2){10}}
\put(95,-50){\line(1,0){10}}
\put(75,-45){$U_1$}
\put(85,-30){\line(-1,-2){10}}
\put(75,-50){\line(1,0){10}}
\put(95,-45){$U_2$}
\put(85,-30){\line(0,-1){20}}
\put(85,-30){\line(1,-1){20}}

\put(110,-45){=}

\put(140,-30){\circle*{2}}
\put(120,-50){\line(1,0){10}}
\put(130,-45){$V_1$}
\put(140,-30){\line(1,-2){10}}
\put(140,-30){\line(-1,-2){10}}
\put(140,-30){\line(0,-1){20}}
\put(140,-50){\circle*{2}}
\put(140,-30){\line(-1,-1){20}}
\put(150,-50){\line(1,0){10}}
\put(150,-45){$V_2$}
\put(140,-30){\line(1,-1){20}}

\put(210,-10){$T_2=$}

\put(245,0){\circle*{2}}
\put(245,0){\line(1,-2){10}}
\put(255,-20){\line(1,0){10}}
\put(235,-15){$V_1$}
\put(245,0){\line(-1,-2){10}}
\put(235,-20){\line(1,0){10}}
\put(255,-15){$V_2$}
\put(245,0){\line(0,-1){20}}
\put(245,0){\line(1,-1){20}}

\end{picture}

\caption{}                 \label{Fig:132}

\end{figure}

With this pictorial convention, we can draw $T_1 *_1 \partial T_2 =
A_3 + A_4 + A_5$ as in Figure \ref{Fig:133}, like A+B+C+D+E+F.

\begin{figure}[h]

\begin{picture}(180,150)(120,-140)

\put(45,0){A}
\put(85,0){\circle*{2}}
\put(65,-20){\line(1,0){40}}
\put(80,-15){$\partial T_2$}
\put(85,-20){\line(0,-1){20}}
\put(85,-40){\circle*{2}}
\put(85,-20){\line(-1,-1){20}}
\put(65,-40){\circle*{2}}
\put(85,0){\line(-1,-1){20}}
\put(85,0){\line(1,-1){20}}

\put(85,-40){\line(-1,-1){20}}
\put(85,-40){\line(1,-1){20}}
\put(65,-60){\line(1,0){40}}
\put(80,-55){$T_1$}

\put(130,-35){+}

\put(145,0){B}
\put(185,0){\circle*{2}}
\put(165,-20){\line(1,0){40}}
\put(180,-15){$\partial T_2$}
\put(185,-20){\line(0,-1){20}}
\put(185,-40){\circle*{2}}
\put(185,-20){\line(1,-1){20}}
\put(205,-40){\circle*{2}}
\put(185,0){\line(-1,-1){20}}
\put(185,0){\line(1,-1){20}}

\put(185,-40){\line(-1,-1){20}}
\put(185,-40){\line(1,-1){20}}
\put(165,-60){\line(1,0){40}}
\put(180,-55){$T_1$}

\put(235,-35){+}

\put(245,0){C}
\put(285,0){\circle*{2}}
\put(245,-40){\line(1,0){20}}
\put(255,-35){${\bar \partial} U_1$}
\put(285,0){\line(1,-2){20}}
\put(285,0){\line(-1,-2){20}}
\put(285,0){\line(0,-1){40}}
\put(285,-40){\circle*{2}}
\put(285,0){\line(-1,-1){40}}
\put(305,-40){\line(1,0){20}}
\put(305,-35){$U_2$}
\put(285,0){\line(1,-1){40}}

\put(285,-40){\line(-1,-1){20}}
\put(285,-40){\line(1,-1){20}}
\put(265,-60){\line(1,0){40}}
\put(280,-55){$T_1$}

\put(30,-115){+}

\put(45,-80){D}
\put(85,-80){\circle*{2}}
\put(45,-120){\line(1,0){20}}
\put(55,-115){$U_1$}
\put(85,-80){\line(1,-2){20}}
\put(85,-80){\line(-1,-2){20}}
\put(85,-80){\line(0,-1){40}}
\put(85,-120){\circle*{2}}
\put(85,-80){\line(-1,-1){40}}
\put(105,-120){\line(1,0){20}}
\put(105,-115){${\bar \partial} U_2$}
\put(85,-80){\line(1,-1){40}}

\put(85,-120){\line(-1,-1){20}}
\put(85,-120){\line(1,-1){20}}
\put(65,-140){\line(1,0){40}}
\put(80,-135){$T_1$}

\put(135,-115){+}

\put(145,-80){E}
\put(185,-80){\circle*{2}}
\put(145,-120){\line(1,0){20}}
\put(155,-115){${\bar \partial} V_1$}
\put(185,-80){\line(1,-2){20}}
\put(185,-80){\line(-1,-2){20}}
\put(185,-80){\line(0,-1){40}}
\put(185,-100){\circle*{4}}
\put(185,-80){\line(-1,-1){40}}
\put(205,-120){\line(1,0){20}}
\put(205,-115){$V_2$}
\put(185,-80){\line(1,-1){40}}

\put(185,-120){\line(-1,-1){20}}
\put(185,-120){\line(1,-1){20}}
\put(165,-140){\line(1,0){40}}
\put(180,-135){$T_1$}

\put(230,-115){+}

\put(245,-80){F}
\put(285,-80){\circle*{2}}
\put(245,-120){\line(1,0){20}}
\put(255,-115){$V_1$}
\put(285,-80){\line(1,-2){20}}
\put(285,-80){\line(-1,-2){20}}
\put(285,-80){\line(0,-1){40}}
\put(285,-100){\circle*{4}}
\put(285,-80){\line(-1,-1){40}}
\put(305,-120){\line(1,0){20}}
\put(305,-115){${\bar \partial} V_2$}
\put(285,-80){\line(1,-1){40}}

\put(285,-120){\line(-1,-1){20}}
\put(285,-120){\line(1,-1){20}}
\put(265,-140){\line(1,0){40}}
\put(280,-135){$T_1$}

\end{picture}

\caption{}                 \label{Fig:133}

\end{figure}

In ${\hat \partial}(T_1 *_1 T_2)$,
${\hat \partial}A_1$ can be drawn as in Figure \ref{Fig:134}.

\begin{figure}[h]

\begin{picture}(180,80)(150,-70)

\put(45,0){1a}
\put(85,0){\circle*{2}}
\put(45,-40){\line(1,0){20}}
\put(55,-35){${\bar \partial} V_1$}
\put(85,0){\line(1,-2){20}}
\put(85,0){\line(-1,-2){20}}
\put(85,0){\line(0,-1){60}}
\put(85,-40){\circle*{4}}
\put(85,0){\line(-1,-1){40}}
\put(105,-40){\line(1,0){20}}
\put(105,-35){$V_2$}
\put(85,0){\line(1,-1){40}}

\put(85,-60){\line(-1,-1){20}}
\put(85,-60){\line(1,-1){20}}
\put(65,-80){\line(1,0){40}}
\put(80,-75){$T_1$}

\put(135,-35){+}

\put(145,0){1b}
\put(185,0){\circle*{2}}
\put(145,-40){\line(1,0){20}}
\put(155,-35){$V_1$}
\put(185,0){\line(1,-2){20}}
\put(185,0){\line(-1,-2){20}}
\put(185,0){\line(0,-1){60}}
\put(185,-40){\circle*{4}}
\put(185,0){\line(-1,-1){40}}
\put(205,-40){\line(1,0){20}}
\put(205,-35){${\bar \partial} V_2$}
\put(185,0){\line(1,-1){40}}

\put(185,-60){\line(-1,-1){20}}
\put(185,-60){\line(1,-1){20}}
\put(165,-80){\line(1,0){40}}
\put(180,-75){$T_1$}

\put(230,-35){+}

\put(245,0){1c}
\put(285,0){\circle*{2}}
\put(245,-40){\line(1,0){20}}
\put(255,-35){$V_1$}
\put(285,0){\line(1,-2){20}}
\put(285,0){\line(-1,-2){20}}
\put(285,0){\line(0,-1){60}}
\put(285,-40){\circle*{4}}
\put(285,0){\line(-1,-1){40}}
\put(305,-40){\line(1,0){20}}
\put(305,-35){$V_2$}
\put(285,0){\line(1,-1){40}}

\put(285,-40){\line(-1,-1){20}}
\put(265,-60){\circle*{2}}

\put(285,-60){\line(-1,-1){20}}
\put(285,-60){\line(1,-1){20}}
\put(265,-80){\line(1,0){40}}
\put(280,-75){$T_1$}

\put(330,-35){+}

\put(345,0){1d}
\put(385,0){\circle*{2}}
\put(345,-40){\line(1,0){20}}
\put(355,-35){$V_1$}
\put(385,0){\line(1,-2){20}}
\put(385,0){\line(-1,-2){20}}
\put(385,0){\line(0,-1){60}}
\put(385,-40){\circle*{4}}
\put(385,0){\line(-1,-1){40}}
\put(405,-40){\line(1,0){20}}
\put(405,-35){$V_2$}
\put(385,0){\line(1,-1){40}}

\put(385,-40){\line(1,-1){20}}
\put(405,-60){\circle*{2}}

\put(385,-60){\line(-1,-1){20}}
\put(385,-60){\line(1,-1){20}}
\put(365,-80){\line(1,0){40}}
\put(380,-75){$T_1$}

\end{picture}

\caption{}                 \label{Fig:134}

\end{figure}

${\hat \partial}A_2$ is as in Figure \ref{Fig:135}.

\begin{figure}[h]

\begin{picture}(180,80)(60,-70)

\put(45,0){2a}
\put(85,0){\circle*{2}}
\put(45,-40){\line(1,0){20}}
\put(55,-35){${\bar \partial} U_1$}
\put(85,0){\line(1,-2){20}}
\put(85,0){\line(-1,-2){20}}
\put(85,0){\line(0,-1){40}}
\put(85,-40){\circle*{2}}
\put(85,0){\line(-1,-1){40}}
\put(105,-40){\line(1,0){20}}
\put(105,-35){$U_2$}
\put(85,0){\line(1,-1){40}}

\put(85,-40){\line(-1,-1){20}}
\put(85,-40){\line(1,-1){20}}
\put(65,-60){\line(1,0){40}}
\put(80,-55){$T_1$}

\put(135,-35){+}

\put(145,0){2b}
\put(185,0){\circle*{2}}
\put(145,-40){\line(1,0){20}}
\put(155,-35){$U_1$}
\put(185,0){\line(1,-2){20}}
\put(185,0){\line(-1,-2){20}}
\put(185,0){\line(0,-1){40}}
\put(185,0){\line(-1,-1){40}}
\put(205,-40){\line(1,0){20}}
\put(205,-35){${\bar\partial} U_2$}
\put(185,0){\line(1,-1){40}}

\put(185,-40){\line(-1,-1){20}}
\put(185,-40){\line(1,-1){20}}
\put(165,-60){\line(1,0){40}}
\put(180,-55){$T_1$}

\end{picture}

\caption{}                 \label{Fig:135}

\end{figure}

${\hat \partial}A_3$ is as in Figure \ref{Fig:136},
where $W_1$ and $W_2$ are as in Figure \ref{Fig:137}.

\begin{figure}[h]

\begin{picture}(180,180)(150,-170)

\put(45,0){3a}
\put(85,0){\circle*{2}}
\put(65,-20){\line(1,0){40}}
\put(80,-15){$T_2$}
\put(85,-20){\line(0,-1){20}}
\put(85,-40){\circle*{2}}
\put(85,-20){\line(1,-1){20}}
\put(105,-40){\circle*{2}}
\put(85,-20){\line(-1,-1){20}}
\put(65,-40){\circle*{2}}
\put(85,0){\line(-1,-1){20}}
\put(85,0){\line(1,-1){20}}

\put(85,-40){\line(-1,-1){20}}
\put(85,-40){\line(1,-1){20}}
\put(65,-60){\line(1,0){40}}
\put(80,-55){$T_1$}

\put(130,-35){+}

\put(145,0){3b}
\put(185,0){\circle*{2}}
\put(165,-20){\line(1,0){40}}
\put(180,-15){$T_2$}
\put(185,-20){\line(0,-1){20}}
\put(185,-40){\circle*{2}}
\put(185,-20){\line(1,-1){20}}
\put(205,-40){\circle*{2}}
\put(185,-20){\line(-1,-1){20}}
\put(165,-40){\circle*{2}}
\put(185,0){\line(-1,-1){20}}
\put(185,0){\line(1,-1){20}}

\put(185,-40){\line(-1,-1){20}}
\put(185,-40){\line(1,-1){20}}
\put(165,-60){\line(1,0){40}}
\put(180,-55){$T_1$}

\put(230,-35){+}

\put(245,0){3c}
\put(285,0){\circle*{2}}
\put(245,-40){\line(1,0){20}}
\put(255,-35){${\bar \partial} W_1$}
\put(285,0){\line(1,-2){20}}
\put(285,0){\line(-1,-2){20}}
\put(285,0){\line(0,-1){60}}
\put(285,-40){\circle*{4}}
\put(285,0){\line(-1,-1){40}}
\put(305,-40){\line(1,0){20}}
\put(305,-35){$W_2$}
\put(285,0){\line(1,-1){40}}

\put(285,-40){\line(-1,-1){20}}
\put(265,-60){\circle*{2}}

\put(285,-60){\line(-1,-1){20}}
\put(285,-60){\line(1,-1){20}}
\put(265,-80){\line(1,0){40}}
\put(280,-75){$T_1$}

\put(330,-35){+}

\put(345,0){3d}
\put(385,0){\circle*{2}}
\put(345,-40){\line(1,0){20}}
\put(355,-35){$W_1$}
\put(385,0){\line(1,-2){20}}
\put(385,0){\line(-1,-2){20}}
\put(385,0){\line(0,-1){60}}
\put(385,-40){\circle*{4}}
\put(385,0){\line(-1,-1){40}}
\put(405,-40){\line(1,0){20}}
\put(405,-35){${\bar \partial} W_2$}
\put(385,0){\line(1,-1){40}}

\put(385,-40){\line(-1,-1){20}}
\put(365,-60){\circle*{2}}

\put(385,-60){\line(-1,-1){20}}
\put(385,-60){\line(1,-1){20}}
\put(365,-80){\line(1,0){40}}
\put(380,-75){$T_1$}

\put(130,-125){+}

\put(145,-90){3e}
\put(185,-90){\circle*{2}}
\put(145,-130){\line(1,0){20}}
\put(155,-125){${\bar \partial} W_1$}
\put(185,-90){\line(1,-2){20}}
\put(185,-90){\line(-1,-2){20}}
\put(185,-90){\line(0,-1){60}}
\put(185,-130){\circle*{4}}
\put(185,-90){\line(-1,-1){40}}
\put(205,-130){\line(1,0){20}}
\put(205,-125){$W_2$}
\put(185,-90){\line(1,-1){40}}

\put(185,-130){\line(1,-1){20}}
\put(205,-150){\circle*{2}}

\put(185,-150){\line(-1,-1){20}}
\put(185,-150){\line(1,-1){20}}
\put(165,-170){\line(1,0){40}}
\put(180,-165){$T_1$}

\put(230,-125){+}

\put(245,-90){3f}
\put(285,-90){\circle*{2}}
\put(245,-130){\line(1,0){20}}
\put(255,-125){$W_1$}
\put(285,-90){\line(1,-2){20}}
\put(285,-90){\line(-1,-2){20}}
\put(285,-90){\line(0,-1){60}}
\put(285,-130){\circle*{4}}
\put(285,-90){\line(-1,-1){40}}
\put(305,-130){\line(1,0){20}}
\put(305,-125){${\bar \partial} W_2$}
\put(285,-90){\line(1,-1){40}}

\put(285,-130){\line(1,-1){20}}
\put(305,-150){\circle*{2}}

\put(285,-150){\line(-1,-1){20}}
\put(285,-150){\line(1,-1){20}}
\put(265,-170){\line(1,0){40}}
\put(280,-165){$T_1$}

\end{picture}

\caption{}                 \label{Fig:136}

\end{figure}

\begin{figure}[h]

\begin{picture}(200,30)(-10,-50)

\put(45,-45){$T_2=$}

\put(85,-30){\circle*{2}}
\put(65,-50){\line(1,0){10}}
\put(65,-45){$W_1$}
\put(85,-30){\line(1,-2){10}}
\put(85,-30){\line(-1,-2){10}}
\put(85,-30){\line(0,-1){20}}
\put(85,-50){\circle*{2}}
\put(85,-30){\line(-1,-1){20}}
\put(95,-50){\line(1,0){10}}
\put(95,-45){$W_2$}
\put(85,-30){\line(1,-1){20}}

\end{picture}

\caption{}                 \label{Fig:137}

\end{figure}

${\hat \partial}A_4$ can be drawn as in Figure \ref{Fig:138}.

\begin{figure}[h]

\begin{picture}(180,60)(150,-50)

\put(45,0){4a}
\put(85,0){\circle*{2}}
\put(45,-40){\line(1,0){20}}
\put(55,-35){${\bar \partial}{\bar \partial} V_1$}
\put(85,0){\line(1,-2){20}}
\put(85,0){\line(-1,-2){20}}
\put(85,0){\line(0,-1){40}}
\put(85,0){\line(-1,-1){40}}
\put(105,-40){\line(1,0){20}}
\put(105,-35){$V_2$}
\put(85,0){\line(1,-1){40}}

\put(85,-40){\line(-1,-1){20}}
\put(85,-40){\line(1,-1){20}}
\put(65,-60){\line(1,0){40}}
\put(80,-55){$T_1$}

\put(135,-35){+}

\put(145,0){4b}
\put(185,0){\circle*{2}}
\put(145,-40){\line(1,0){20}}
\put(155,-35){${\bar \partial} V_1$}
\put(185,0){\line(1,-2){20}}
\put(185,0){\line(-1,-2){20}}
\put(185,0){\line(0,-1){40}}
\put(185,0){\line(-1,-1){40}}
\put(205,-40){\line(1,0){20}}
\put(205,-35){${\bar \partial} V_2$}
\put(185,0){\line(1,-1){40}}

\put(185,-40){\line(-1,-1){20}}
\put(185,-40){\line(1,-1){20}}
\put(165,-60){\line(1,0){40}}
\put(180,-55){$T_1$}

\put(230,-35){+}

\put(245,0){4c}
\put(285,0){\circle*{2}}
\put(245,-40){\line(1,0){20}}
\put(255,-35){${\bar \partial} V_1$}
\put(285,0){\line(1,-2){20}}
\put(285,0){\line(-1,-2){20}}
\put(285,0){\line(0,-1){40}}
\put(285,0){\line(-1,-1){40}}
\put(305,-40){\line(1,0){20}}
\put(305,-35){${\bar \partial} V_2$}
\put(285,0){\line(1,-1){40}}

\put(285,-40){\line(-1,-1){20}}
\put(285,-40){\line(1,-1){20}}
\put(265,-60){\line(1,0){40}}
\put(280,-55){$T_1$}

\put(330,-35){+}

\put(345,0){4d}
\put(385,0){\circle*{2}}
\put(345,-40){\line(1,0){20}}
\put(355,-35){$V_1$}
\put(385,0){\line(1,-2){20}}
\put(385,0){\line(-1,-2){20}}
\put(385,0){\line(0,-1){60}}
\put(385,-40){\circle*{4}}
\put(385,0){\line(-1,-1){40}}
\put(405,-40){\line(1,0){20}}
\put(405,-35){${\bar \partial}{\bar \partial} V_2$}
\put(385,0){\line(1,-1){40}}

\put(385,-40){\line(-1,-1){20}}
\put(385,-40){\line(1,-1){20}}
\put(365,-60){\line(1,0){40}}
\put(380,-55){$T_1$}

\end{picture}

\caption{}                 \label{Fig:138}

\end{figure}

And ${\hat \partial}A_5$ can be drawn as in Figure \ref{Fig:139}.

\begin{figure}[h]

\begin{picture}(180,180)(150,-170)

\put(45,0){5a}
\put(85,0){\circle*{2}}
\put(45,-40){\line(1,0){20}}
\put(55,-35){${\bar \partial}{\bar \partial} W_1$}
\put(85,0){\line(1,-2){20}}
\put(85,0){\line(-1,-2){20}}
\put(85,0){\line(0,-1){60}}
\put(85,-40){\circle*{4}}
\put(85,0){\line(-1,-1){40}}
\put(105,-40){\line(1,0){20}}
\put(105,-35){$W_2$}
\put(85,0){\line(1,-1){40}}

\put(85,-60){\line(-1,-1){20}}
\put(85,-60){\line(1,-1){20}}
\put(65,-80){\line(1,0){40}}
\put(80,-75){$T_1$}

\put(135,-35){+}

\put(145,0){5b}
\put(185,0){\circle*{2}}
\put(145,-40){\line(1,0){20}}
\put(155,-35){${\bar \partial} W_1$}
\put(185,0){\line(1,-2){20}}
\put(185,0){\line(-1,-2){20}}
\put(185,0){\line(0,-1){60}}
\put(185,-40){\circle*{4}}
\put(185,0){\line(-1,-1){40}}
\put(205,-40){\line(1,0){20}}
\put(205,-35){${\bar \partial} W_2$}
\put(185,0){\line(1,-1){40}}

\put(185,-60){\line(-1,-1){20}}
\put(185,-60){\line(1,-1){20}}
\put(165,-80){\line(1,0){40}}
\put(180,-75){$T_1$}

\put(230,-35){+}

\put(245,0){5c}
\put(285,0){\circle*{2}}
\put(245,-40){\line(1,0){20}}
\put(255,-35){${\bar \partial} W_1$}
\put(285,0){\line(1,-2){20}}
\put(285,0){\line(-1,-2){20}}
\put(285,0){\line(0,-1){60}}
\put(285,-40){\circle*{4}}
\put(285,0){\line(-1,-1){40}}
\put(305,-40){\line(1,0){20}}
\put(305,-35){${\bar \partial} W_2$}
\put(285,0){\line(1,-1){40}}

\put(285,-60){\line(-1,-1){20}}
\put(285,-60){\line(1,-1){20}}
\put(265,-80){\line(1,0){40}}
\put(280,-75){$T_1$}

\put(330,-35){+}

\put(345,0){5d}
\put(385,0){\circle*{2}}
\put(345,-40){\line(1,0){20}}
\put(355,-35){$W_1$}
\put(385,0){\line(1,-2){20}}
\put(385,0){\line(-1,-2){20}}
\put(385,0){\line(0,-1){60}}
\put(385,-40){\circle*{4}}
\put(385,0){\line(-1,-1){40}}
\put(405,-40){\line(1,0){20}}
\put(405,-35){${\bar \partial}{\bar \partial} W_2$}
\put(385,0){\line(1,-1){40}}

\put(385,-60){\line(-1,-1){20}}
\put(385,-60){\line(1,-1){20}}
\put(365,-80){\line(1,0){40}}
\put(380,-75){$T_1$}

\put(35,-135){+}

\put(45,-100){5e}
\put(85,-100){\circle*{2}}
\put(45,-140){\line(1,0){20}}
\put(55,-135){${\bar \partial} W_1$}
\put(85,-100){\line(1,-2){20}}
\put(85,-100){\line(-1,-2){20}}
\put(85,-100){\line(0,-1){60}}
\put(85,-140){\circle*{4}}
\put(85,-100){\line(-1,-1){40}}
\put(105,-140){\line(1,0){20}}
\put(105,-135){$W_2$}
\put(85,-100){\line(1,-1){40}}

\put(85,-140){\line(-1,-1){20}}
\put(65,-160){\circle*{2}}

\put(85,-160){\line(-1,-1){20}}
\put(85,-160){\line(1,-1){20}}
\put(65,-180){\line(1,0){40}}
\put(80,-175){$T_1$}

\put(135,-135){+}

\put(145,-100){5f}
\put(185,-100){\circle*{2}}
\put(145,-140){\line(1,0){20}}
\put(155,-135){$W_1$}
\put(185,-100){\line(1,-2){20}}
\put(185,-100){\line(-1,-2){20}}
\put(185,-100){\line(0,-1){60}}
\put(185,-140){\circle*{4}}
\put(185,-100){\line(-1,-1){40}}
\put(205,-140){\line(1,0){20}}
\put(205,-135){${\bar \partial} W_2$}
\put(185,-100){\line(1,-1){40}}

\put(185,-140){\line(-1,-1){20}}
\put(165,-160){\circle*{2}}

\put(185,-160){\line(-1,-1){20}}
\put(185,-160){\line(1,-1){20}}
\put(165,-180){\line(1,0){40}}
\put(180,-175){$T_1$}

\put(230,-135){+}

\put(245,-100){5g}
\put(285,-100){\circle*{2}}
\put(245,-140){\line(1,0){20}}
\put(255,-135){${\bar \partial} W_1$}
\put(285,-100){\line(1,-2){20}}
\put(285,-100){\line(-1,-2){20}}
\put(285,-100){\line(0,-1){60}}
\put(285,-140){\circle*{4}}
\put(285,-100){\line(-1,-1){40}}
\put(305,-140){\line(1,0){20}}
\put(305,-135){$W_2$}
\put(285,-100){\line(1,-1){40}}

\put(285,-140){\line(1,-1){20}}
\put(305,-160){\circle*{2}}

\put(285,-160){\line(-1,-1){20}}
\put(285,-160){\line(1,-1){20}}
\put(265,-180){\line(1,0){40}}
\put(280,-175){$T_1$}

\put(330,-135){+}

\put(345,-100){5h}
\put(385,-100){\circle*{2}}
\put(345,-140){\line(1,0){20}}
\put(355,-135){$W_1$}
\put(385,-100){\line(1,-2){20}}
\put(385,-100){\line(-1,-2){20}}
\put(385,-100){\line(0,-1){60}}
\put(385,-140){\circle*{4}}
\put(385,-100){\line(-1,-1){40}}
\put(405,-140){\line(1,0){20}}
\put(405,-135){${\bar \partial} W_2$}
\put(385,-100){\line(1,-1){40}}

\put(385,-140){\line(1,-1){20}}
\put(405,-160){\circle*{2}}

\put(385,-160){\line(-1,-1){20}}
\put(385,-160){\line(1,-1){20}}
\put(365,-180){\line(1,0){40}}
\put(380,-175){$T_1$}

\end{picture}

\caption{}                 \label{Fig:139}

\end{figure}

Then we can get the pairs which are canceled as follows;
A and 1c, B and 1d, C and 2a, D and 2b, E and 1a, F and 1b, 3a and 3b,
3c and 5e, 3d and 5f, 3e and 5g, 3f and 5h, 4b and 4c, 5b and 5c.
4a, 4d, 5a and 5d are 0, because they have ${\bar \partial} ^2$ in it.
(Since 4b and 4c cancel each other, ${\hat \partial}A_4 = 0$.)

And every bonsai in Figures \ref{Fig:133} - \ref{Fig:139} ais in one
of those pairs. So $T_1 *_2 T_2 =0$.

When $T_1$ is the one-vertex bonsai: Since $T_1 *_1 T_2 = A_3 + A_5$,
it is the sum of A, B, E and F.
And since $T_1 *_1 T_2 = A_1 + A_5$,
we have ${\hat \partial}(T_1 *_1 T_2)
={\hat \partial}A_1 + {\hat \partial}A_5$, and it is the sum of the bonsais
in Figure \ref{Fig:134} and Figure \ref{Fig:139}.
So as above, we have $T_1 *_1 T_2 = 0$.

\subsubsection{
When $\partial T_2 \neq 0$ and $deg(T_2) = m-2$ }
When $T_1$ is not the one-vertex bonsai: We have $\partial(\partial T_2) = 0$
and $deg(\partial T_2) = m-1$. So we have $T_1 *_1 \partial T_2 = A_3 + A_5$
and ${\hat \partial}(T_1 *_1 T_2) = {\hat \partial}(A_1 + A_2 +A_3+A_4 +A_5)$.
So the $T_1 *_2 T_2$ is the sum of the bonsais A, B, E and F of Figure
\ref{Fig:133} and the bonsais in Figures \ref{Fig:134} - \ref{Fig:136}
and \ref{Fig:138} - \ref{Fig:139}(2a and 2b are 0,
since the valences at the roots are $m+1$).
So as in the above case, $T_1 *_2 T_2 = 0$.

When $T_1$ is not the one-vertex bonsai: $T_1 *_1 \partial T_2 = A_5$
and ${\hat \partial}(T_1 *_1 T_2) = {\hat \partial}(A_1 + A_5)$.
So $T_1 *_2 T_2$ is the sum of the bonsais E and F of the Figure \ref{Fig:133}
and the bonsais in Figures \ref{Fig:134} and \ref{Fig:139}.

\subsubsection{
When $\partial T_2 \neq 0$ and $deg(T_2) = m-1$}

In this case and others, we will just write down what $T_1 *_1
\partial T_2$ and ${\hat \partial}(T_1 *_1 T_2)$ are. In each case,
as above, the bonsais in $T_1 *_2 T_2$ all cancel out.

In this case, $\partial (\partial T_2) =0$ and $deg(\partial T_2)=m$.

When $T_1$ is not the one-vertex bonsai:
$T_1 *_1 \partial T_2 = A_3 + A_5$ and
${\hat \partial}(T_1 *_1 T_2)={\hat \partial}(A_1 + A_3 +A_5) $.

When $T_1$ ia the one-vertex bonsai: $T_1 *_1 \partial T_2 = A_5$
and ${\hat \partial}(T_1 *_1 T_2) = {\hat \partial}(A_1 +A_5)$.

\subsubsection{
When $\partial T_2 = 0$ and $deg(T_2) \leq m-2$}

In this case, $\partial T_2 =0$ and $deg(T_2) \leq m-2$.

When $T_1$ is not the one-vertex bonsai:
$T_1 *_1 \partial T_2 = 0$ and
${\hat \partial}(T_1 *_1 T_2)={\hat \partial}(A_3 + A_4 +A_5) $.

When $T_1$ ia the one-vertex bonsai: $T_1 *_1 \partial T_2 = 0$
and ${\hat \partial}(T_1 *_1 T_2) = {\hat \partial}(A_5)$.

\subsubsection{
When $\partial T_2 = 0$ and $deg(T_2) = m-1$}

In this case, $\partial T_2 =0$ and $deg(T_2)=m-1$.

When $T_1$ is not the one-vertex bonsai:
$T_1 *_1 \partial T_2 = A_3 + A_5$ and
${\hat \partial}(T_1 *_1 T_2)={\hat \partial}(A_3 +A_5) $.

When $T_1$ ia the one-vertex bonsai: $T_1 *_1 \partial T_2 = 0$
and ${\hat \partial}(T_1 *_1 T_2) = {\hat \partial}(A_5)$.

\subsubsection{When $deg(T_2) = m$}

In this case, $\partial T_2 =0$ and $deg(T_2) = m $.

When $T_1$ is not the one-vertex bonsai:
$T_1 *_1 \partial T_2 = 0$ and
${\hat \partial}(T_1 *_1 T_2)={\hat \partial}(A_3 +A_5) $.

When $T_1$ ia the one-vertex bonsai: $T_1 *_1 \partial T_2 = 0$
and ${\hat \partial}(T_1 *_1 T_2) = 0$.

\subsection{$T_1 *_2 T_2$ for general $T_2$}

Now let us consider the case where $T_2$ is a general $m$-bonsai,
not only a corolla. First, let us give an expression of a general
$m$-bonsai as a concatenation of corollas. When we have a general
clear-edged $m$-bonsai $T$ as in Figure \ref{Fig:140}, we first
enumerate the non-tip vertices of $T$ in traversing order (cf.
Section 7), as in Figure \ref{Fig:141}, and for each non-tip vertex
numbered $i$, denote the corollas attached to that vertex as
$T_{i1},T_{i2},...,T_{im_i}$ from the left, and denote the given
bonsai as $T(T_{11},...,T_{1m_1};...;T_{k1},...,T_{km_k})$, where
$k$ is the number of non-tip vertices of $T$.

\begin{figure}[h]

\begin{picture}(160,90)(-70,-80)

\put(0,0){\circle*{4}}

\put(0,0){\line(-1,-1){40}}
\put(-40,-40){\circle*{4}}
\put(0,0){\line(0,-1){40}}
\put(0,-40){\circle*{4}}
\put(0,0){\line(1,-1){40}}
\put(40,-40){\circle*{4}}

\put(0,0){\line(-1,-2){10}}
\put(-10,-20){\circle*{4}}
\put(0,0){\line(1,-2){10}}
\put(10,-20){\circle*{4}}
\put(0,0){\line(2,-1){20}}
\put(20,-10){\circle*{4}}

\put(-40,-40){\circle*{4}}
\put(-40,-40){\line(-1,-1){20}}
\put(-60,-60){\circle*{4}}
\put(-40,-40){\line(0,-1){20}}
\put(-40,-60){\circle*{4}}
\put(-40,-40){\line(1,-1){20}}
\put(-20,-60){\circle*{4}}

\put(0,-40){\line(0,-1){20}}
\put(0,-60){\circle*{4}}

\put(40,-40){\circle*{4}}
\put(40,-40){\line(-1,-1){20}}
\put(20,-60){\circle*{4}}
\put(40,-40){\line(0,-1){20}}
\put(40,-60){\circle*{4}}
\put(40,-40){\line(1,-1){20}}
\put(60,-60){\circle*{4}}

\put(20,-60){\circle*{4}}
\put(20,-60){\line(-1,-1){20}}
\put(0,-80){\circle*{4}}
\put(20,-60){\line(0,-1){20}}
\put(20,-80){\circle*{4}}

\put(40,-60){\line(0,-1){20}}
\put(40,-80){\circle*{4}}

\end{picture}

\caption{}              \label{Fig:140}

\end{figure}

\begin{figure}[h]

\begin{picture}(160,160)(-70,-150)

\put(0,3){1}
\put(0,0){\circle*{4}}

\put(0,0){\line(-1,-1){40}}
\put(-40,-40){\circle*{4}}
\put(0,0){\line(0,-1){40}}
\put(0,-40){\circle*{4}}
\put(0,0){\line(1,-1){40}}
\put(40,-40){\circle*{4}}

\put(0,0){\line(-1,-2){10}}
\put(-10,-20){\circle*{4}}
\put(0,0){\line(1,-2){10}}
\put(10,-20){\circle*{4}}
\put(0,0){\line(2,-1){20}}
\put(20,-10){\circle*{4}}

\put(-40,-37){2}

\put(-40,-40){\circle*{4}}
\put(-40,-40){\line(-1,-1){20}}
\put(-60,-60){\circle*{4}}
\put(-40,-40){\line(0,-1){20}}
\put(-40,-60){\circle*{4}}
\put(-40,-40){\line(1,-1){20}}
\put(-20,-60){\circle*{4}}

\put(2,-37){3}

\put(0,-40){\line(0,-1){20}}
\put(0,-60){\circle*{4}}

\put(42,-37){4}

\put(40,-40){\circle*{4}}
\put(40,-40){\line(-1,-1){20}}
\put(20,-60){\circle*{4}}
\put(40,-40){\line(0,-1){20}}
\put(40,-60){\circle*{4}}
\put(40,-40){\line(1,-1){20}}
\put(60,-60){\circle*{4}}

\put(22,-57){5}

\put(20,-60){\circle*{4}}
\put(20,-60){\line(-1,-1){20}}
\put(0,-80){\circle*{4}}
\put(20,-60){\line(0,-1){20}}
\put(20,-80){\circle*{4}}

\put(42,-57){6}

\put(40,-60){\line(0,-1){20}}
\put(40,-80){\circle*{4}}

\put(-100,-100){$T_{11}=$}
\put(-70,-100){\circle*{4}}

\put(-50,-100){$T_{12}=T_{13}=T_{14}=$}
\put(40,-90){\circle*{4}}
\put(40,-90){\line(0,-1){20}}
\put(40,-110){\circle*{4}}

\put(60,-100){$T_{21}=$}

\put(100,-90){\circle*{4}}
\put(100,-90){\line(-1,-1){20}}
\put(80,-110){\circle*{4}}
\put(100,-90){\line(0,-1){20}}
\put(100,-110){\circle*{4}}
\put(100,-90){\line(1,-1){20}}
\put(120,-110){\circle*{4}}

\put(130,-100){$T_{31}=$}

\put(160,-90){\circle*{4}}
\put(160,-90){\line(0,-1){20}}
\put(160,-110){\circle*{4}}

\put(-100,-130){$T_{41}=T_{42}=$}
\put(-30,-130){\circle*{4}}

\put(10,-130){$T_{43}=$}
\put(40,-120){\circle*{4}}
\put(40,-120){\line(0,-1){20}}
\put(40,-140){\circle*{4}}

\put(60,-130){$T_{51}=$}

\put(100,-120){\circle*{4}}
\put(100,-120){\line(-1,-1){20}}
\put(80,-140){\circle*{4}}
\put(100,-120){\line(0,-1){20}}
\put(100,-140){\circle*{4}}

\put(130,-130){$T_{61}=$}

\put(160,-120){\circle*{4}}
\put(160,-120){\line(0,-1){20}}
\put(160,-140){\circle*{4}}

\end{picture}

\caption{}              \label{Fig:141}

\end{figure}

In Figure \ref{Fig:141}, the given bonsai $T$ is
\begin{eqnarray}
T(T_{11},T_{12},T_{13},T_{14};T_{21};T_{31};
   T_{41},T_{42},T_{43};T_{51};T_{61}).
\end{eqnarray}
Now let $T=T(T_{11},...,T_{1m_1};...;T_{k1},...,T_{km_k})$ and just
for convenience of algebra, let us donote $T$ as
$T(T_1,T_2,...,T_m)$, where $T_1 = T_{11}, T_2 = T_{12},...
T_k=T_{km_k}$. Then we have $\partial T = \sum_{i}T(T_1,...,{\bar
\partial}T_i,...,T_m)$ and $S * T = \sum_{i}T(T_1,...,S *
T_i,...,T_m)$. So we have, by the definitions of ${\hat \partial}$,
${\bar \partial}$ and $*$,

\begin{eqnarray}
 &S *_1 T = S* \partial T + {\hat \partial}(S * T)  \\
=& \sum_i         T(T_1,...,S * {\bar \partial}T_i,...,T_m)         \notag \\
 &+\sum_{i \neq j}T(T_1,...,{\bar \partial}T_i,...,S * T_j,...,T_m) \notag \\
 &+\sum_i         T(T_1,...,{\hat \partial}(S * T_i),...,T-m)       \notag \\
 &+\sum_{i \neq j}T(T_1,...,{\bar \partial}T_i,...,S * T_j,...,T_m) \notag \\
=& \sum_i         T(T_1,...,S * {\bar \partial}T_i,...,T_m)         \notag \\
 &+\sum_i         T(T_1,...,{\hat \partial}(S * T_i),...,T-m)       \notag \\
=& \sum_i         T(T_1,...,S *_1 T_i,...,T_m).                        \notag
\end{eqnarray}

Similarly, we have

\begin{eqnarray}
 &S *_2 T = S*_1 \partial T + {\hat \partial}(S *_1 T)  \\
=& \sum_i         T(T_1,...,S *_1 {\bar \partial}T_i,...,T_m)         \notag \\
 &+\sum_{i \neq j}T(T_1,...,{\bar \partial}T_i,...,S *_1 T_j,...,T_m) \notag \\
 &+\sum_i         T(T_1,...,{\hat \partial}(S *_1 T_i),...,T-m)       \notag \\
 &+\sum_{i \neq j}T(T_1,...,{\bar \partial}T_i,...,S *_1 T_j,...,T_m) \notag \\
=& \sum_i         T(T_1,...,S *_1 {\bar \partial}T_i,...,T_m)         \notag \\
 &+\sum_i         T(T_1,...,{\hat \partial}(S *_1 T_i),...,T-m)       \notag \\
=& \sum_i         T(T_1,...,S *_2 T_i,...,T_m)                        \notag \\
=& 0.
\end{eqnarray}

So we have

\begin{thm}
Mod 2, for any clear-edged $m$-bonsai $T_1$ and $T_2$, we have $T_1
*_2 T_2 = 0$.
\end{thm}

\section{Further Direction}

In the next paper of the author, we will investigate a
generalization of the $m$-bonsai Hopf algebra, its differentials and
cohomology groups. Also we will investigate some possibility of
generalization of the appending operation $*$.


\begin{thebibliography}{Abcd 99}

\bibitem[BK]{BK} C. Bergbauer, D. Kreimer, The Hopf algebra of rooted trees in
Epstein-Glaser renormalization, hep-th/0403207

\bibitem[CK]{CK} A. Connes and D. Kreimer, Hopf Algebras, Renormalization
and Noncommutative Geometry, {\em Eur. Phys. J. C7(1999) 697-708},
hep-th/9808042

\bibitem[CK2]{CK2} A. Connes and D. Kreimer, Insertion and Elimination:
the doubly infinite Lie algebra and Feynman graphs, hep-th/0201157

\bibitem[G]{G} M. Gerstenhaber,
The Cohomology Structure of an Associative Ring,
{\em The Annals of Mathematics, Second Series,
Vol. 78, Issue 2 (Sep.,1963) 267-288}

\bibitem[Ha]{Ha} D. Harrivel, Planar Binary Trees and Perturbative Calculus of
Observables in Classical Field Theory, AP/0410050

\bibitem[Kr]{Kr}  D. Kreimer, On the Hopf algebra structure of
pertubative quantum field theories,{\em Adv. Theor. Math. Phys. 2(1998)
303-334 }, q-alg/9707029

\bibitem[KS]{KS} H. Kajiura, J. Stasheff,  Homotopy algebras inspired by
classical open-closed string field theory, QA/0410291

\bibitem[MSS]{MSS} M. Markl, S. Shnider and J. Stasheff,
{\em Operads in Algebra, Topology and Physics}, American
Mathematical Society, 2002

\bibitem[PS]{PS} L. Pachter, B. Sturmfels, The Mathematics of Phylogenomics,
math.ST/0409132

\bibitem[S]{S} R. P. Stanley, {\em Enumerative Combinatorics Volume 2},
Cambridge University Press, 1999

\end{thebibliography}
\end{document}